\documentclass[a4paper,fleqn,usenatbib]{mnras}

\usepackage[T1]{fontenc}
\usepackage{ae,aecompl}

\usepackage{amsmath}
\usepackage{grffile}

\def\HI{{\mathrm{H}_{\rm I}}}
\def\H2{{\mathrm{H}_2}}
\def\GIZMO{{\sc GIZMO}}
\def\MUSIC{{\sc MUSIC}}
\def\MassiveFIRE{{\sc MassiveFIRE}}
\def\sSFR{{\rm{sSFR}}}

\usepackage{url}
\usepackage{times}

\usepackage{graphicx}

\usepackage{amssymb,amsmath}

\usepackage{color}

\let\oldhref\href
\renewcommand{\href}[2]{\oldhref{#1}{\hbox{#2}}}

\newcommand{\acknowledgements}{\begin{small}\section*{Acknowledgments}\end{small}}
\newcommand{\tableline}{\hline}

\title[MassiveFIRE: Massive Galaxies at the Cosmic Noon]{Colours, Star formation Rates, and Environments of Star forming and Quiescent Galaxies at the Cosmic Noon}

\author[R. Feldmann et al.]{
Robert Feldmann$^{1, 2}$\thanks{E-mail: feldmann@physik.uzh.ch (RF)},
Eliot Quataert$^{2}$, 
Philip F. Hopkins$^{3}$,\newauthor\phantom{x}
Claude-Andr\'{e} Faucher-Gigu\`{e}re$^{4}$,
and Du\v{s}an Kere\v{s}$^{5}$
\\
$^{1}$Institute for Computational Science, University of Zurich, Zurich CH-8057, Switzerland\\
$^{2}$Department of Astronomy \& Theoretical Astrophysics Center, University of California, Berkeley, CA 94720-3411, USA\\
$^{3}$TAPIR 350-17, California Institute of Technology, Pasadena, CA 91125, USA\\
$^{4}$Department of Physics and Astronomy and CIERA, Northwestern University, Evanston, IL 60208, USA\\
$^{5}$Center for Astrophysics and Space Sciences, University of California, San Diego, CA 92093, USA
}

\begin{document} 

\maketitle 

\begin{abstract}
We analyse the star formation rates (SFRs), colours, and dust extinctions of galaxies in massive ($10^{12.5}-10^{13.5}$ $M_\odot$) haloes at $z\sim{}2$ in high-resolution, cosmological zoom-in simulations as part of the \emph{Feedback in Realistic Environments} (FIRE) project. The simulations do not model feedback from active galactic nuclei (AGN) but reproduce well the observed relations between stellar and halo mass and between stellar mass and SFR. 
About half (a third) of the simulated massive galaxies (massive central galaxies) at $z\sim{}2$ have broad-band colours classifying them as `quiescent', and the fraction of quiescent centrals is steeply decreasing towards higher redshift, in agreement with observations. The progenitors of $z\sim{}2$ quiescent central galaxies are, on average, more massive, have lower specific SFRs, and reside in more massive haloes than the progenitors of similarly massive star forming centrals.
The simulations further predict a morphological mix of galaxies that includes disk-dominated, irregular, and early-type galaxies. However, our simulations do not reproduce the reddest of the quiescent galaxies observed at $z\sim{}2$. We also do not find evidence for a colour bimodality, but are limited by our modest sample size.
In our simulations, the star formation activity of central galaxies of moderate mass ($M_{\rm star}\sim{}10^{10}-10^{11}$ $M_\odot$) is affected by a combination of two distinct physical processes. Outflows powered by stellar feedback result in a short-lived ($<100$ Myr), but almost complete, suppression of star formation activity after which many galaxies quickly recover and continue to form stars at normal rates. In addition, galaxies residing in slowly growing haloes tend to experience a moderate reduction of their SFRs (`cosmological starvation'). The relative importance of these processes and AGN feedback is uncertain and will be explored in future work.
\end{abstract}

\begin{keywords}
galaxies: formation -- galaxies: evolution -- galaxies: high-redshift -- galaxies: star formation -- galaxies:haloes
\end{keywords}

\section{Introduction}
\label{sect:introduction}

Galaxies in the nearby Universe have a bimodal distribution of their colours (e.g., \citealt{Strateva2001, Blanton2003, Baldry2004}) and stellar age indicators \citep{Kauffmann2003, Thomas2010b}. The two peaks correspond to star forming galaxies (or `blue cloud' or `main sequence' galaxies) and quiescent galaxies (or `red sequence' galaxies), respectively. The former include galaxies with blue colours, young stellar ages, and high sSFRs (e.g., \citealt{Brinchmann2004b}). Most local galaxies of moderately low to intermediate mass ($M_{\rm star}\sim{}10^{8}-10^{10}$ $M_\odot$) belong to this class (e.g. \citealt{Yang2009, Peng2010}). The second class of galaxies, quiescent galaxies, form stars at low rates, have red optical colours, light-averaged stellar ages well in excess of 3 Gyr \citep{Thomas2010b}, and dominate the massive end of the stellar mass function.

Star forming and quiescent galaxies are observed across cosmic time out to $z\sim{}3$ (e.g., \citealt{Marchesini2014, Tomczak2014, Man2016, Martis2016}), or perhaps even $z\sim{}4$ \citep{Muzzin2013a, Straatman2014a}. Interestingly, while local massive galaxies are typically quiescent, star forming galaxies make up a large fraction ($\sim{}30-50$\%) of the massive galaxy population at $z\sim{}1.5-2.5$ \citep{Brammer2011f, Muzzin2013a, Tomczak2014, Martis2016}. However, these high redshift galaxies differ from their low redshift counterparts in many ways. For instance, star forming galaxies at $z\gtrsim{}1$ have an order of magnitude higher SFR per unit stellar mass than local galaxies (e.g. 
\citealt{Noeske2007d, Daddi2007a, Magdis2010, Reddy2012a, Pannella2015}), are more gas-rich \citep{Daddi2010a, Tacconi2013a, Saintonge2013a, Scoville2016, Aravena2016, Seko2016}, and are smaller at fixed stellar mass (e.g., \citealt{Williams2010c, VanderWel2014}). Quiescent galaxies at high redshift are even more compact than their low redshift counterparts (e.g., \citealt{Daddi2005, Trujillo2006, Bezanson2009, Damjanov2011, Conselice2014, VanderWel2014, VanDokkum2015}), signifying that, at late times, such galaxies either grow substantially (e.g, via minor merging; \citealt{Naab2007, Naab2009, Bezanson2009, Feldmann2010a, Oser2010, Wellons2016}) or that large galaxies preferentially join the quiescent population (e.g., \citealt{Valentinuzzi2010, Carollo2013}).

The presence of massive, quiescent galaxies at early times poses a serious challenge for theoretical models. At present, there is no consensus on how star formation in these galaxies is suppressed, although various suggestions abound in the literature. The responsible processes may differ depending on galaxy type (e.g., central vs. satellite galaxies), environment (e.g., field vs. cluster), and redshift, thus complicating the analysis. 

A popular quenching scenario ties the formation of quiescent galaxies to feedback from active galactic nuclei perhaps connected to galaxy mergers (AGN; \citealt{DiMatteo2005b, Springel2005, Hopkins2006b, Hopkins2008a, Choi2014b}). Large box cosmological simulations with different models for AGN feedback are able to reproduce the massive end of the observed stellar mass function and result in a large fraction of massive, quiescent galaxies at $z\lesssim{}1$ \citep{Vogelsberger2014, Schaye2015, Trayford2016}. Although promising, the modelling of AGN feedback in cosmological simulations is clearly still in its beginning. Furthermore, various problems matching the observations remain (e.g., \citealt{Ragone-Figueroa2013b, Hahn2015}) and evidence for a causal connection between morphological change, AGN feedback, and quenching is difficult to interpret (e.g., \citealt{Alexander2012, Fabian2012, Kormendy2013}).

Aside from AGN feedback a number of other physical processes have been suggested to play a role in reducing the star formation activity of massive galaxies. These include the formation of a hot gas atmosphere around massive galaxies (`halo quenching', \citealt{Keres2005, Dekel2006, Cattaneo2006b, Gabor2012a}), with long-term maintenance possibly aided by AGN feedback coupled to the hot gas \citep{Dekel2006}; a reduced accretion rate onto haloes (`cosmological starvation', \citealt{Feldmann2015, Feldmann2016}); or star formation and stellar feedback driven gas depletion (e.g., `wet compaction', \citealt{Dekel2014, Tacchella2016}). In addition, processes specific to dense environments may affect predominantly satellite galaxies, such as ram-pressure stripping of the cold ISM \citep{Gunn1972, Abadi1999, Cen2014a, Bahe2015a}, stripping of the tenuous hot atmosphere (`starvation'; \citealt{Larson1980, Balogh2000, Kawata2008, McCarthy2008, VanDenBosch2008, Feldmann2011e, Bahe2013}), and frequent tidal interactions with members of galaxy groups and clusters (`harassment'; \citealt{Farouki1981, Moore1996}).

Numerical simulations are a valuable tool to decipher how star forming galaxies transform into quiescent ones (e.g. see \citealt{Somerville2015} for a recent review). In simulations, the evolution of individual galaxies and their properties can be directly traced across cosmic time. Furthermore, it is straightforward to analyse the properties of separate galaxy populations, e.g., centrals and satellites. Of course, simulations come with their own challenges. In particular, the large dynamic range of the physical processes involved in galaxy formation requires the use of sub-grid models and introduces systematic uncertainties. It is thus important to cross-validate simulations against available observations. Fortunately, state-of-the-art hydrodynamical simulations reproduce many properties of observed galaxies thanks to efforts in modelling stellar feedback more accurately and increasing numerical resolution (e.g., \citealt{Feldmann2010a, Guedes2011, Hopkins2011c, Agertz2013, Hopkins2014, Ceverino2015}).

Theoretical models predict that star formation in galaxies (especially at higher redshift) is largely driven by the accretion of gas from the intergalactic medium (e.g., \citealt{Keres2005, Dekel2009a, Bouche2010, Dave2012a, Feldmann2013, Lilly2013c, SanchezAlmeida2014b, Schaye2015}). Stellar feedback has been shown to change the shape of the star formation histories of galaxies (shifting star formation to later times and increasing variability, e.g., \citealt{Schaye2010, Hopkins2014}) and to disrupt the tight relation between galaxy growth and halo growth \citep{VandeVoort2011a, Faucher-Giguere2011}. However, the qualitative trend remains that - all else being equal - faster growing haloes harbour faster growing galaxies \citep{Rodriguez-Puebla2016, Feldmann2016}. A link between star formation and halo growth would offer a physical explanation for the empirical correlation between galaxy colours and halo formation time assumed in age-matching \citep{Hearin2013e}. However, it does not by itself explain the bimodal sSFR distribution.

In this paper, we analyse properties related to star formation, including sSFRs, stellar masses, and galaxy colours from a new sample of massive galaxies (\MassiveFIRE{}; \citealt{Feldmann2016}) as part of the Feedback In Realistic Environments (FIRE) project\footnote{See the FIRE project web site at: \url{http://fire.northwestern.edu}.} \citep{Hopkins2014}. In particular, we aim to quantify the differences between star forming and quiescent galaxies of moderate mass ($\sim{}10^{10}-10^{11}$ $M_\odot$) at the cosmic noon ($z\sim{}2$). The simulations presented in this paper are run with the same code, adopt the identical physics modelling, and use a similar resolution as the simulations reported in \cite{Hopkins2014} and \cite{Faucher-Giguere2015} but target galaxies residing in massive haloes at $z\sim{}2$.

The FIRE approach of modelling galactic star formation and stellar feedback has been validated against observational data in a number of publications. The tests include the stellar-to-halo-mass relation (SHMR; \citealt{Hopkins2014}) and the stellar mass -- metallicity relation \citep{Ma2016} of $\leq{}L_*$ galaxies at $z=0$ and their high redshift progenitors. Other checks include the $\HI$ content of galaxy haloes at both low and high redshifts  \citep{Faucher-Giguere2015, Faucher-Giguere2016, Hafen2016}, the properties of sub-millimetre galaxies \citep{Narayanan2015}, the formation of giant star forming clumps in $z\sim{}1-2$ gas rich disks \citep{Oklopcic2016}, and the X-ray and S-Z signals arising from the hot gas haloes surrounding massive galaxies \citep{VandeVoort2016}. We note that none of the simulations used in these works include AGN feedback. This is an intentional choice since it enables us to identify which aspects of the observed galaxy population can be understood using stellar physics alone.

The outline of this paper is as follows.  The details of the simulation suite, including the set-up, sample selection, and post-processing are introduced in \S\ref{sect:sample}. The subsequent sections discuss the colours of massive galaxies at $z\sim{}1.7-4$ and the different pathways to suppressing star formation (\S\ref{sect:GalColors}), the importance of dust extinction (\S\ref{sect:Dust}), the stellar mass -- SFR relation (\S\ref{sect:SFRrel}), the role of the environment (\S\ref{sect:environ}), and the relation between stellar and halo mass (\S\ref{sect:SHMR}). We summarise our findings and conclude in the final section.

\section{Methodology}
\label{sect:sample}

\subsection{Setup of the simulations}
\label{sect:Setup}

We created initial conditions for \MassiveFIRE{} with the multi-scale initial conditions tool \MUSIC{} \citep{Hahn2011}. Our sample is drawn from a (144 Mpc)$^3$ comoving box with $\Omega_{\rm matter}=0.2821$, $\Omega_\Lambda=1-\Omega_{\rm matter}=0.7179$, $\sigma_8=0.817$, $n_{\rm s}=0.9646$, and $H_0=69.7$ km s$^{-1}$ Mpc$^{-1}$ \citep{Hinshaw2013}.

\begin{figure*}
\begin{tabular}{cc}
\includegraphics[width=85mm]{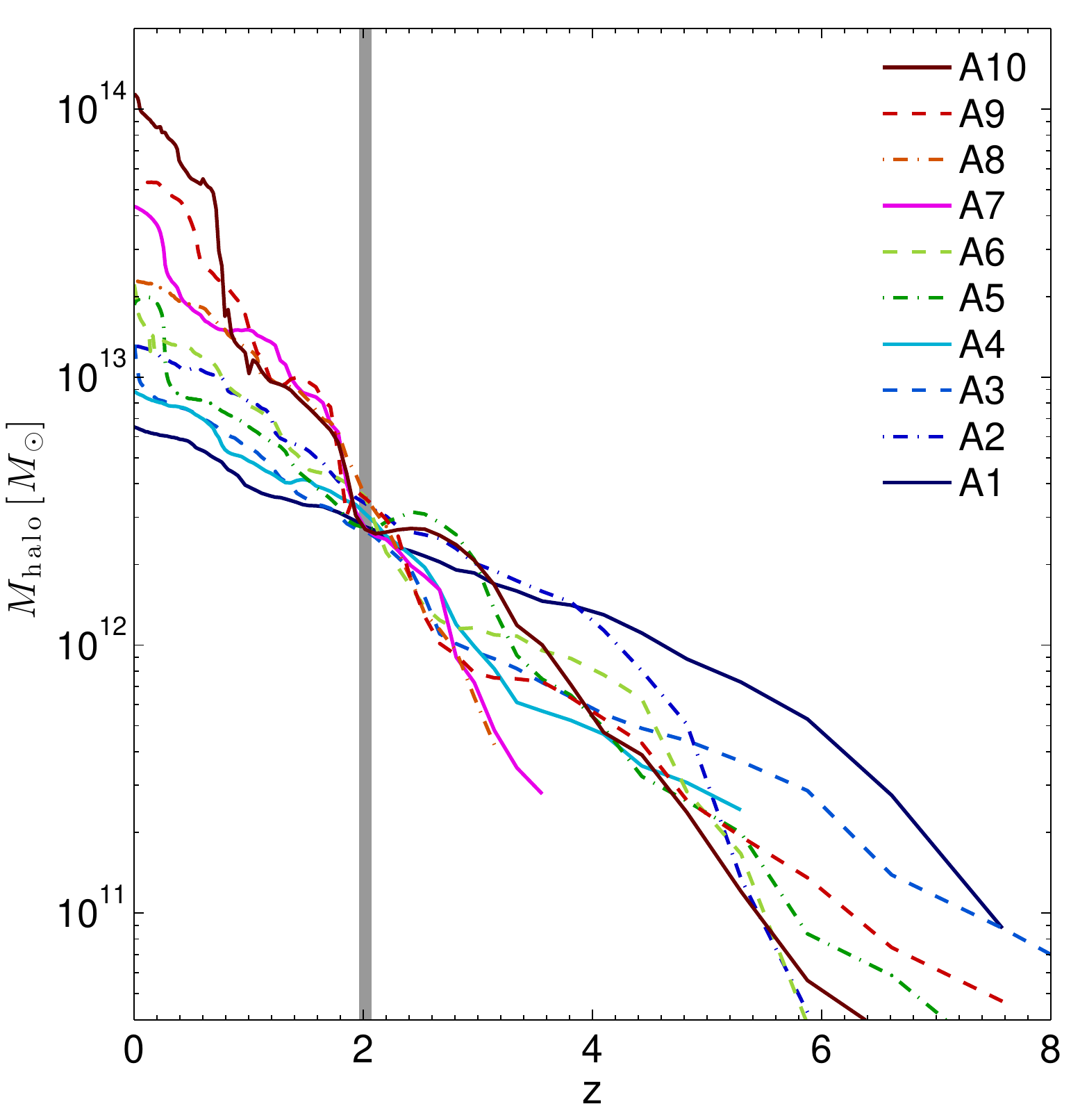} &
\includegraphics[width=85mm]{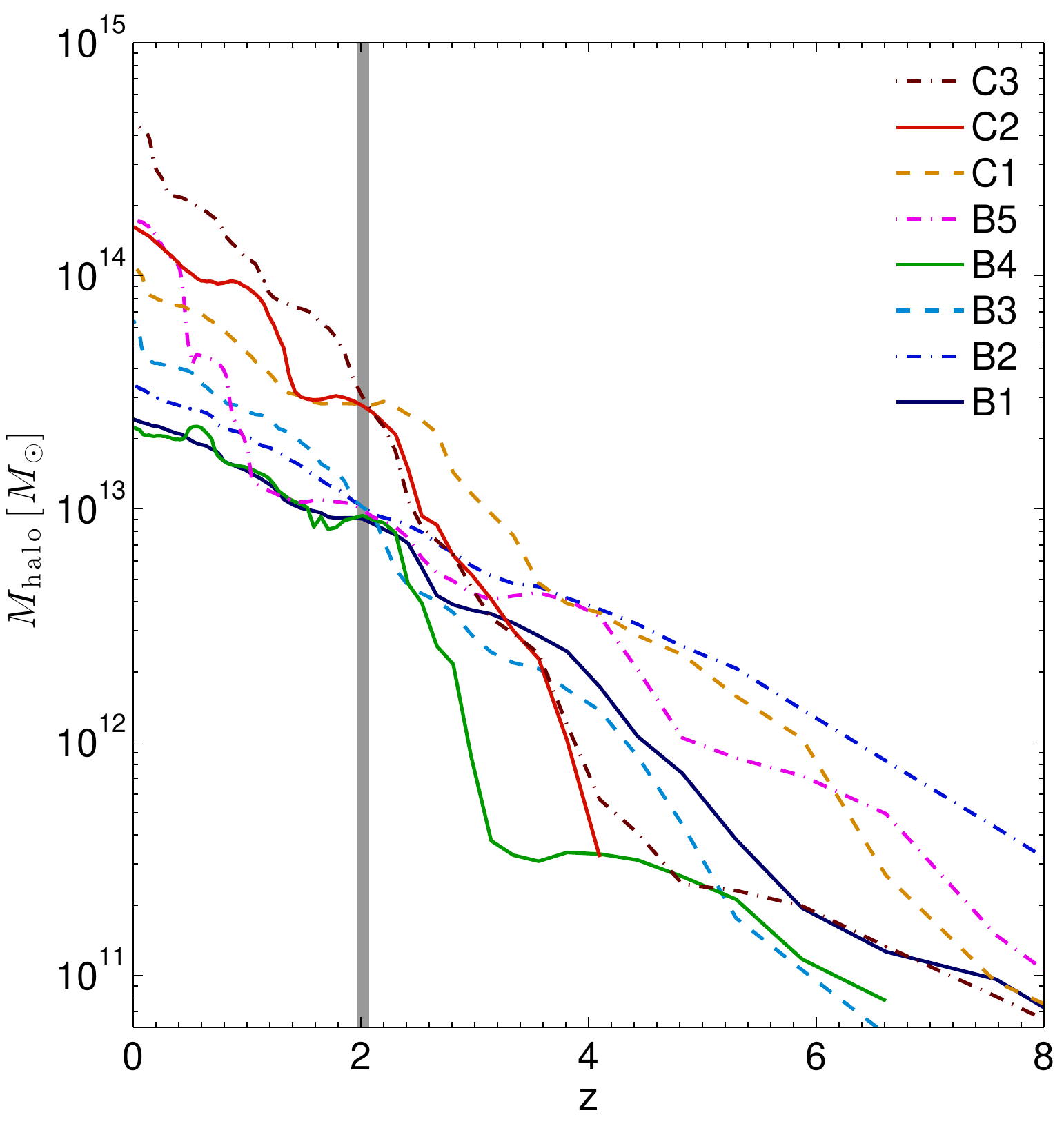} \\
\includegraphics[width=85mm]{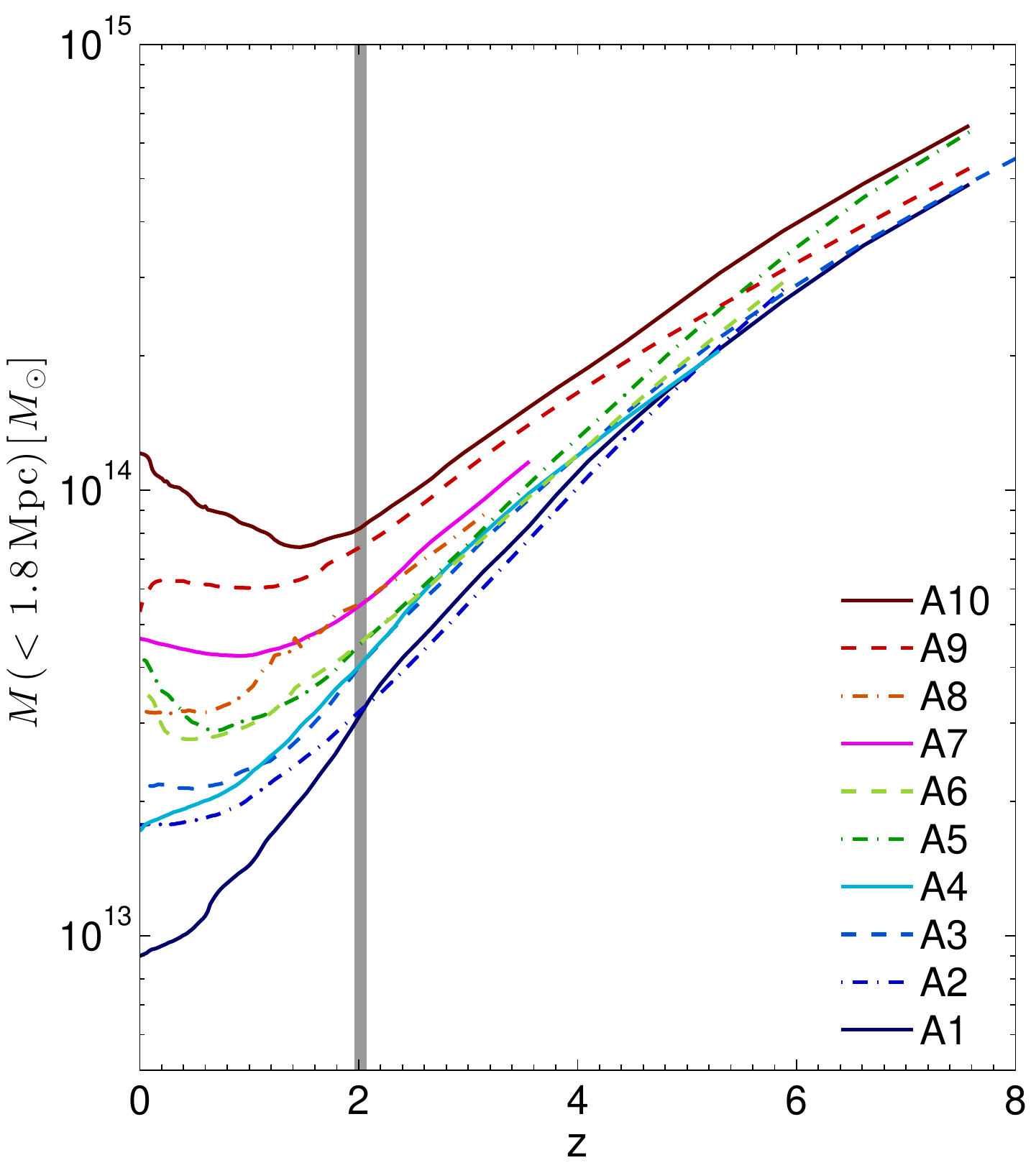} &
\includegraphics[width=85mm]{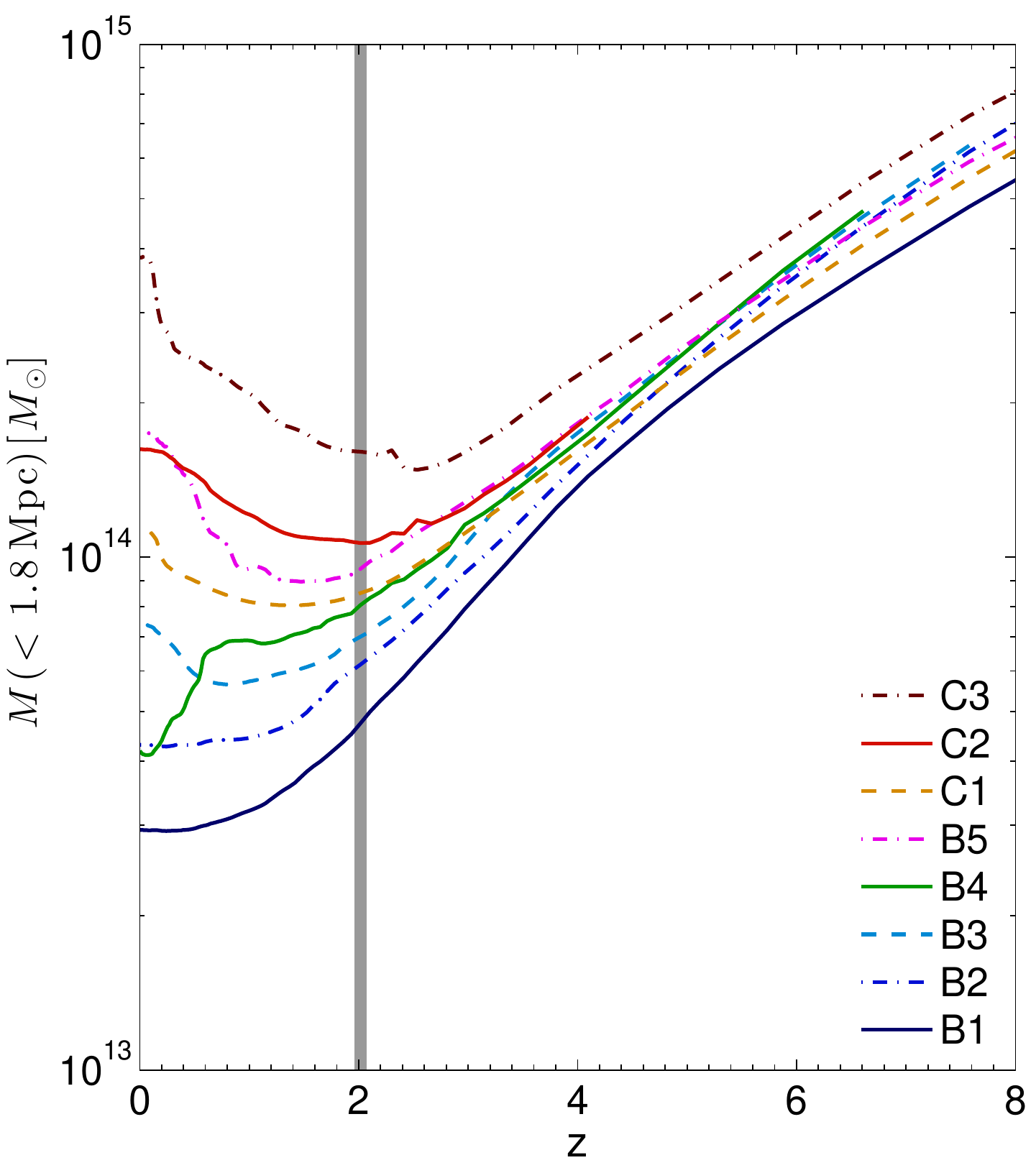} \\
\end{tabular}
\caption{\emph{Redshift evolution of halo mass (top panels) and local overdensity (bottom panels) of the haloes selected by this study based on a lower resolution N-body simulation.} The masses of the selected haloes at $z=2$ fall into one of the following ranges: $2.5-3.6\times{}10^{12}$ $M_\odot$ (top left panel, series A), $0.8-1.2\times{}10^{13}$ $M_\odot$ (lower 5 lines at $z=2$ in the top right panel, series B), or $2.5-3.6\times{}10^{13}$ $M_\odot$ (upper 3 lines at $z=2$ in the top right panel, series C). The grey band in each panel indicates the selection redshift $z=2$.
From each mass bin we select haloes from low, medium, and high density environments, with no other selection criteria. The simulations reported here span a large range of halo growth histories and environmental densities.}
\label{fig:Selection}
\end{figure*}

From a low-resolution DM only run, we select isolated haloes that fall into the following three mass bins at $z=2$: (i) $2.5-3.6\times{}10^{12}$ $M_\odot$, (ii) $0.9-1.1\times{}10^{13}$ $M_\odot$, and (iii) $2.5-3.6\times{}10^{13}$ $M_\odot$. For each such halo, we compute the mass contained within a radius of 1.8 proper Mpc as a measure of local environmental density. For the low (intermediate) mass bin, we select a total of 10 (5) haloes, two haloes (one halo) each from the $5$, $25$, $50$, $75$, and $95$th percentile of the distribution of local environmental densities. For the most massive bin we select 3 haloes, one each from the $5$, $50$, and $95$th percentile. Overall, we thus select 18 `primary' haloes in 3 narrow mass ranges and with a variety of local environmental densities. No other selection criteria are used. We show the evolution of virial masses and environmental overdensities for the selected haloes in Fig~\ref{fig:Selection}. 

Initial conditions for our `zoom-in' runs follow the standard procedure from \cite{Hahn2011} with a convex hull surrounding all particles within $3\times{}R_{\rm vir}$ at $z=2$ of the chosen halo defining the Lagrangian high-resolution region. 

The particle masses at the default high resolution (HR) are $m_{\rm DM}=1.7\times{}10^5$ $M_\odot$ and $m_{\rm gas}=3.3\times{}10^4$ $M_\odot$, respectively. Star particles that form during a simulation have a mass equal to the gas particle they are spawned from. Masses are eight times larger at medium resolution (MR). The gravitational softening lengths for DM and star particles are fixed at 143 parsec (pc) and 21 pc (physical), respectively, for both HR and MR runs. The gravitational softening lengths for gas particles are adaptive and reach a minimum value of 9 pc in the dense interstellar medium. When weighted by instantaneous SFR, the gas softening lengths in our simulation volumes average to about $30$ pc.

The \MassiveFIRE{} simulation suite is summarised in Table~\ref{tab:Over}. The full name of each simulation reported here is MFz2\_{\it Series}{\it Number}, with {\it Series} being one of A, B, C, or Cm and {\it Number} ranging from 1 to the number of runs in each series. In the remainder of this paper we use the short form {\it Series}{\it Number} to denote the runs.

\begin{table*}
\begin{center}
\begin{tabular}{cccccccccc}
\tableline\tableline
 Name & Number & $M_{\rm 180m}(z=2)$  & $z_{\rm end}$ & resolution & $m_{\rm gas}$           & $m_{\rm DM}$           & $\epsilon_{\rm gas}$ & $\epsilon_{\rm star}$ & $\epsilon_{\rm DM}$  \\
  of series          &    of runs             & ($\log_{10} M_\odot$) &                    &                  & ($M_\odot$) & ($M_\odot$) & (pc)             & (pc) & (pc)  \\  \tableline  \vspace{-0.4cm}  \\ 
\tableline
A & 10          & $12.5$   & 1.7 & HR  & $3.3\times{}10^4$     & $1.7\times{}10^5$       & 9   & 21 & 143 \\
B & 5            & $13.0$   & 1.7 & HR  & $3.3\times{}10^4$     & $1.7\times{}10^5$   & 9   & 21 & 143 \\
C & 1            & $13.5$   & 2 & HR  & $3.3\times{}10^4$     & $1.7\times{}10^5$       & 9   & 21 & 143 \\
Cm & 2            & $13.5$   & 1.7, 2 & MR  & $2.7\times{}10^5$     & $1.4\times{}10^6$       & 9   & 21 & 143 \\
\tableline
\end{tabular}
\caption{\emph{Simulations in this paper.} The simulations are grouped into series A, B, C, and Cm (first column). The number of runs in each series is listed in the second column and subsequent columns offer additional information about each series. The third column lists the mass of the targeted, primary halo in a given run. Column 4 reports the redshift at which the simulation is stopped. Column 5 through 10 list the numerical resolution. Columns 6 and 7 state the mass resolution of gas and dark matter particles, while columns 8 though 10 list the gravitational softening lengths for gas, star, and dark matter particles (the gas softening is fully adaptive and the minimal softening length is quoted). All \MassiveFIRE{} runs use identical FIRE code and physics \citep{Hopkins2014}.}
\label{tab:Over}
\end{center}
\end{table*}

\begin{table*}
\begin{center}
\begin{tabular}{lccccccccr}
\tableline\tableline
Name & Central/Satellite & $\log_{10} M_{\rm 180m}$  & $R_{\rm halo}$ & $\log_{10} M_{\rm star}$ & SFR & $\log_{10}$ sSFR & U-V & V-J & $f_{\rm Q}$ \\
          &  & ($\log_{10} M_\odot$)            & (kpc)                 &  ($\log_{10} M_\odot$)    & ($M_\odot$ yr$^{-1}$) & ($\log_{10}$ yr$^{-1}$) & (mag) & (mag) & (\%) \\  \tableline  \vspace{-0.4cm}  \\ 
\tableline
A1:0 & Central & 12.38 & 144.3 & 10.40 & 12.5 & -9.19 & 1.205 & 0.711 & 2 \\
A2:0 & Central & 12.48 & 156.0 & 10.54 & 13.6 & -9.23 & 1.316 & 0.849 & 4 \\
A3:0 & Central & 12.38 & 144.9 & 10.04 & 6.4 & -9.18 & 1.037 & 0.690 & 0 \\
A4:0 & Central & 12.46 & 153.5 & 10.33 & 6.2 & -9.24 & 1.242 & 0.632 & 98 \\
A5:0 & Central & 12.38 & 144.6 & 10.25 & 14.3 & -8.90 & 0.951 & 0.271 & 2 \\
A6:0 & Central & 12.44 & 151.9 & 10.41 & 0.6 & -10.43 & 1.488 & 0.742 & 100 \\
A7:0 & Central & 12.41 & 148.7 & 10.29 & 1.2 & -9.80 & 1.246 & 0.529 & 60 \\
A8:0 & Central & 12.56 & 166.4 & 10.08 & 0.2 & -9.98 & 1.183 & 0.409 & 40 \\
A9:0 & Central & 12.48 & 156.6 & 10.00 & 0.1 & -11.09 & 1.269 & 0.420 & 100 \\
A9:1 & Central & 12.16 & 122.6 & 10.23 & 3.8 & -9.45 & 1.330 & 0.857 & 52 \\
A10:0 & Central & 12.53 & 162.4 & 10.43 & 6.8 & -9.30 & 0.984 & 0.262 & 8 \\
B1:0 & Central & 12.93 & 221.0 & 10.91 & 30.5 & -9.19 & 1.246 & 0.819 & 0 \\
B2:0 & Central & 12.97 & 227.5 & 10.88 & 0.8 & -10.77 & 1.406 & 0.585 & 100 \\
B3:0 & Central & 13.00 & 232.6 & 10.83 & 12.3 & -9.41 & 1.073 & 0.630 & 0 \\
B3:1 & Satellite & 11.54 & 30.1 & 9.75 & 3.4 & -9.33 & 0.925 & 0.179 & 0 \\
B4:0 & Central & 12.94 & 222.6 & 10.55 & 61.3 & -8.39 & 0.626 & 0.120 & 0 \\
B4:1 & Central & 12.25 & 130.6 & 9.78 & 0.4 & -9.82 & 1.164 & 0.282 & 30 \\
B4:2 & Central & 12.06 & 113.0 & 10.03 & 0.3 & -10.37 & 1.280 & 0.536 & 100 \\
B4:3 & Satellite & 12.03 & 70.7 & 10.11 & 13.2 & -8.90 & 1.188 & 0.757 & 14 \\
B5:0 & Central & 12.97 & 227.8 & 10.75 & 22.8 & -9.13 & 1.368 & 1.010 & 0 \\
B5:1 & Satellite & 11.41 & 39.1 & 9.84 & 0.0 & -12.57 & 1.411 & 0.477 & 100 \\
Cm1:0 & Central & 13.45 & 330.1 & 11.73 & 160.6 & -9.36 & 1.117 & 0.543 & 4 \\
Cm1:1 & Satellite & 11.55 & 75.2 & 10.01 & 25.6 & -8.57 & 0.890 & 0.331 & 0 \\
C2:0 & Central & 13.42 & 321.4 & 11.14 & 11.1 & -9.38 & 0.992 & 0.548 & 0 \\
C2:1 & Satellite & 11.91 & 24.8 & 10.17 & 51.3 & -8.78 & 0.876 & 0.499 & 0 \\
C2:2 & Satellite & 12.08 & 50.6 & 10.36 & 22.8 & -9.00 & 1.165 & 0.633 & 18 \\
C2:3 & Satellite & 11.31 & 30.7 & 9.81 & 0.0 & -13.33 & 1.259 & 0.295 & 96 \\
Cm3:0 & Central & 13.46 & 331.8 & 11.38 & 183.8 & -8.98 & 1.140 & 0.606 & 8 \\
Cm3:1 & Central & 12.99 & 231.8 & 11.24 & 153.6 & -8.90 & 1.297 & 0.802 & 48 \\
Cm3:2 & Satellite & 12.73 & 75.7 & 10.96 & 55.3 & -9.16 & 1.247 & 0.656 & 90 \\
Cm3:3 & Central & 12.13 & 119.5 & 10.35 & 47.2 & -8.57 & 0.973 & 0.678 & 0 \\
Cm3:4 & Satellite & 11.95 & 89.7 & 10.52 & 37.7 & -8.90 & 1.227 & 0.915 & 0 \\
Cm3:5 & Central & 11.91 & 100.8 & 10.19 & 32.0 & -8.62 & 1.069 & 0.934 & 0 \\
Cm3:6 & Satellite & 11.83 & 95.0 & 10.15 & 31.5 & -8.39 & 0.822 & 0.095 & 0 \\
Cm3:7 & Central & 11.67 & 84.0 & 9.94 & 0.4 & -10.26 & 1.136 & 0.251 & 30 \\
Cm3:8 & Satellite & 11.50 & 34.9 & 10.22 & 27.2 & -8.82 & 1.064 & 0.569 & 0 \\
Cm3:9 & Satellite & 11.47 & 26.1 & 10.16 & 20.3 & -8.97 & 1.099 & 0.765 & 0 \\
Cm3:10 & Satellite & 11.04 & 15.4 & 10.00 & 0.2 & -10.80 & 1.090 & 0.243 & 4 \\
\tableline
\end{tabular}
\caption{\emph{Properties of simulated galaxies at $z=2$.} The name of the galaxy is provided in the first column. The next column shows whether the galaxy is a central or a satellite galaxy at this redshift. Columns 4 and 5 provide the logarithm of the halo mass and the halo radius, respectively. Columns 6 lists the stellar mass within $0.1\,R_{\rm halo}$. Column 7 and 8 report the SFR and the logarithm of the sSFR within 5 kpc. The next two columns list the median (for 50 random lines of sight) rest-frame U-V and V-J colours. The galaxy would be classified as quiescent according to a colour-colour criterion \citep{Whitaker2011d} for the fraction of random lines of sight provided in the last column.}
\label{tab:GalPropsz20}
\end{center}
\end{table*}

\subsection{Modelling of physical processes} 
\label{sect:PhysicsNumerics}

All simulations here use the identical \emph{Feedback In Realistic Environments} (FIRE) source code, physics, and parameters from \cite{Hopkins2014}. This is the same code used in the FIRE simulations published in previous work \citep{Hopkins2014, Onorbe2015, Chan2015a, Faucher-Giguere2015, Feldmann2016}.

For convenience, we briefly review the most important details of the simulations. The simulations are run with the gravity-hydrodynamics code \GIZMO\footnote{A public version of \GIZMO{} is available at \url{http://www.tapir.caltech.edu/~phopkins/Site/GIZMO.html}} \citep{Hopkins2015a}, in Pressure-energy Smoothed Particle Hydrodynamics (P-SPH) mode, an improved SPH method which conserves energy, entropy, momentum and overcomes some of the problems of traditional SPH methods related to fluid mixing instabilities \citep{Agertz2007, Hopkins2013b}. It also includes improved treatments of artificial viscosity \citep{Cullen2010}, conductivity \citep{Price2008}, and higher-order kernels \citep{Dehnen2012a}. 

Gas cools according to the combination of free-free, photo-ionisation/recombination, Compton, photo-electric, metal-line, molecular, and fine-structure processes, calculated from $10-10^{10}\,$K, and self-consistently accounting for 11 separately tracked species (H, He, C, N, O, Ne, Mg, Si, S, Ca, Fe) each with their own yield tables directly associated with the different stellar mass return mechanisms below. Star formation occurs according to a sink-particle prescription, only in self-gravitating, dense, self-shielding molecular gas. Specifically, gas which is locally self-gravitating (according to the automatically adaptive criterion developed in \cite{Hopkins2013f} from simulations of star-forming regions) and has density in excess of $n > 5\,{\rm cm^{-3}}$, is assigned a SFR $\dot{\rho} = f_{\rm mol}\,\rho/t_{\rm ff}$ where $t_{\rm ff}$ is the free-fall time and $f_{\rm mol}$ is the self-shielding molecular fraction calculated following \citep{Krumholz2011c}. As shown in several previous papers \citep{Hopkins2013f, Hopkins2014}, stellar feedback leads naturally to a self-regulating SF efficiency of $\sim1\%$ per free-fall time in both dense gas and on galaxy scales. Note that, because of the self-gravity criterion, the mean density at which star formation occurs is much higher ($\sim{}100$ cm$^{-3}$ for the resolution adopted in this work).

Once formed, each star particle acts as a single stellar population with given mass, metallicity, and age; all relevant feedback quantities are directly tabulated as a function of time from the stellar population models in STARBURST99 with a Kroupa IMF  \citep{Leitherer1999a}, without subsequent adjustment or fine-tuning. The simulations include several different stellar feedback mechanisms, including (1) local and long-range momentum flux from radiative pressure, (2) energy, momentum, mass and metal injection from SNe and stellar winds, and (3) photo-ionisation and photo-electric heating. We follow \cite{Wiersma2009b} and include mass recycling from Type-II SNe, Type-Ia SNe, and stellar winds.

\begin{figure*}
\begin{tabular}{cc}
\includegraphics[width=87mm]{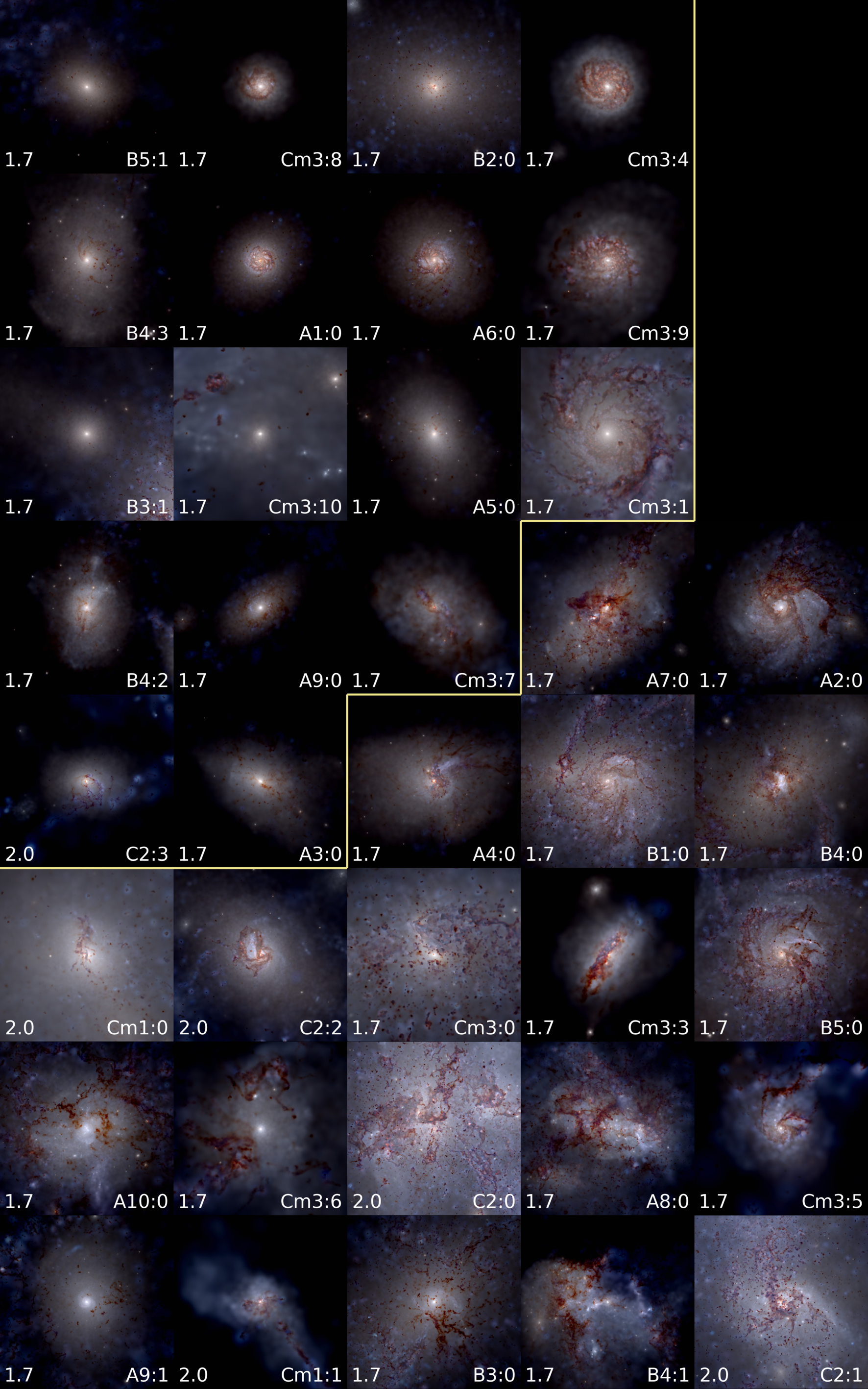} &
\includegraphics[width=87mm]{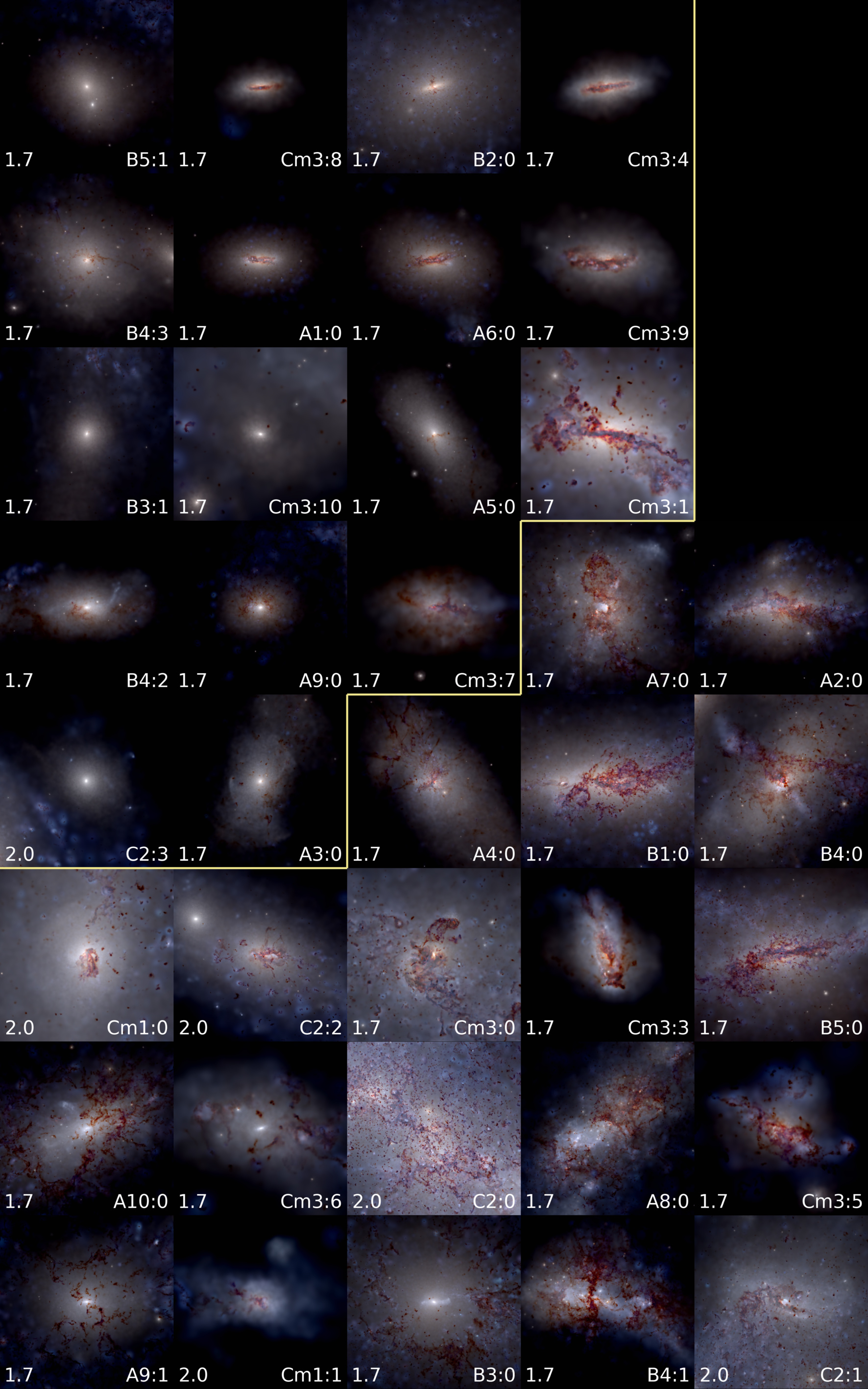}
\end{tabular}
\caption{\emph{Restframe U, V, J band composite images of the $z\sim{}2$ \MassiveFIRE{} galaxies.} We show face-on (left) and edge-on (right) projections for the final snapshot (i.e., either $z=2$ or $z=1.7$) of each galaxy. The image stamps are ordered according to their rest-frame V--J (x-axis) and U--V (y-axis) colours as described in the text. The yellow boundary separates galaxies classified as quiescent and star forming, respectively, based on the \protect\citep{Whitaker2011d} criterion. Each image stamp has a size of $30\times{}30$ kpc$^2$ centred on the target galaxy and shows rest-frame U, V, and J surface brightness mapped to blue, green, and red channels with the same normalisation relative to the Sun and with a fixed dynamic range of 1000. Our sample includes galaxies with a large variety of morphologies, including large star forming disk galaxies (e.g., the galaxy in the 5th row, 4th column), massive, quiescent early-type galaxies (e.g., the galaxy in the top row, 3rd column), and irregular, star forming galaxies (e.g., the galaxy in the bottom row, 4th column).}
\label{fig:UVJ_images}
\end{figure*}

\begin{figure*}
\begin{tabular}{cc}
\includegraphics[width=87mm]{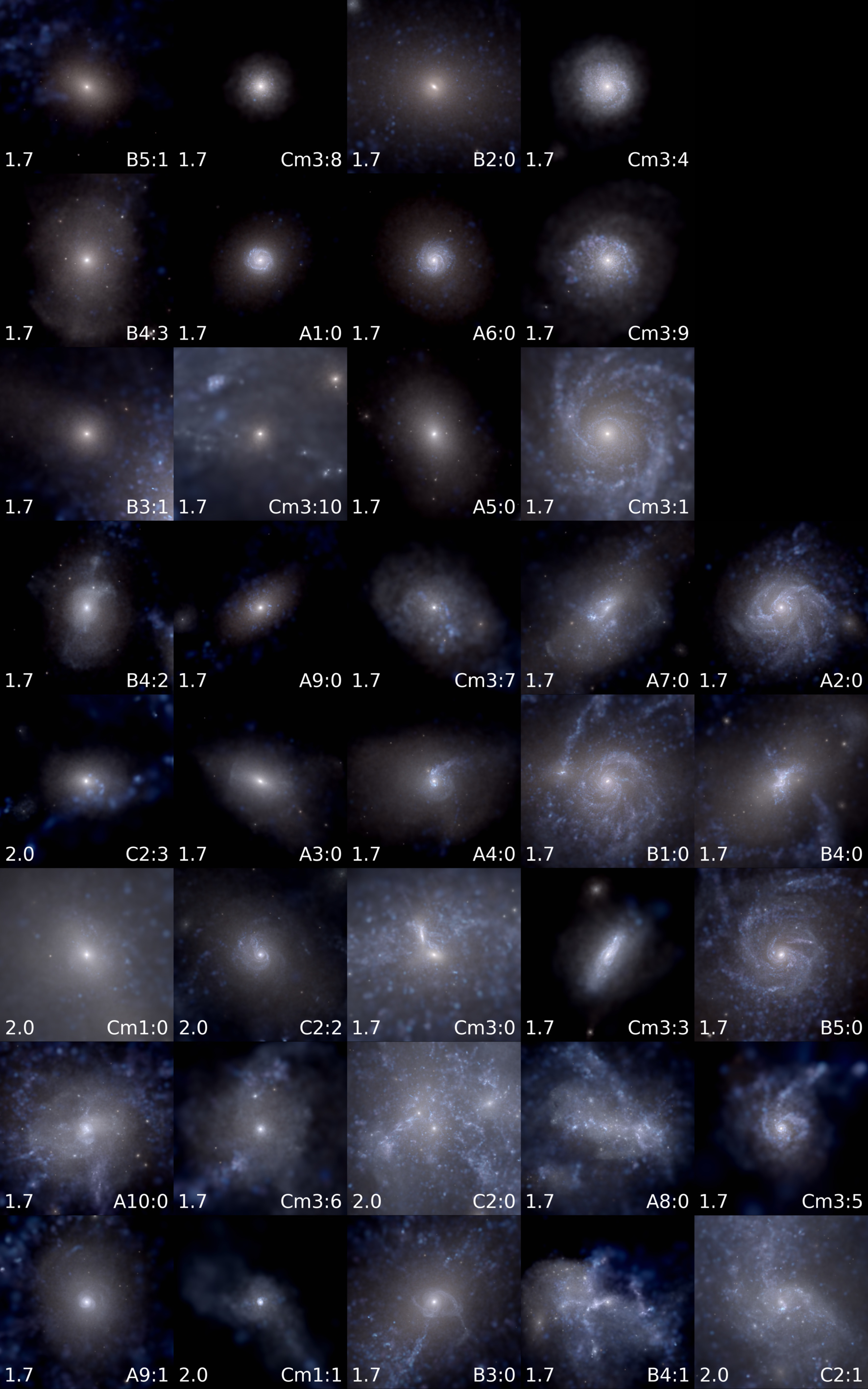} &
\includegraphics[width=87mm]{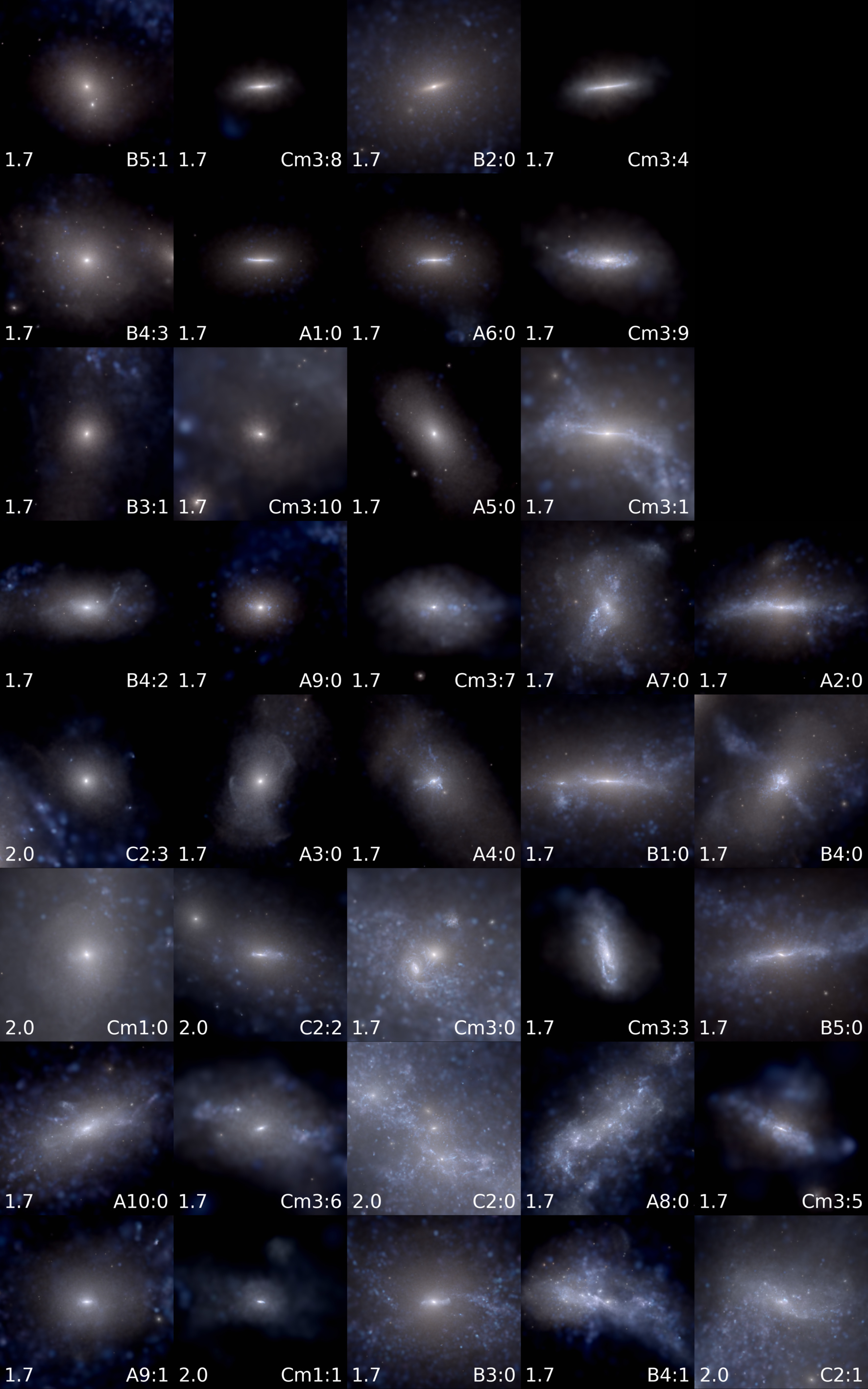}
\end{tabular}
\caption{\emph{Restframe U, V, J band images as in Fig.~\ref{fig:UVJ_images} but without dust (panel ordering is the same).} Visual inspection reveals that $\sim{}1/3$ of the galaxies in our sample harbor an extended disk of young stars.}
\label{fig:UVJ_images_nd}
\end{figure*}

\subsection{Sample selection}
\label{sect:SampleSelection}
 
From each simulation, we analysed about 45 snapshots evenly spaced in cosmic time between $z=10$ to $z=1.7$. We identified haloes and their dark matter (DM), gas, and stellar content with the help of the Amiga Halo Finder (AHF;  \citealt{Gill2004, Knollmann2009}). We selected haloes that, at simulation redshift $z=2.5$, (i) are isolated, (ii) harbour a stellar component exceeding $10^{10}$ $M_\odot$ within $R_{\rm halo}$, and (iii) are not significantly contaminated with lower resolution DM particles (less than $1\%$ by mass). Each selected halo was followed forward and backward in time by linking it to its most massive progenitor and descendent halo from the previous and next snapshot, respectively.

Our selection resulted in 37 haloes that host galaxies with stellar masses ranging from $\sim{}10^{10}$ $M_\odot$ to $\sim{}5\times{}10^{11}$ $M_\odot$ at $z=2$. Of the selected haloes, 21 are isolated at the final snapshot, while 16 are sub-haloes. In other words, 21 of the 37 selected massive galaxies are centrals and 16 are satellite galaxies. Each zoom-in region also contains a large number of lower mass ($M_*<10^{10}$ $M_\odot$) galaxies. These galaxies are not the focus of this work and their properties are discussed elsewhere (e.g., \citealt{Sparre2015}). We list global properties of the selected \MassiveFIRE{} galaxies and their haloes at $z=2$ in Table~\ref{tab:GalPropsz20} and at $z=1.7$ in the Appendix. 

\subsection{Postprocessing}
\label{sect:PostProcessing}

\paragraph*{Masses and SFRs:} The stellar mass of each galaxy is defined as the stellar mass within a sphere of radius $0.1$ $R_{\rm halo}$ centred on the galaxy. We mask the contribution of satellite galaxies by subtracting the stellar mass contained within sub-haloes located at a distance $2$ kpc $\leq{} r \leq{} R_{\rm halo}$ from the centre of the parent halo. In a few cases (during major mergers), this removal reduces the stellar mass within $0.1$ $R_{\rm halo}$ by up to $50\%$, but typically the correction is small. We measure SFRs and specific SFRs in 5 kpc radii to approximately mimic aperture based flux measurements \citep{Whitaker2011d, Schreiber2015b}. Gas masses are also measured with 5 kpc radii and include all gas phases. The SFR is computed as the average number of stars formed within the past $10^8$ years unless specified otherwise. 

\paragraph*{Measurements of environment:} The local environmental density of a galaxy is defined as the average density within a spherical shell with boundaries at $R=R_{\rm halo}$ and $R=5\,R_{\rm halo}$. Compared with the definition used to select haloes (\S\ref{sect:Setup}) the mass of the parent halo of the galaxy is excluded and, thus, does not affect the estimate of the environmental density. Furthermore, by defining the local environment in terms of $R_{\rm halo}$, we can compare the value of the  environmental density for haloes with very different masses. In particular, if haloes of different masses were completely self-similar with identical radial density profile out to $5\,R_{\rm halo}$ then the definition above would return the same environmental density for each of them. 
We use the Hill radius of the halo of a galaxy as a second measure of environment. The Hill radius quantifies the extent of the gravitational sphere of influence of a halo in the presence of external perturbing bodies. We compute the Hill radius of halo $j$ with mass $M_j$ based on the approximate formula $R_{\rm Hill, j} = \min_{i\neq{}j} R_i [M_j / (3 M_i)]^{1/3}$, where $R_i$ is the proper distance (between halo centres) to an isolated halo $i$ of mass $M_i$. Haloes $i$ with $M_i<M_j/3$ lie inside the Hill sphere they produce and are excluded from the analysis.

\paragraph*{Galaxy colours and images:}
We obtain images and fluxes in the rest-frame Johnson U, V, and Cousins J broad-bands by ray-tracing the light emission from stellar particles and including the attenuation by intervening dust, but not scattering, following \cite{Hopkins2014}. Each stellar particle is treated as a single stellar population of given age, mass, and metallicity. We use the isochrone synthesis model by \citet{Bruzual2003} with Padova stellar tracks to compute the luminosities of star particles in each band. The initial stellar mass function \citep{Chabrier2003} has lower and upper cut-off masses of 0.1 $M_\odot$ and 100 $M_\odot$, respectively. The dust opacity in each band is computed for an SMC dust composition \citep{Pei1992}, scaled linearly with the metallicity of the gas ($Z$) and normalised such that the optical depth in the Johnson B band is 0.78 for a $10^{21}$ cm$^{-2}$ column of hydrogen atoms with $Z=0.02$ \citep{Pei1992}. U-V and V-J colours are measured within (projected) circular apertures of 5 kpc radius centred on each galaxy with the contribution from satellite galaxies masked.

\paragraph*{Growth rates:}
Growth rates at $z\sim{}2$ are computed for the cold baryonic mass ($M_{\rm bar}=M_{\rm star}+M_\HI+M_\H2$ within $0.1\,R_{\rm halo}$) of each galaxy in our sample and for the DM mass of their haloes ($M_{\rm DM}$). The procedure is briefly described in \cite{Feldmann2016} and it follows largely \cite{McBride2009}. $M_{\rm bar}$ and $M_{\rm DM}$ are fitted with a modified exponential $\propto{}(1+z)^{\rm \beta}{\rm e}^{-\gamma{}z}$ \citep{Tasitsiomi2004} over the redshift range $z\sim{}2-7$. The specific growth rate at redshift $z_0$ is then computed as ${\rm d}\ln{}M/{\rm d}t = [\beta/(1+z_0) - \gamma]\,{\rm d}z/{\rm d}t$.

\section{Galaxy Properties}
\label{sect:GalProps}

We show composite images in rest-frame U, V, and J band filters of the \MassiveFIRE{} sample at $z\sim{}1.7-2$ in Fig.~\ref{fig:UVJ_images}. The panels are ordered according to the rest-frame U--V and V--J colours. Fig.~\ref{fig:UVJ_images} thus mimics a colour-colour diagram with quiescent galaxies at the top left and star forming galaxies at the bottom and at the right. In particular, we first sort galaxies based on their restframe U--V colour and split them into 8 bins (from bottom to top). In each U--V bin we sort galaxies according the restframe V--J colour (from left to right). The yellow line separates star forming from quiescent galaxies based on the criterion that a galaxy is quiescent if $U-V>1.2$, $V-J<1.4$, and $U - V > 0.88 \times{} (V - J) + 0.59$ \citep{Whitaker2011d}. 

Overall, the galaxies show a large diversity in morphologies, colours, and dust abundances. The sample includes dusty, star forming disk galaxies, dusty irregular galaxies, dust-poor early-type galaxies, and merging/interacting galaxies. Visually, the dust abundance is lower in quiescent galaxies but perhaps somewhat surprisingly many galaxies classified as quiescent contain a significant amount of dust (discussed in \S\ref{sect:GalColors}).

In Fig.~\ref{fig:UVJ_images}, face-on (edge-on) projections\footnote{The face-on direction is defined as parallel to the angular momentum vector of star particles with an age less than 200 Myr located within 2.5 kpc from the centre of the galaxy. When necessary, we increase the maximum age in steps of 200 Myr until we end up with at least 100 such star particles.} of the simulated galaxies are shown in the left (right) panels. The edge-on view reveals that many galaxies in our sample (about 1/3 of the star forming galaxies by visual inspection) have a well-defined disk component. This should be compared with observations which indicate that about 30\%-80\% of $z\sim{}2$ galaxies are rotation supported and/or have low S\'{e}rsic indices \citep{ForsterSchreiber2009, Jones2010, VanderWel2014b, Wisnioski2015}.

Quiescent galaxies are in about equal parts centrals and satellites, while 65\% of the star forming galaxies are centrals and only 35\% are satellites. This difference is not unexpected as satellite galaxies may experience strong environmental forces, such as ram pressure and tidal stripping, that affect star formation. However, given our limited sample, this is not a highly significant difference ($p=0.14$ according to a one-tailed two proportion test).

\subsection{Galaxy colours}
\label{sect:GalColors}

\begin{figure}
\begin{tabular}{cc}
\includegraphics[width=85mm]{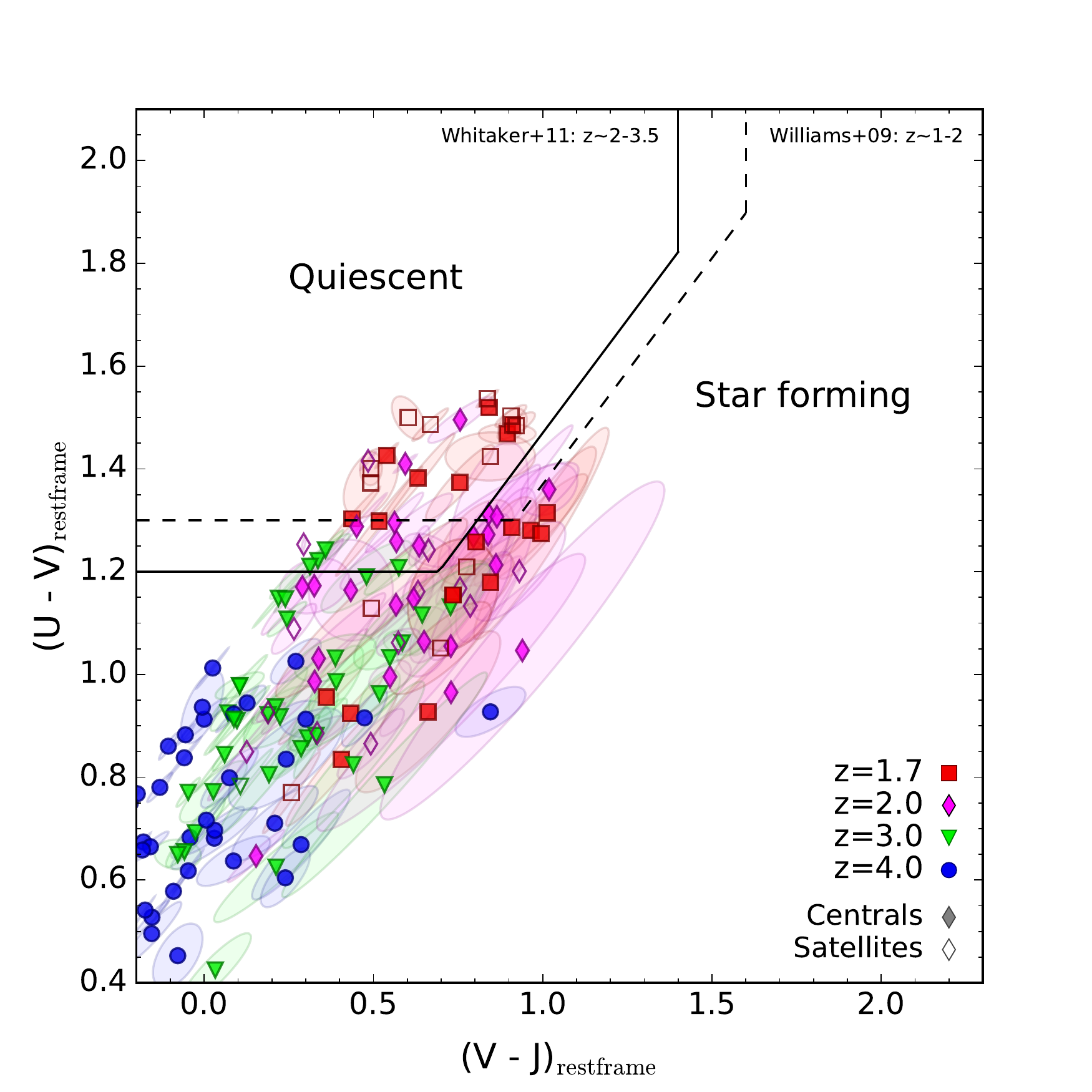}
\end{tabular}
\caption{\emph{Restframe U--V vs. V--J colours of the \MassiveFIRE{} sample at $z=4, 3, 2,$ and 1.7.} U--V and V--J colour are averaged over 50 random projections of each galaxy. Coloured ellipses show 1-$\sigma$ deviations around the mean. Filled and empty symbols distinguish central from satellite galaxies. The solid line (our adopted choice for the remainder of this paper; \protect\citealt{Whitaker2011d}) and the dashed line \protect\citep{Williams2009g} are used to empirically separate quiescent galaxies from star forming galaxies.}
\label{fig:UVJ_z}
\end{figure}

\begin{figure*}
\begin{tabular}{cc}
\includegraphics[width=85mm]{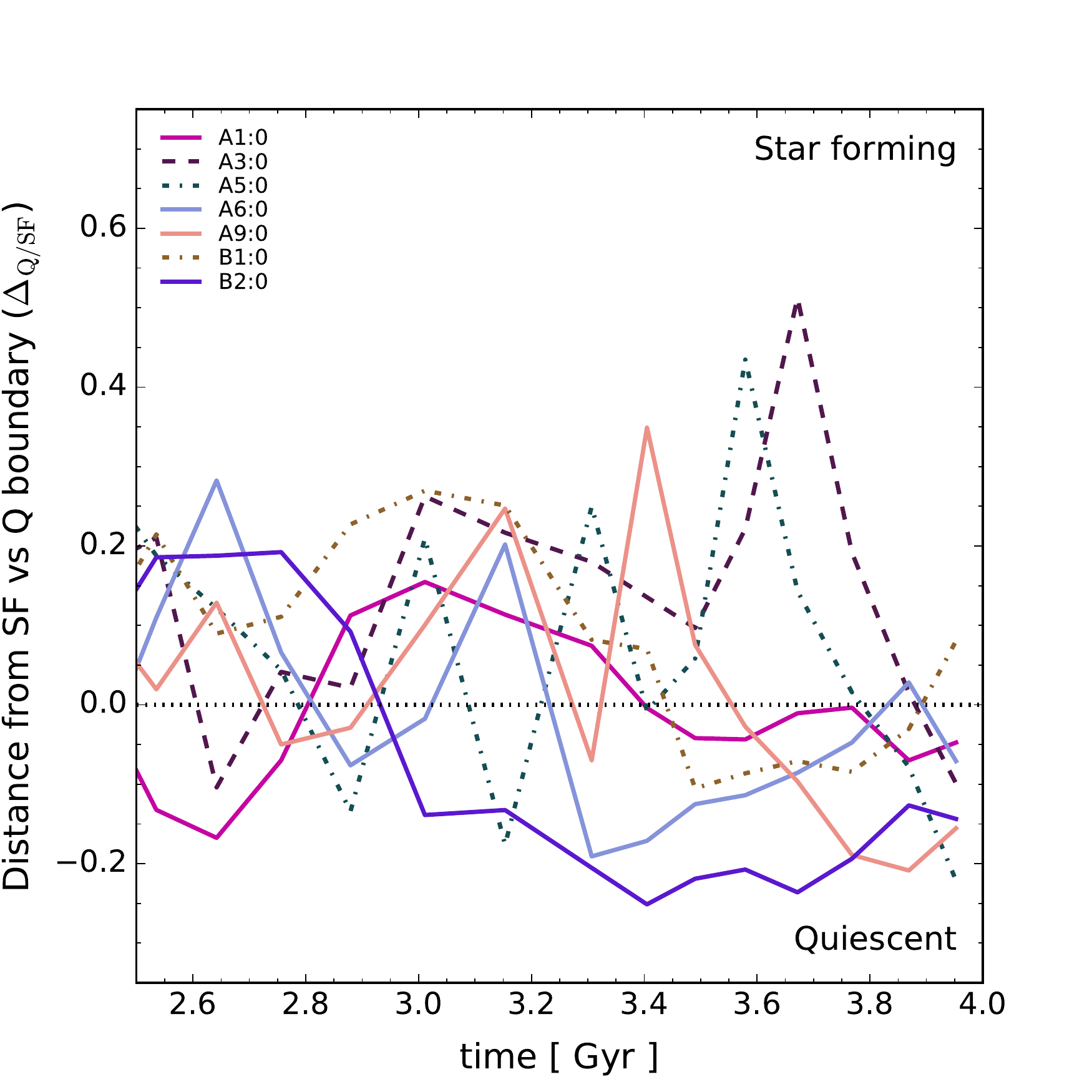} & \includegraphics[width=85mm]{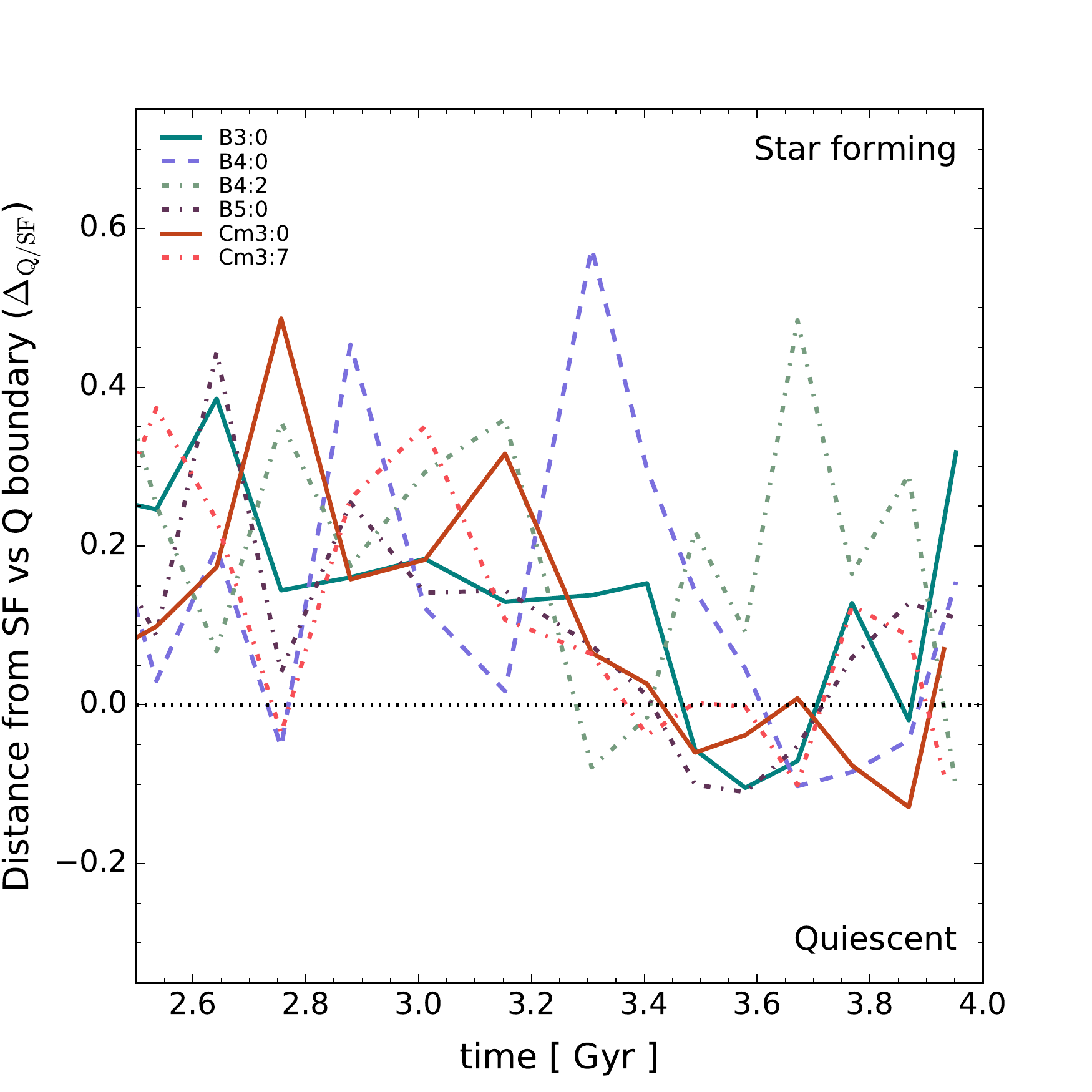}
\end{tabular}
\caption{\emph{Median distance from the boundary separating star forming and quiescent galaxies in Fig.~\ref{fig:UVJ_z} as a function of time.} The panels show the subset of central galaxies in our sample that spend some time in the quiescent regime. Galaxies that lie above (below) the dotted line are classified as star forming (quiescent). The time sampling is about 100 Myr, comparable to the timescale for intrinsic colour evolution. Many of our galaxies are quiescent for only a brief period of time ($<100$ Myr) but several remain quiescent for many hundreds of Myr.}
\label{fig:DQSF}
\end{figure*}

Fig.~\ref{fig:UVJ_images_nd} shows dust-free images of our sample of galaxies, arranged as in Fig.~\ref{fig:UVJ_images}. The edge-on views in Fig.~\ref{fig:UVJ_images} give the impression that disk galaxies have large scale heights and disturbed morphologies. However, if we remove the dust extinction we see that young stars are often arranged in a well-defined disk with a scale height significantly shallower than that of the dust. Most galaxies show the presence of an extended and more spherically distributed stellar component of older stars, but often at low surface brightness. In some cases stellar feedback is able to clear the central few kpc of a galaxy of dust and gas.

Interestingly, a small fraction of these high redshift galaxies lack a well-defined centre, e.g., B4:1, A8:0. These galaxies are typically fast growing, highly star forming galaxies with blue U--V colours.

Colour-colour diagrams have proved to be an efficient means of separating galaxies with below average sSFRs from more typical, star forming galaxies (e.g. \citealt{Labbe2005, Wuyts2007, Williams2009g, Whitaker2011d, Brammer2011f}). In Fig.~\ref{fig:UVJ_z} we plot the average (over 50 random lines of sight) U--V and V--J rest-frame colours of our sample at $z=4, 3, 2,$ and 1.7. The figure highlights that the simulation at $z\lesssim{}2$ include both galaxies classified as quiescent and as star forming according to the empirical UVJ criterion.

For a more quantitative analysis, we select a subset of galaxies from the UltraVISTA survey \citep{McCracken2012, Muzzin2013d} with similar stellar masses and redshifts. The detailed analysis is provided in Appendix~\ref{sect:CompUltraVISTA}. While the colours of the simulated and observed galaxies generally overlap, there are significant differences. In particular, unlike observations, our simulations lack galaxies with U--V rest-frame colours above 1.6. We note that the UltraVISTA catalog is not mass-complete down to $10^{10}$ $M_\odot$ at $z=2$ and that galaxies with U--V $>1.6$ are typically more massive ($M_{\rm star}>10^{11}$ $M_\odot$, \citealt{Williams2010c}). However, the mass dependence alone is unlikely to fully explain this difference.This may imply that our simulations lack an important ingredient, e.g., AGN feedback. We also note that star forming galaxies in our simulations lie, on average, somewhat closer to the star forming vs. quiescent separation line than star forming galaxies in UltraVISTA.

A bimodality in colour-colour space has been observed out to $z\sim{}2$ (e.g., \citealt{Williams2009g, Whitaker2011d, Muzzin2013d, Martis2016}), and it has been suggested that the bimodality exists at even higher redshifts (e.g., \citep{Brammer2009, Whitaker2011d, Muzzin2013d, Tomczak2014}). We compute the distance from the star forming / quiescent boundary \citep{Whitaker2011d} and test for multimodality with the dip test \citep{Hartigan1985a} by selecting one line of sight for each galaxy. The dip test does not reveal evidence for a non-unimodal distribution. However, as we show quantitatively in Appendix \ref{sect:TestMultiModal}, the size of our sample is too small to detect a colour bimodality with statistical significance even if one were present.

Fig.~\ref{fig:UVJ_z} also shows that galaxy colours become redder with time. This result is a consequence of two independent processes. First, galaxies with constant or declining star formation rates become redder as the stellar population ages (e.g., \citealt{Wuyts2007}). Secondly, the dust extinction of many galaxies increases with time between $z=4$ and $z=1.7$ resulting in a diagonal shift towards redder U--V and V--J colours. 

The colours of most galaxies depend on the particular line-of-sight. However, as the diagonal separation line of the UVJ criterion is approximately parallel to the dust reddening vector, the classification of galaxies into quiescent / star forming is typically much less affected by a change in viewing angle. The Euclidian distance to the separation line, $\Delta_{\rm Q/SF}$, is thus a useful quantity that only weakly depends weakly on the chosen line-of-sight for a given galaxy. Negative (positive) values of $\Delta_{\rm Q/SF}$ correspond to the quiescent (star forming) region of the colour--colour diagram.

Fig.~\ref{fig:DQSF} shows the separation from the quiescent vs. star forming boundary, $\Delta_{\rm Q/SF}$, for the 13 of our 21 central galaxies that become quiescent by $z\sim{}1.7$ or that are quiescent for two or more consecutive simulation snapshots. As is clear from the figure, the quiescent / star forming classification of many galaxies varies both on long and on short time scales. Some galaxies, e.g., B2:0, become and remain quiescent for hundreds of Myr, while others, e.g., B4:2, are in the quiescent region of the UVJ colour diagram for only a brief amount of time ($\lesssim{}100$ Myr).

\begin{figure}
\begin{tabular}{c}
\includegraphics[width=85mm]{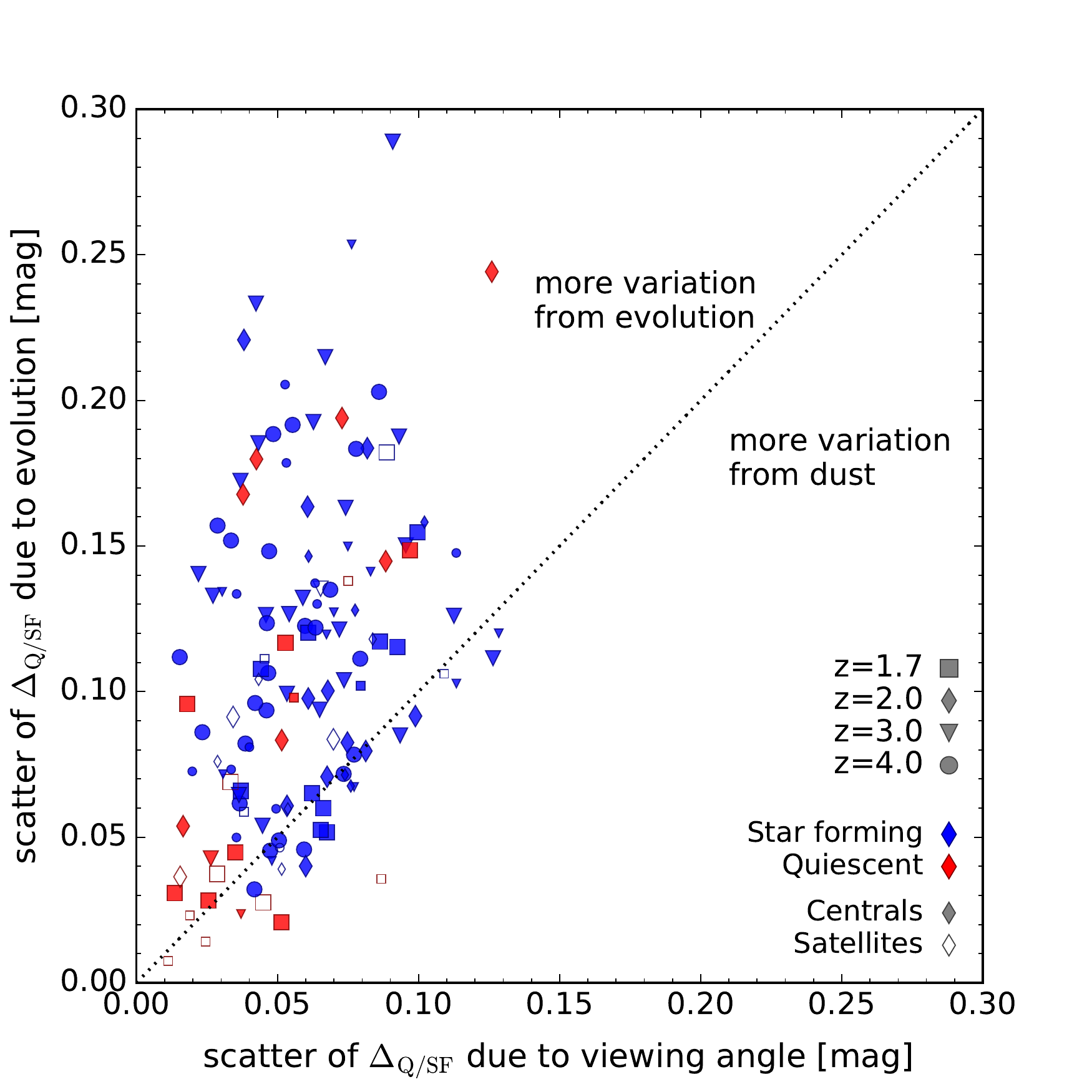}
\end{tabular}
\caption{\emph{Scatter of $\Delta_{\rm Q/SF}$ caused by line-of-sight variations and by evolution over 100 Myr timescales.} Red (blue) symbols denote lines of sight that place a given galaxy in the quiescent (star forming) part of the UVJ diagram. Large and small symbols show HR runs (series A, B, and C) and MR runs (series Cm), respectively. Evolutionary processes lead to a larger scatter of $\Delta_{\rm Q/SF}$ ($\sim{}0.03-0.25$ mag) than viewing angle variations ($\sim{}0.02-0.1$ mag). Evolutionary changes are linked to colour variations of the underlying stellar population while the line-of-sight variations are driven by a non-homogeneous dust distribution.}
\label{fig:UVJ_scat}
\end{figure}

In Fig.~\ref{fig:UVJ_scat} we compare the scatter in $\Delta_{\rm Q/SF}$ arising from viewing angle variations (standard deviation of $\Delta_{\rm Q/SF}$ over 50 random sightlines) with the scatter caused by evolutionary processes (standard deviation of $\Delta_{\rm Q/SF}$ sampled at 3 times separated by 100 Myr). The scatter of $\Delta_{\rm Q/SF}$ caused by evolutionary processes dominates the the scatter caused by viewing angle variations.

Without dust, variations in colours with viewing angle are negligible, yet the variations with time remain similar. In other words, evolutionary color changes are caused by bursty star formation and not by changes in the dust distribution. For star forming galaxies, time variations of $\Delta_{\rm Q/SF}$ based on intrinsic colours are actually often somewhat \emph{larger} than time variations based on dust-reddened colours. Hence, for star forming galaxies in our sample, dust preferentially reddens regions with young stellar populations, i.e., sites of recent star formation.

\begin{figure*}
\begin{tabular}{cc}
\includegraphics[width=85mm]{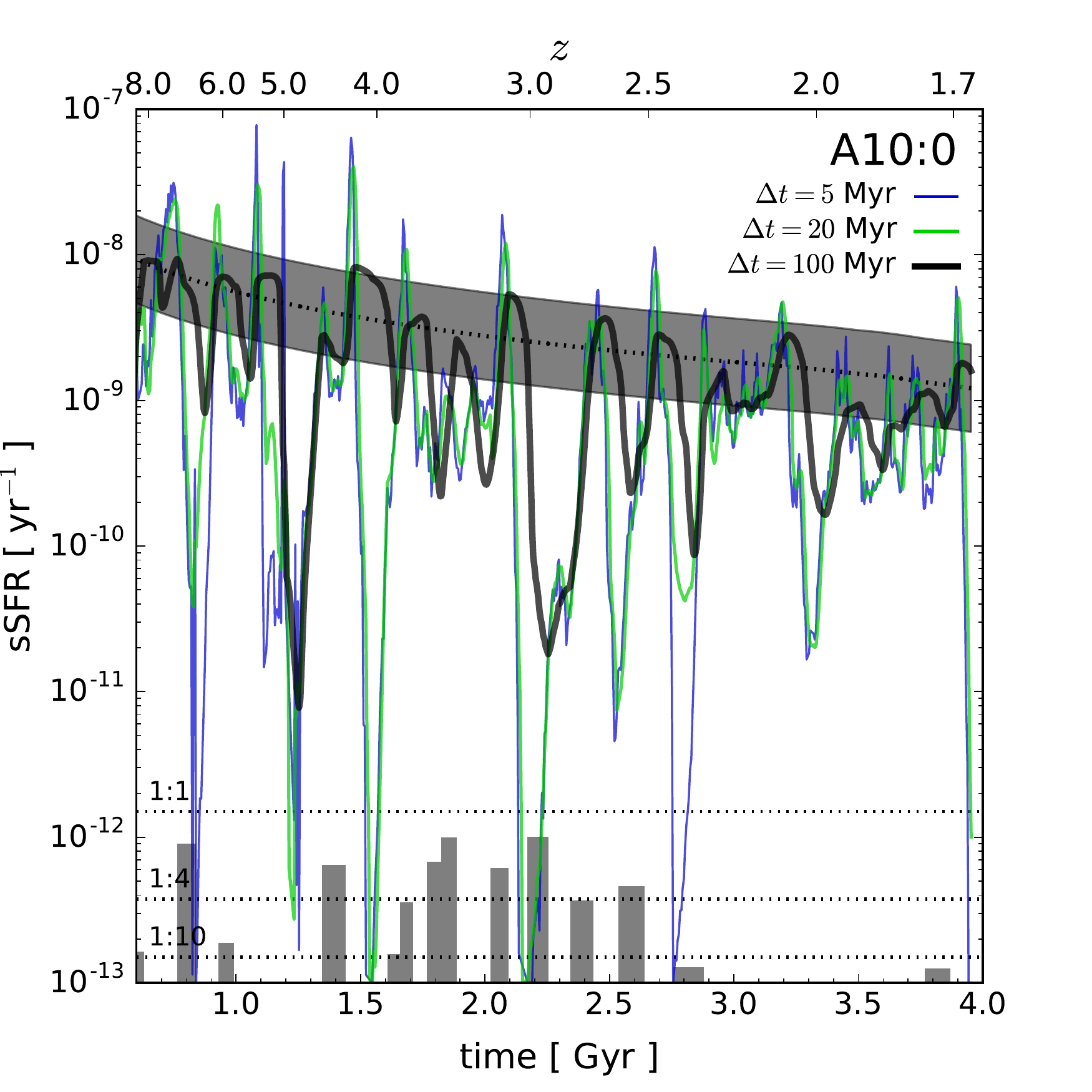} & 
\includegraphics[width=85mm]{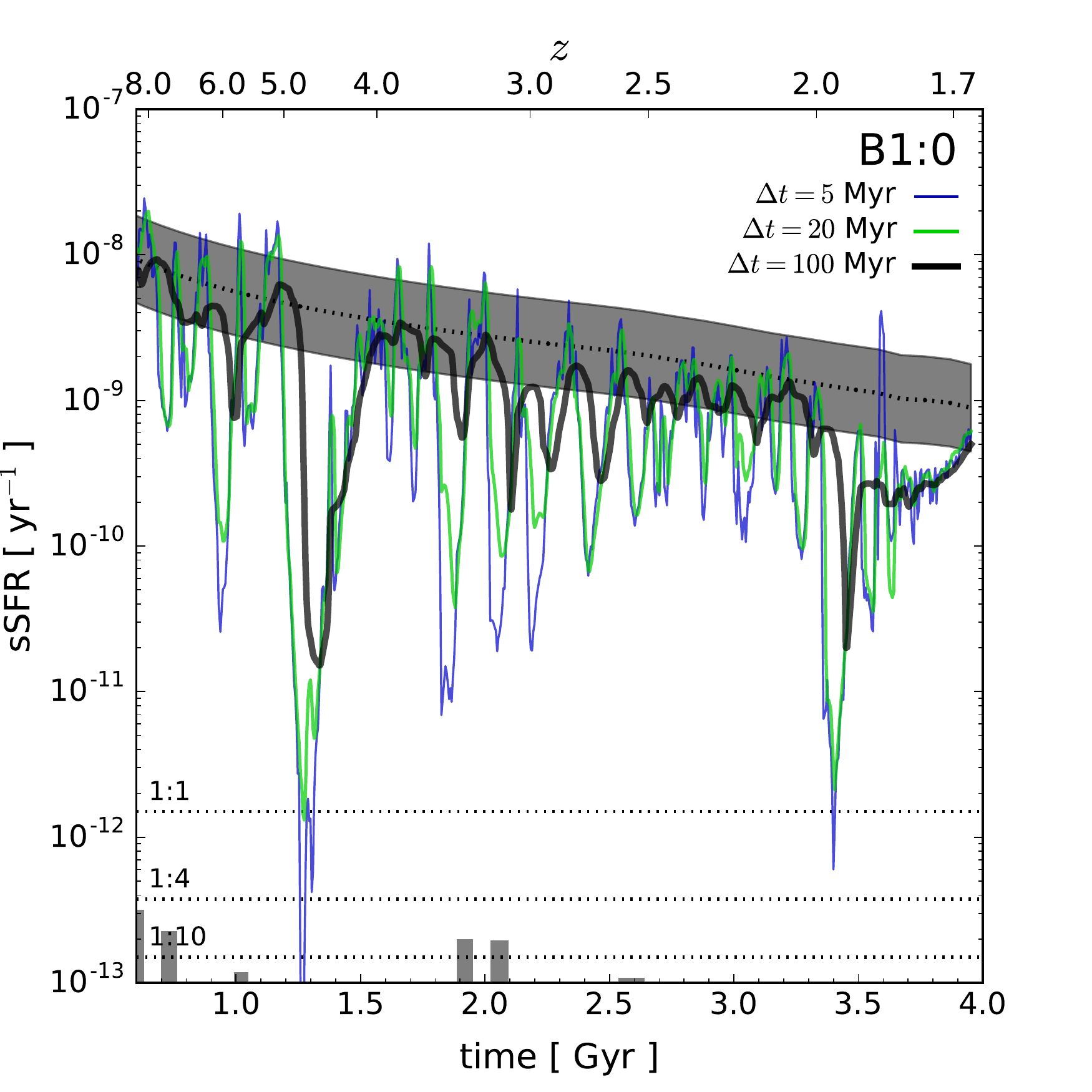} \\
\includegraphics[width=85mm]{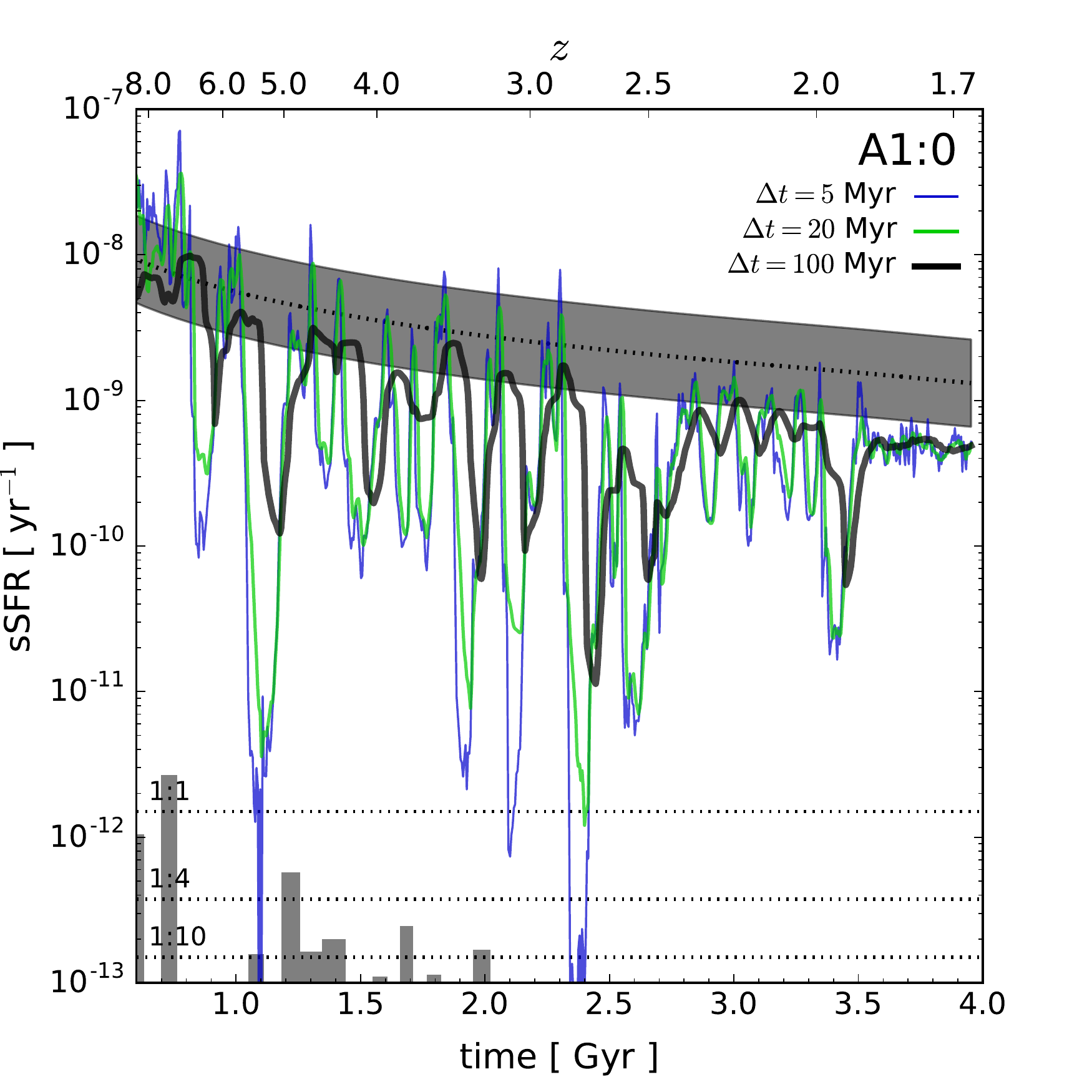} & 
\includegraphics[width=85mm]{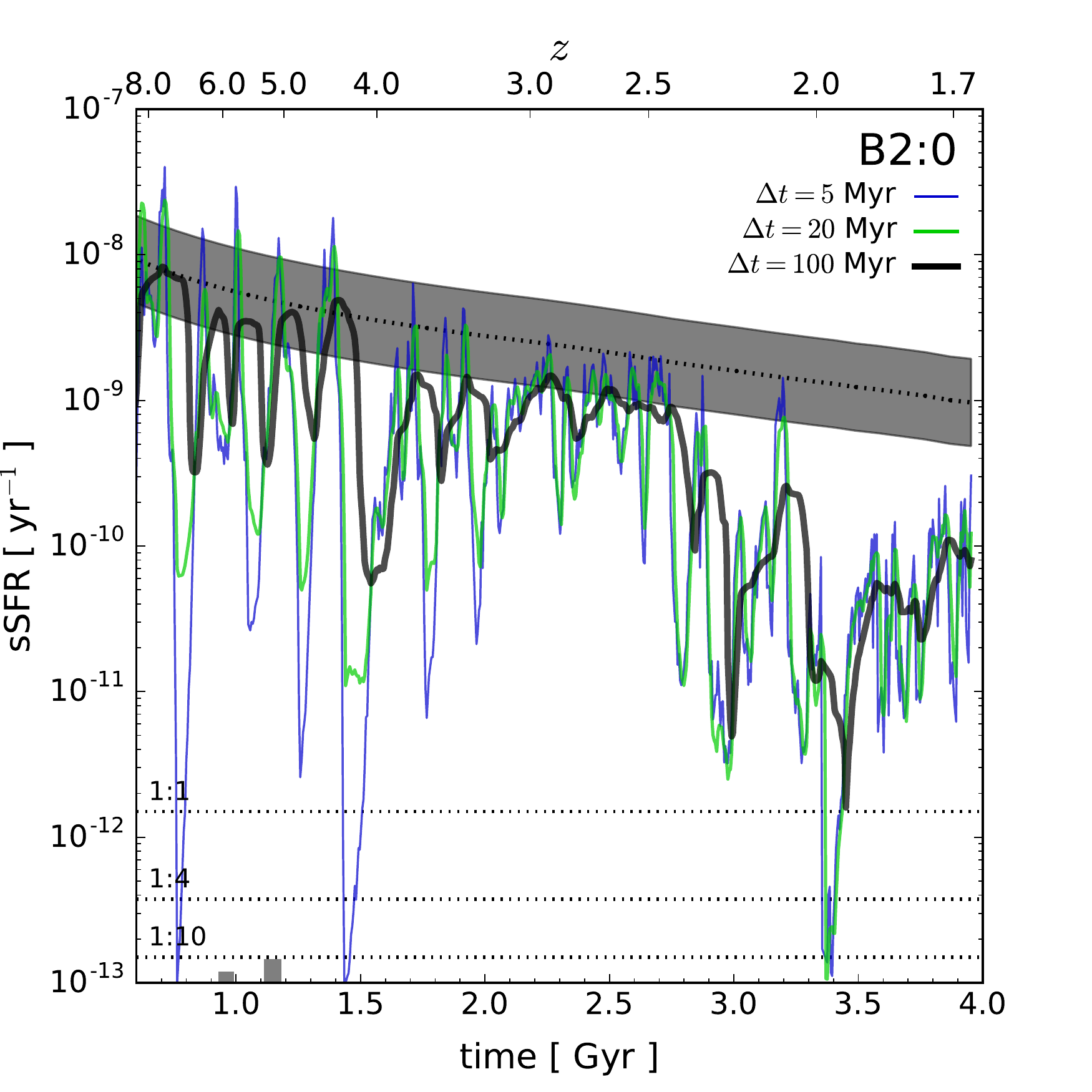} \\
\end{tabular}
\caption{\emph{Evolution of the sSFR for galaxies classified as star forming (top row) and quiescent (bottom row) based on their U--V and V--J colours.} Thick, medium-thick, and thin lines show the sSFRs averaged over the past 100 Myr, 20 Myr, and 5 Myr, respectively. Grey bands and dotted lines indicate typical sSFRs of observed star forming galaxies at comparable redshifts and with similar stellar masses \citep{Schreiber2015b}. The bars at the bottom of each panel reveal how much mergers contribute to the change in stellar mass between two snapshots, see text. The horizontal dotted lines indicate merger-to-galaxy mass ratios of 1:1, 1:4, and 1:10. The simulated galaxies have very complex star formation histories with large short-term variability that is often not directly related to galaxy mergers. Star forming galaxies tend to evolve along the star forming sequence but with frequent deviations by a factor of a few upwards (star bursts) and by orders of magnitude downwards (short-term suppression of star formation). Quiescent galaxies also show large variability but the average SFRs tend to fall below the level expected from the star forming sequence.}
\label{fig:sSFR_evolution_examples}
\end{figure*}

To better understand the different timescales over which galaxies have colours that place them in the quiescent regime, Fig.~\ref{fig:sSFR_evolution_examples} shows the sSFRs for a representative subset of the central galaxies in our sample. Specifically, we measure the sSFR of 4 galaxies (two are classified as star forming and the other two as quiescent at $z=1.7$ based on their U--V and V--J colours) and their main progenitors between $z=1.7$ and $z=8$. This figure highlights a number of important results.

First, it shows that high-redshift galaxies have complex star formation histories with large amounts of short-term variability. The amount of variability is stronger for SFR estimators with shorter tracer lifetimes. In particular, the 5 Myr average sSFRs of many galaxies show bursts of a factor of a few above the star forming sequence. These short star bursts are typically followed by a strong, but short-lived ($\lesssim{}100$ Myr), suppression of the star formation activity. SFRs averaged over longer lifetimes, e.g., 100 Myr, show still a significant, albeit reduced, amount of variability. With this longer tracer, individual star forming galaxies approximately co-evolve with the star forming sequence.  

Fig.~\ref{fig:sSFR_evolution_examples} also clarifies the role of mergers in suppressing the star formation activity in galaxies. Specifically, we compare the stellar mass that is accreted during a given snapshot with the stellar mass of the central galaxy (all masses are measured within the central 5 kpc) at the given time, see grey bars in Fig.~\ref{fig:sSFR_evolution_examples}. Ratios of $\gtrsim{}1:4$ suggest significant merger events. We find that many of the starbursts and star formation suppression events do not coincide with major mergers. We leave a more detailed study of the triggers of starbursts in massive, high-z galaxies for future work.

\begin{figure}
\begin{tabular}{c}
\includegraphics[width=80mm]{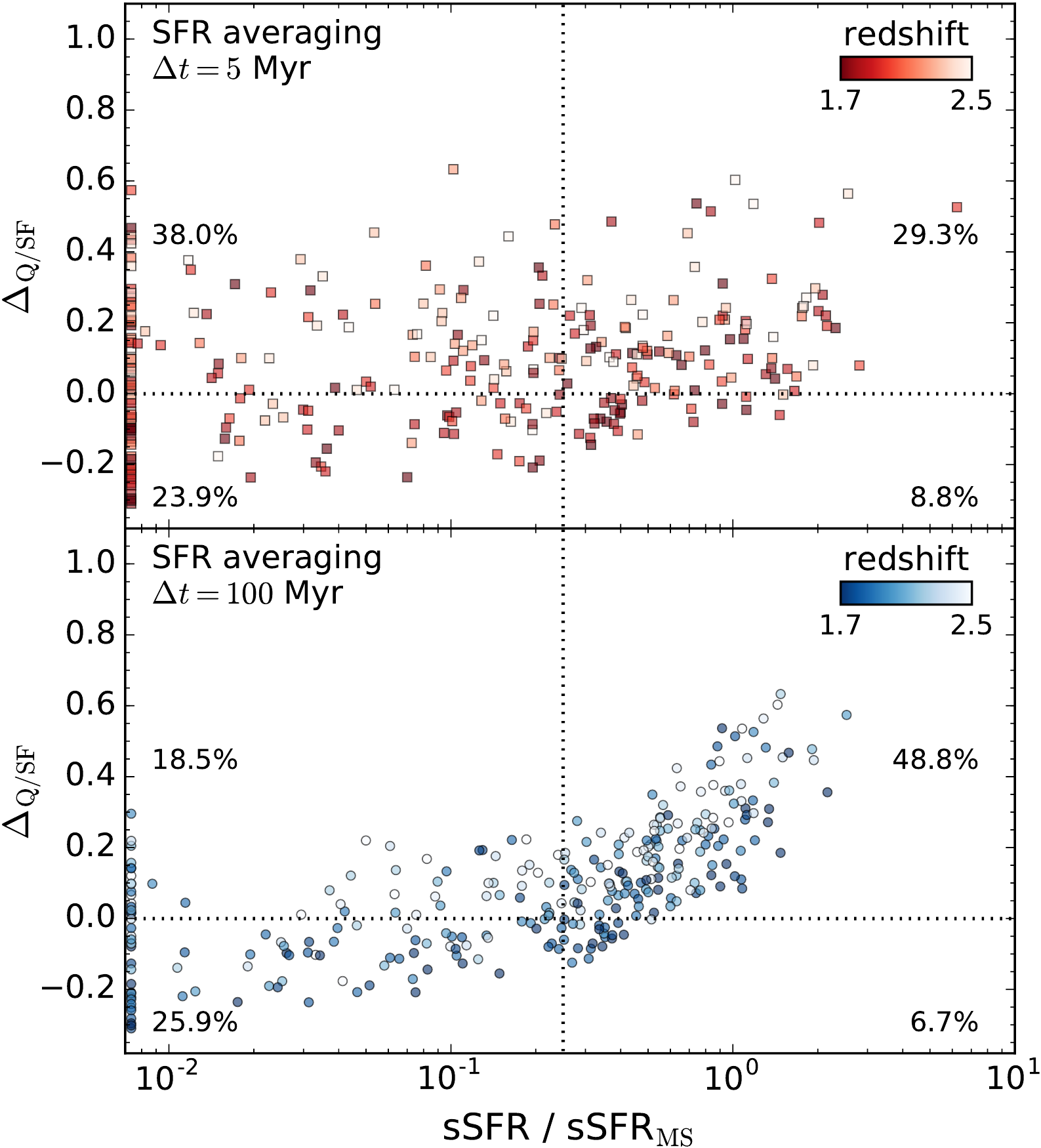}
\end{tabular}
\caption{\emph{Distance from the boundary separating star forming and quiescent galaxies in colour space, $\Delta_{\rm Q/SF}$, versus the offset from the star forming sequence.} SFRs are measured over the past 5 Myr (top panel) or past 100 Myr (bottom panel). Only galaxies with $M_{\rm star}>10^9$ $M_\odot$ at $z=1.7-2.5$ are included. Galaxies with sSFRs of a factor of 140 or more below the star forming sequence are shown at the left boundary in each panel. The dotted lines divide the space into 4 quadrants. The horizontal line separates quiescent from star forming galaxies based on colours, while the vertical line indicates a sSFR that is a factor 4 below the star forming sequence. Percentage values indicate the fraction of galaxies in each of the four quadrants. A colour-based classification into star forming and quiescent galaxies corresponds well to a classification based on sSFRs measured on $\sim{}100$ Myr timescales (bottom panel). In contrast, near-instantaneous sSFRs vary on short timescales and are thus poor predictors of galaxy colours (top panel).}
\label{fig:colorSSFR} 
\end{figure}

In Fig.~\ref{fig:colorSSFR}, we directly compare how the star forming / quiescent classification based on colours relates to a division based on sSFRs. In particular, we show $\Delta_{\rm Q/SF}$ vs the sSFR offset from the star forming sequence (i.e., the locus of `main sequence' galaxies). The figure demonstrates that it matters how SFRs are measured -- the results are remarkably different if  near instantaneous SFRs or SFRs averaged over 100 Myr are used.

The bottom panel shows our default case of averaging SFRs over 100 Myr (roughly comparable to SFRs based on FUV, \citealt{Sparre2015}). Quiescent galaxies have sSFRs that are a factor 4 or more below the star forming sequence. In contrast, star forming galaxies tend to have sSFRs that are at most a factor of a few above or below the star forming sequence and that are strongly correlated with $\Delta_{\rm Q/SF}$.

In the top panel of Fig.~\ref{fig:colorSSFR} we show the corresponding results if near instantaneous SFRs are used (here 5 Myr averaging time; roughly comparable to SFRs based on H-$\alpha$ luminosity). In this case, the tight correlation between $\Delta_{\rm Q/SF}$ and the sSFR offset from the star forming sequence is almost lost. SFRs of star forming galaxies are highly variable on short timescales (see Fig.~\ref{fig:sSFR_evolution_examples}) resulting in slightly more than half of such galaxies having sSFRs a factor 4 or more below the star forming sequence.

In short, classifying galaxies as star forming / quiescent based on colours is roughly equivalent to a classification based on sSFRs \emph{if sufficiently long SFR averaging timescales are used}.

\subsection{Dust Extinction}
\label{sect:Dust}

Dust reddening and extinction can strongly affect the observed colours of galaxies. The `true' dust extinction $A_{\rm V}$ is computed as the difference between the observed and the intrinsic rest-frame V magnitude of a given galaxy, see \S\ref{sect:PostProcessing}. Note that we do not model dust extinction as a simple screen as we find this to overestimate (for a given fixed amount of dust mass) the reddening. Our models do not include a separate circumstellar dust component that is present in local star forming galaxies \citep{Calzetti2000, Wild2011}, and perhaps also at $z\sim{}2$ (e.g., \citealt{ForsterSchreiber2009, Yoshikawa2010, Kashino2013, Price2014}).

\begin{figure*}
\begin{tabular}{cc}
\includegraphics[width=85mm]{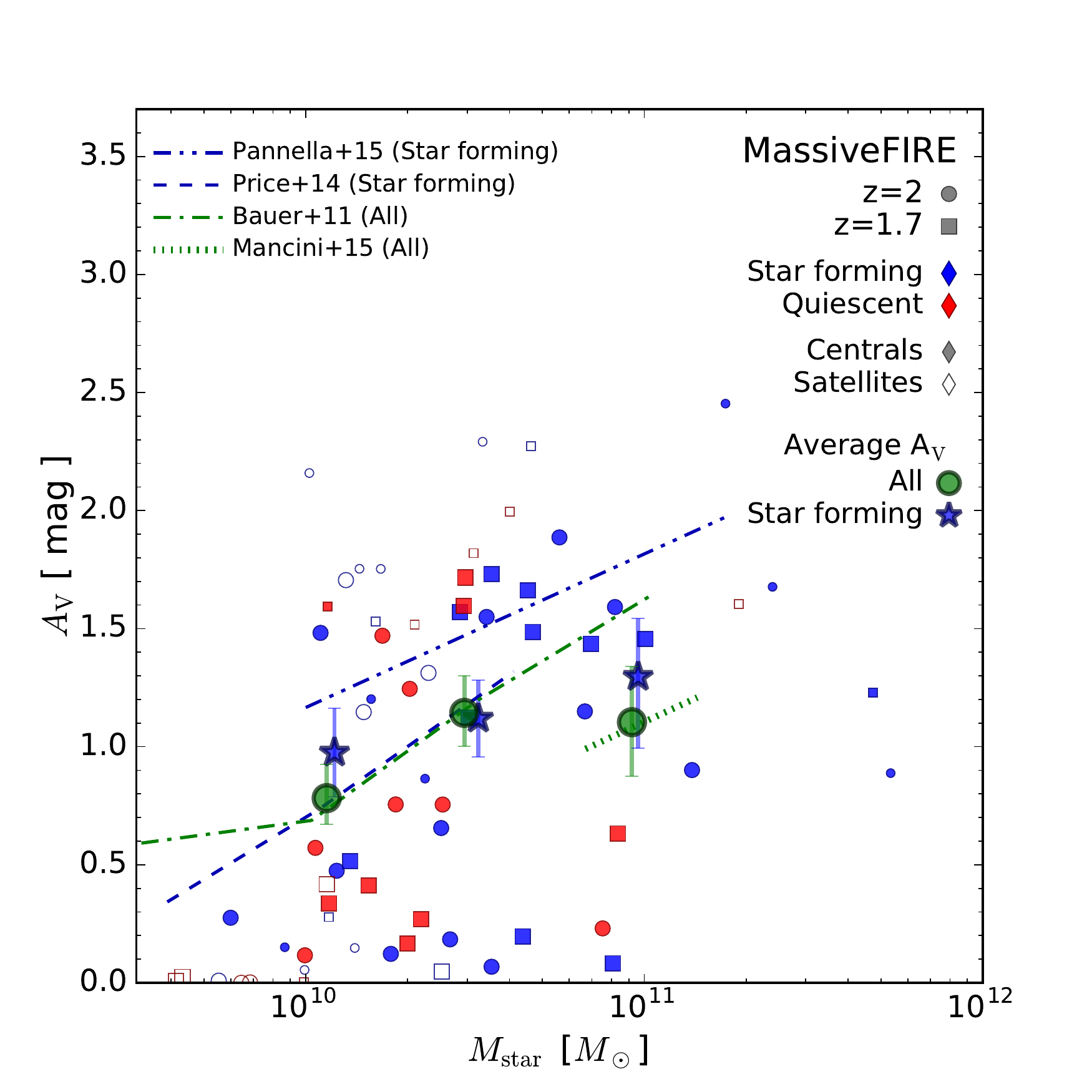} & \includegraphics[width=85mm]{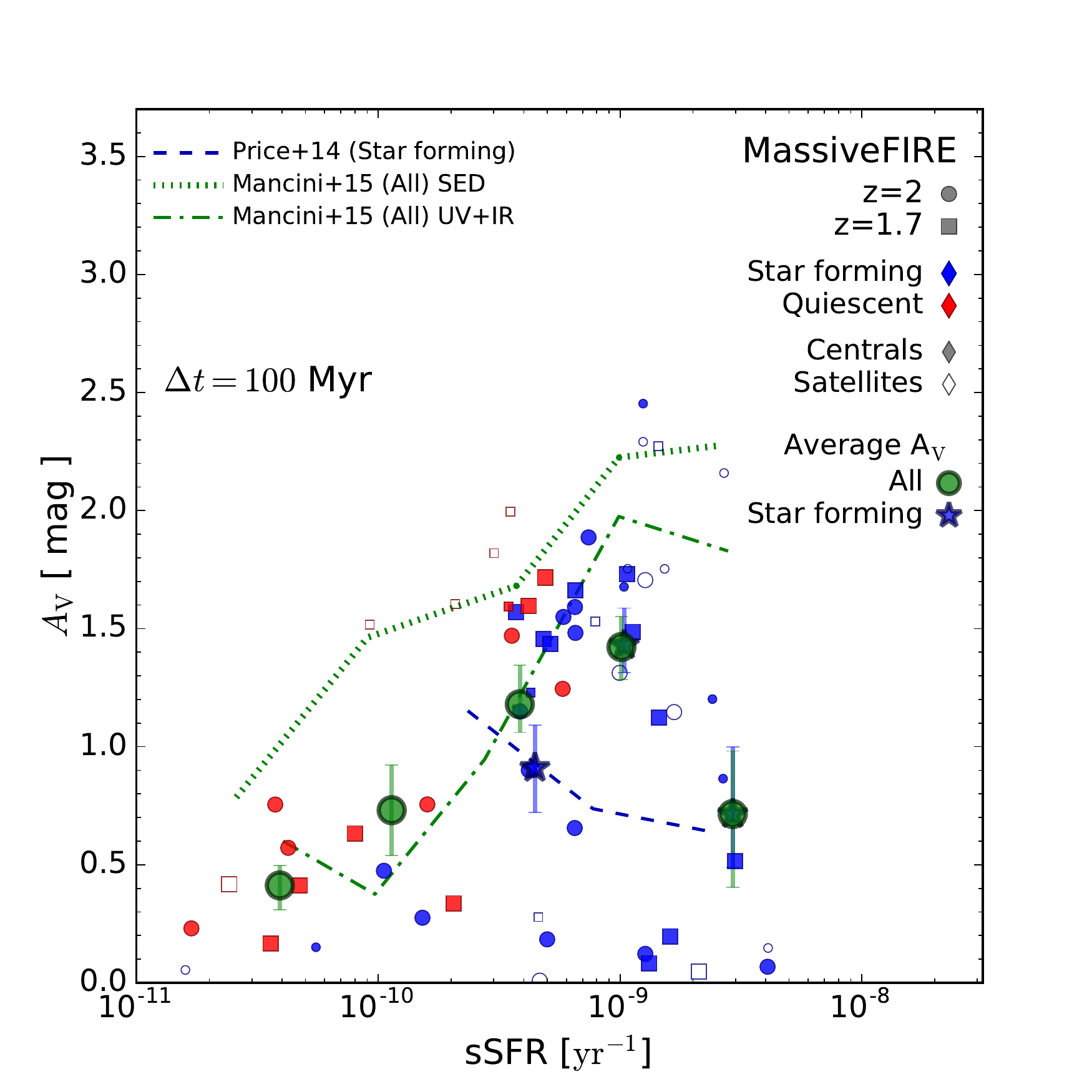}
\end{tabular}
\caption{\emph{$A_{\rm V}$ vs. stellar mass (left) and specific SFR averaged over 100 Myr (right).} Symbols are the same as in Fig.~\ref{fig:UVJ_scat}, except for the large circles (stars) that show the mean $A_{\rm V}$ in bins of $M_{\rm star}$ and sSFR for all galaxies (for star forming galaxies) in our sample. Lines show measurements of the dust extinction at $z\sim{}1.5-2$ \protect\citep{Bauer2011, Price2014, Mancini2015, Pannella2015}. The amount of dust extinction increases with stellar mass and sSFR, perhaps saturating or reaching a maximum for galaxies near the star forming sequence (sSFR $\sim{}10^{-9}$ yr$^{-1}$ at $z\sim{}2$). Overall, our simulations predict levels of dust extinction in broad agreement with observations.}
\label{fig:MstarAV}
\end{figure*}

In Fig.~\ref{fig:MstarAV} we show the dust extinction of our simulated galaxies as function of stellar mass and sSFR. The level of dust extinction in our sample matches estimates for mass-complete\footnote{Stellar masses reported by \protect\citep{Bauer2011, Mancini2015, Pannella2015} are lowered by 0.24 dex to convert from a Salpeter to a Chabrier IMF \protect\citep{Santini2012}.} samples of $z\sim{}1.5-2$ galaxies \citep{Bauer2011, Price2014, Mancini2015, Pannella2015, Dunlop2016}. Furthermore, our simulations predict a scaling of $A_{\rm V}$ with $M_{\rm star}$ and with sSFR in broad agreement with observations. In particular, the amount of dust extinction increases with increasing stellar mass, with increasing SFR (not shown), and with increasing sSFR (for galaxies on and below the star forming sequence). 

Some of our quiescent galaxies contain substantial dust masses. Such galaxies tend to have sSFRs that place them only somewhat below the star forming sequence, but most of their current star formation is dust-enshrouded and the stellar light is dominated by an old stellar population. We also find a significant fraction of galaxies with sSFR (averaged over 100 Myr) near the main sequence that show low amounts of extinction ($A_{\rm V}<0.5$). These galaxies have actually low levels of instantaneous star formation. Dust extinction is more strongly correlated with the sSFR averaged over shorter timescales; we thus predict that an observationally derived correlation between $A_{\rm V}$ and sSFR based on tracers with long life times, e.g., FUV or IR derived estimates, will show a significant amount of scatter.

When considering different galaxy types, we find that quiescent satellites are typically the least dust obscured (the 16, 50, and 84th percentiles of $A_{\rm V}$/mag are 0, 0.41, and 1.78, respectively), followed by quiescent centrals (0.22, 0.6, 1.55), star forming centrals (0.24, 1.14, 1.81), and then star forming satellites (0.05, 1.35, 2.17). The $A_{\rm V}$ distributions of the latter two are statistically indistinguishable (according to Kolmogorov-Smirnov and Anderson-Darling tests). This finding appears to be broadly consistent with the observation that star forming satellite galaxies in groups and clusters have similar properties as star forming galaxies of the same stellar mass in the field (e.g., \citealt{Balogh2004, Park2007, Peng2010, McGee2010, Wijesinghe2012}).

We see a clear trend of $A_{\rm V}$ increasing with time (the median $A_{\rm V}$ for the galaxies in our sample scales approximately as $0.43(t/{\rm Gyr}) - 0.50$) as both the galaxy masses and metallicities increase - broadly in agreement with observations of Lyman break galaxies at high redshift (e.g., \citealt{Steidel1999, Bouwens2009}).

\subsection{The star forming sequence}
\label{sect:SFRrel}

SFRs of non-quiescent galaxies are positively correlated with galaxy masses (e.g., \citealt{Daddi2007a, Whitaker2012b, Speagle2014}). The relation is approximately linear over a large range in stellar masses, i.e., the sSFR is not a strong function of stellar mass, except perhaps at the high mass end. The normalisation of the stellar mass -- sSFR relation evolves strongly with redshift. To zeroth order, the sSFRs of galaxies evolves similar to the specific halo accretion rates (sHAR) of the parent dark matter haloes hosting these galaxies (e.g., \citealt{Dekel2009a, Yang2013, Lilly2013c, Rodriguez-Puebla2016}). However, in more detail there are significant differences. 

For instance, the normalisation of the sSFR evolves faster at $z<1$ and slower at $z>1$ compared with the sHAR (e.g., \citealt{Weinmann2011b}). Furthermore, the sSFR is typically larger than the sHAR by a factor of 2 (even after accounting for gas recycling from stellar mass loss) and also shows a mildly different residual trend with mass \citep{Lilly2013c}. Stellar feedback and subsequent recycling likely play a crucial role in offsetting the evolution of the sSFR of galaxies from the zeroth order expectation set by the halo growth. The stellar mass -- sSFR relation (and its evolution with redshift) thus places important constraints on galaxy formation simulations.

\begin{figure}
\begin{tabular}{c}
\includegraphics[width=80mm]{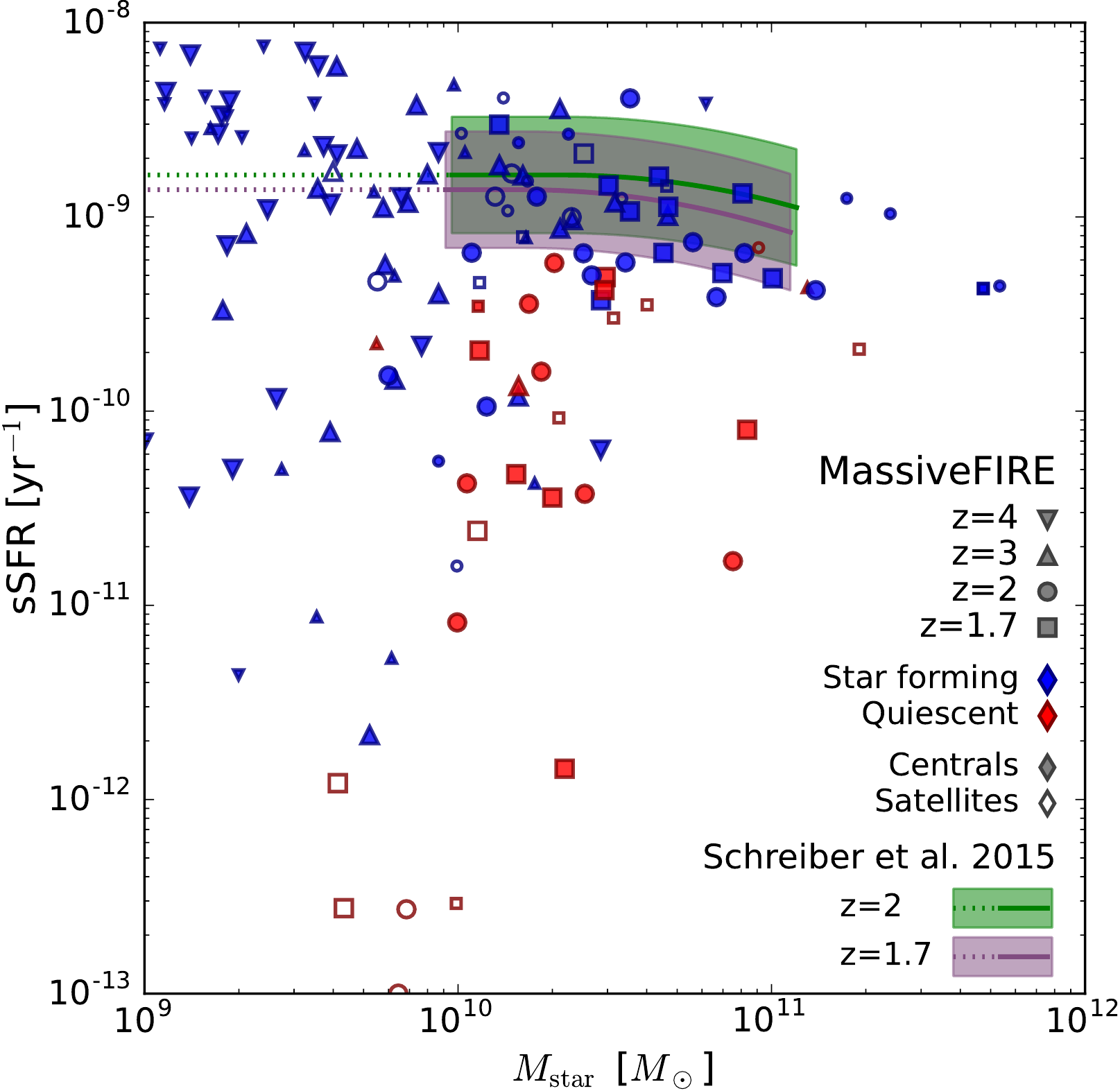}
\end{tabular}
\caption{\emph{Specific star formation rate (sSFR) vs stellar mass.} We split galaxies into quiescent and star forming (red vs. blue symbols) as well as centrals and satellites (filled vs. empty symbols). The different symbol shapes denote redshifts 4, 3, 2, and 1.7, see legend. 
Large and small symbols show HR runs (series A, B, and C) and MR runs (series Cm), respectively. 
The shaded areas show the observed stellar mass -- sSFR relation at $z=1.7-2$ with a 0.3 dex scatter \protect\citep{Schreiber2015b}. Corrections to account for IMF differences are made. Dotted lines extrapolate these measurements to lower stellar masses. Massive galaxies identified as quiescent (star forming) according to their U--V and V--J colours generally lie below (fall onto) the observed star forming sequence.}
\label{fig:Mstar_sSFR}
\end{figure}

Fig.~\ref{fig:Mstar_sSFR} shows the relation between stellar mass and sSFR, the so-called `main sequence' of star formation, for the galaxies in our sample at redshifts $z=4$, 3, 2, and 1.7. Clearly, galaxies at those redshifts show a large range in SFRs (here averaged over 100 Myr). However, much of this variation arises from the subset of quiescent galaxies and from galaxies with $M_{\rm star}<10^{10}$ $M_\odot$ at $z\gtrsim{}3$. Massive galaxies that are star forming at $z=1.7-2$ have a rather narrow spread of SFRs at a given stellar mass, i.e., they form a star forming sequence with a typical sSFR of $\sim{}1-2$ Gyr$^{-1}$. Quiescent galaxies have a large range of (lower) sSFRs $\sim{}0.001-0.5$ Gyr$^{-1}$.

We perform a linear regression to the stellar mass -- sSFR relation of all star forming, central galaxies at $z=1.7$ and $z=2$ that we simulated at HR resolution. Effectively, given the stellar mass selection of our sample, this limits the probed stellar mass range to $\sim{}10^{10}-10^{11}$ $M_\odot$. The dispersion of the sSFR at given $M_{\rm star}$ is $0.35^{+0.08}_{-0.04}$ dex. The confidence interval ($1\sigma$) is obtained from bootstrapping. The logarithmic slope of the stellar mass -- sSFR relation is consistent with zero within the 1-$\sigma$ confidence interval. The normalisation is $\log_{10}{\rm sSFR [yr^{-1}]}=-9.16^{+0.07}_{-0.11}$ for $\log_{10}M_{\rm star}[M_\odot]=10.5$. Our estimates are in similar to the measurement of the stellar mass -- sSFR relation based on deep HST and Herschel photometry \citep{Schreiber2015b}, although perhaps with a $\sim{}0.2$ dex offset towards lower sSFRs. For $z=1.7-2$ galaxies, these authors report a dispersion of $\sim{}0.3$ dex, a vanishing slope, and a normalisation of $\sim{}10^{-9}$ yr$^{-1}$. Similar values for the scatter were reported in other works (e.g., \citealt{Rodighiero2010, Whitaker2012b, Speagle2014, Salmon2015b, Shivaei2015}).

Reproducing the stellar mass -- sSFR relation has long been a challenge for numerical simulations and semi-analytical models (e.g., \citealt{Daddi2007a, Furlong2015a, Somerville2015, Sparre2015a, Johnston2015, Tomczak2016}). In particular, simulations calibrated to reproduce $z=0$ observations often struggle to match the normalisation of the stellar mass -- sSFR relation at $z\sim{}1-2$. At such intermediate redshifts, the observed sSFRs of galaxies deviate noticeably from their specific halo accretion rates (e.g. \citealt{Dave2008, Sparre2015a}) and hydrodynamical simulations typically underpredict the sSFR by up to 0.5 dex (\citealt{Sparre2015a}). The simulations presented in this paper fair somewhat better in matching the normalisation, slope, and scatter of the observed stellar mass -- sSFR relation at $z\sim{}1.7-2$. In addition, \citet{Hopkins2014} demonstrate that FIRE simulations with smaller halo masses ($\sim{}10^{9}-10^{12}$ $M_\odot$ at $z=0$) reproduce the relation at redshifts $z=0-2$ as well.

\cite{Sparre2015a} suggest that the difference between the simulated and the observed  stellar mass -- sSFR relation could be reduced if star formation histories were very bursty. Indeed, the FIRE approach of modelling stellar feedback predicts strongly fluctuating SFRs (\citealt{Hopkins2014, Sparre2015, Muratov2015}, and \S\ref{sect:GalColors}). Also, as shown explicitly in \cite{Hopkins2014}, strong outflows driven by stellar feedback at high $z$ lower the SFR at early times and increase the SFRs at subsequent times when the expelled gas falls back (see also \citealt{Dave2011a, Narayanan2015, Wang2015}; Angl\'es-Alc\'azar et al. in prep). We thus speculate that efficient stellar feedback is responsible for bringing the predictions of \MassiveFIRE{} into better agreement with observations at $z\sim{}1.7-2$.

\begin{figure*}
\begin{tabular}{c}
\includegraphics[width=170mm]{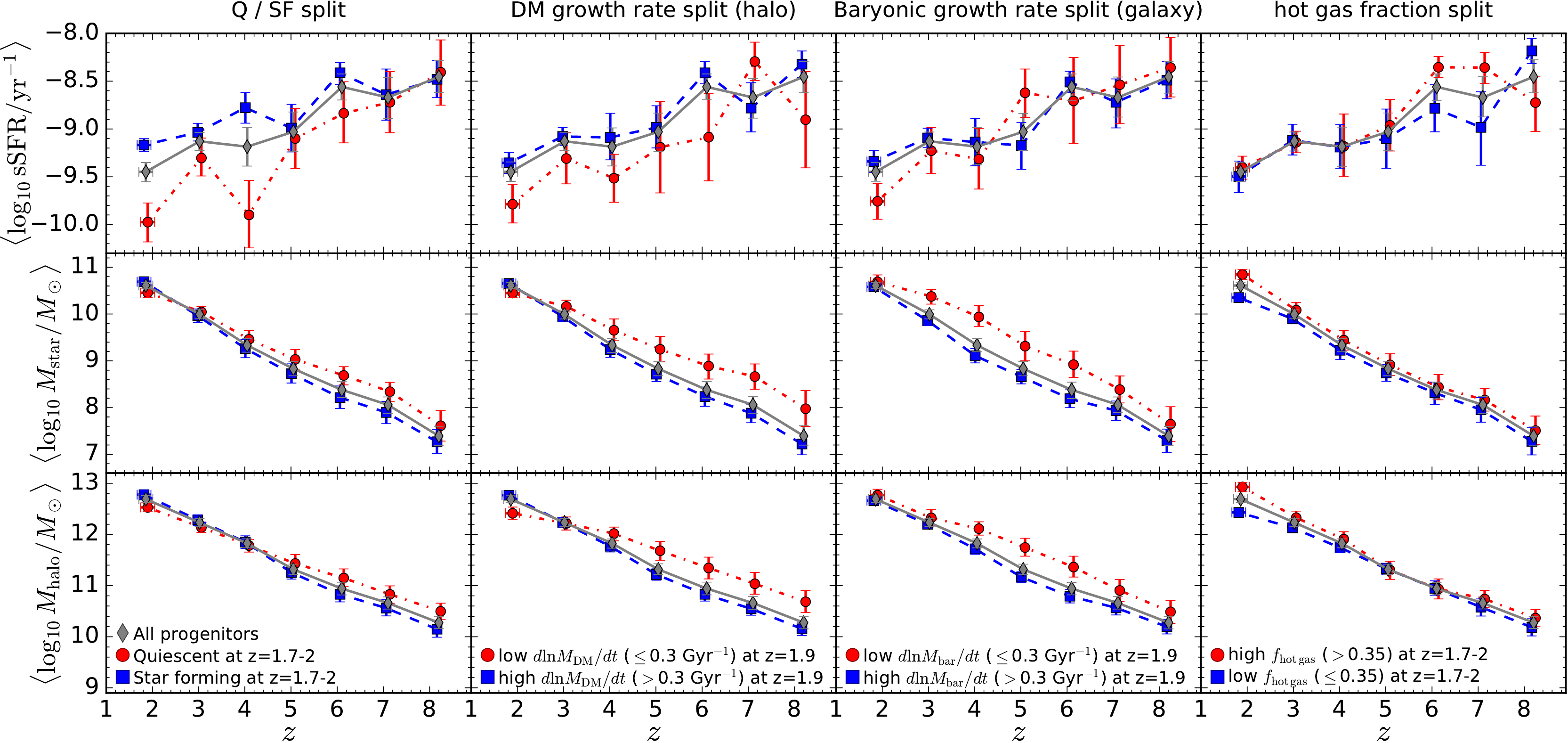}
\end{tabular}
\caption{\emph{Specific SFR, stellar masses, and halo masses of massive galaxies at $z\sim{}2$ and their progenitors.} We show average properties for galaxies with $M_{\rm halo}(z=2)>10^{12}$ $M_\odot$ in our sample, split by their $z=1.7-2$ `quiescent' status (1st column), growth rates of the DM mass of their haloes (2nd column), growth rate of their baryonic mass (3rd column), or their hot gas fractions within $R_{\rm halo}$ (4th column), see text for details. Rows show the evolution of the average sSFRs (top row), stellar masses (middle row), and halo masses (bottom row) of the selected galaxies and their main progenitors. The circular and square symbols are shifted by $\Delta{}z=\pm{}0.04$ for visualisation purposes. Error bars are computed via bootstrapping. The progenitor galaxies of $z\sim{}2$ galaxies that are quiescent or that have low halo / baryonic growth rates are typically more massive at higher redshift, reside in more massive haloes, and have lower sSFR than the progenitor galaxies of $z\sim{}2$ galaxies that are star forming or that have high halo / baryonic growth rates.}
\label{fig:sSFR_evolution}
\end{figure*}

We can calculate the dispersion of the stellar mass -- sSFR relation for the higher redshift (and lower mass) progenitors of $z\sim{}2$ galaxies. For the sub-sample of galaxies that are classified as star forming at their final snapshot ($z\sim{}1.7$ or $z=2$), we find that the scatter at $z=3$ is very similar ($0.29^{+0.09}_{-0.03}$). However, the scatter increases at higher $z$, e.g., it is $0.59^{+0.20}_{-0.10}$ at $z=4$, not surprising given that the progenitors are, at those early times, small dwarfs with bursty star formation \cite{Sparre2015}. We showed in \S\ref{sect:GalColors} that almost all \MassiveFIRE{} galaxies classified as quiescent at $z\sim{}1.7$ are identified as star forming at $z\geq{}3$. Interestingly, the dispersion of the stellar mass -- sSFR relation for these galaxies is generally high and does not strongly evolve with redshift ($0.70^{+0.20}_{-0.09}$ at $z=1.7-2$, $0.84^{+0.38}_{-0.18}$ at $z=3$, and $0.72^{+0.21}_{-0.04}$ at $z=4$). Apparently, galaxies at $z=3$ already ``know'' about their fate one Gyr later.

In Fig~\ref{fig:sSFR_evolution} we investigate this `memory' effect in more detail using all galaxies in our sample that reside in moderately massive haloes ($M_{\rm halo}>10^{12}$ $M_\odot$).
Specifically, we divide our sample into various subsamples and compare the evolution of the sSFRs, stellar masses, and halo masses of both the full sample and each of the subsamples. The first subsamples are created by dividing galaxies into star forming and quiescent based on their rest-frame UVJ colours at the final simulation snapshot ($z=1.7-2$). In addition, we compare galaxies that reside in slowly growing haloes at $z\sim{}2$ with those residing in quickly growing haloes. A specific growth rate (of the DM mass of the halo) of 0.3 Gyr$^{-1}$ at $z=1.9$ is chosen as the dividing line \citep{Feldmann2016}. Third, we distinguish galaxies that grow their baryonic component (the sum of stellar mass and $\HI+\H2$ mass within $0.1 R_{\rm halo}$) at a specific rate below or above 0.3 Gyr$^{-1}$ at $z=1.9$. Furthermore, we compute the hot gas fraction $f_{\rm hot\,gas}=M_{\rm hot\,gas}/(f_{\rm bar}\,M_{\rm 180m})$, where $M_{\rm hot\,gas}$ is the mass of halo gas with a temperature above $2.5\times{}10^5$ K and $f_{\rm bar}=0.163$ is the universal baryon fraction. We then divide galaxies depending on whether the hot gas fractions of their haloes are above or below a certain threshold. The median hot gas fraction of our sample is 0.35 and we adopted this value as our default threshold but discuss other choices as well.

Starting with the star forming and quiescent galaxy sub-samples, Fig~\ref{fig:sSFR_evolution} shows that the average sSFRs of the galaxies in the two sub-samples differ by a factor $\sim{}4-7$ at $z\sim{}2$ while the average stellar and halo masses are relatively similar. In addition, even at $z\sim{}4$, the progenitors of $z=2$ quiescent galaxies have average sSFRs that are lower than those of the progenitors of star forming galaxies. This demonstrates that sSFRs at $z>2$ are not entirely driven by short-term physics but that they are also tied to the growth rate of their parent haloes.

Furthermore, the haloes harbouring $z\sim{}2$ quiescent galaxies tend to grow more slowly than the haloes of star forming galaxies at late times, in agreement with an earlier analysis of \MassiveFIRE{} simulated galaxies \citep{Feldmann2016}. Also, quiescent galaxies assemble much of their stellar mass earlier than star forming galaxies \emph{of the same final stellar mass}. Systematic differences in the stellar growth history of star forming and quiescent galaxies of same mass may have important practical implications for the accuracy of abundance matching techniques, as discussed in \cite{Clauwens2016}.

Dividing galaxies based on the growth rates of their haloes leads to a very similar overall picture. In short, galaxies that reside in slowly growing haloes at $z\sim{}2$ tend to have lower sSFRs than galaxies residing in quickly growing haloes. This is consistent with the picture of `cosmological starvation', i.e., a link between the star formation rates of galaxies and the growth rates of their haloes. We stress that the halo growth rate is computed using only the dark matter mass of the halo, i.e., the growth rate should primarily be determined by the gravitational growth of cosmological structure. Consequently, it is plausible that a low halo growth rate \emph{causes} a reduced star formation activity of galaxies instead of merely being correlated with it.

A qualitatively similar picture is obtained again if galaxies are divided based on the growth of their baryonic component. Galaxies that grow quickly tend to have higher sSFRs than galaxies that grow slowly. Furthermore, the stellar and halo mass growth histories differ. Unsurprisingly, galaxies that grow quickly (slowly) at late times tend to be less (more) massive and reside in less (more) massive haloes earlier on.

An interesting question is whether the reduced star formation activity in galaxies with low halo growth rates is related to the build-up of a hot gas halo. We therefore also split our sample into galaxies residing in haloes with high ($>0.35$) and low ($\leq{}0.35$) hot gas fractions of their haloes. We do not find a significant difference in their average sSFRs. This results also holds if we reduce the value of the critical hot gas fraction to 0.28 (the first quartile of the distribution of hot gas fractions) or increase it to 0.45 (the third quartile). However, galaxies with a higher hot gas fraction tend to be more massive and reside in more massive halo across the whole redshift range. Finally, we also increase the minimum temperature defining gas as `hot' from $2.5\times{}10^5$ K to $10^6$ K. Again we find that galaxies with high and low hot gas fractions have comparable sSFRs. However, the higher temperature threshold increases the average stellar and halo masses of galaxies with high hot gas fractions compared to those with low hot gas fractions. We conclude that the fraction of hot gas in moderately massive haloes of $z\gtrsim{}2$ galaxies is not a strong predictor of star formation activity. This conclusion could change once feedback from AGN is included.

\subsection{Gas fractions and depletion times}
\label{sect:Gas}

\begin{figure*}
\begin{tabular}{cc}
\includegraphics[width=80mm]{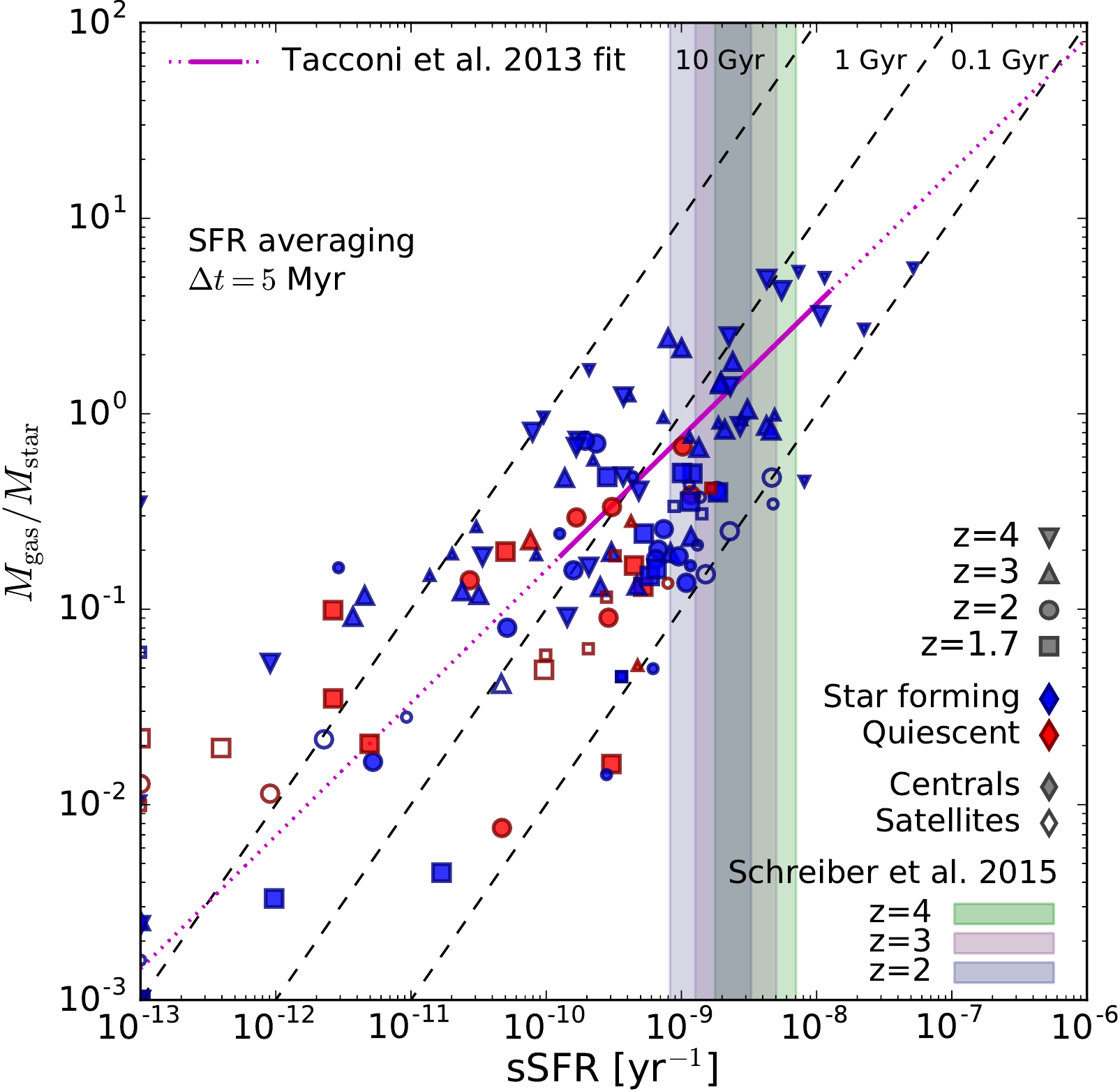} & \includegraphics[width=80mm]{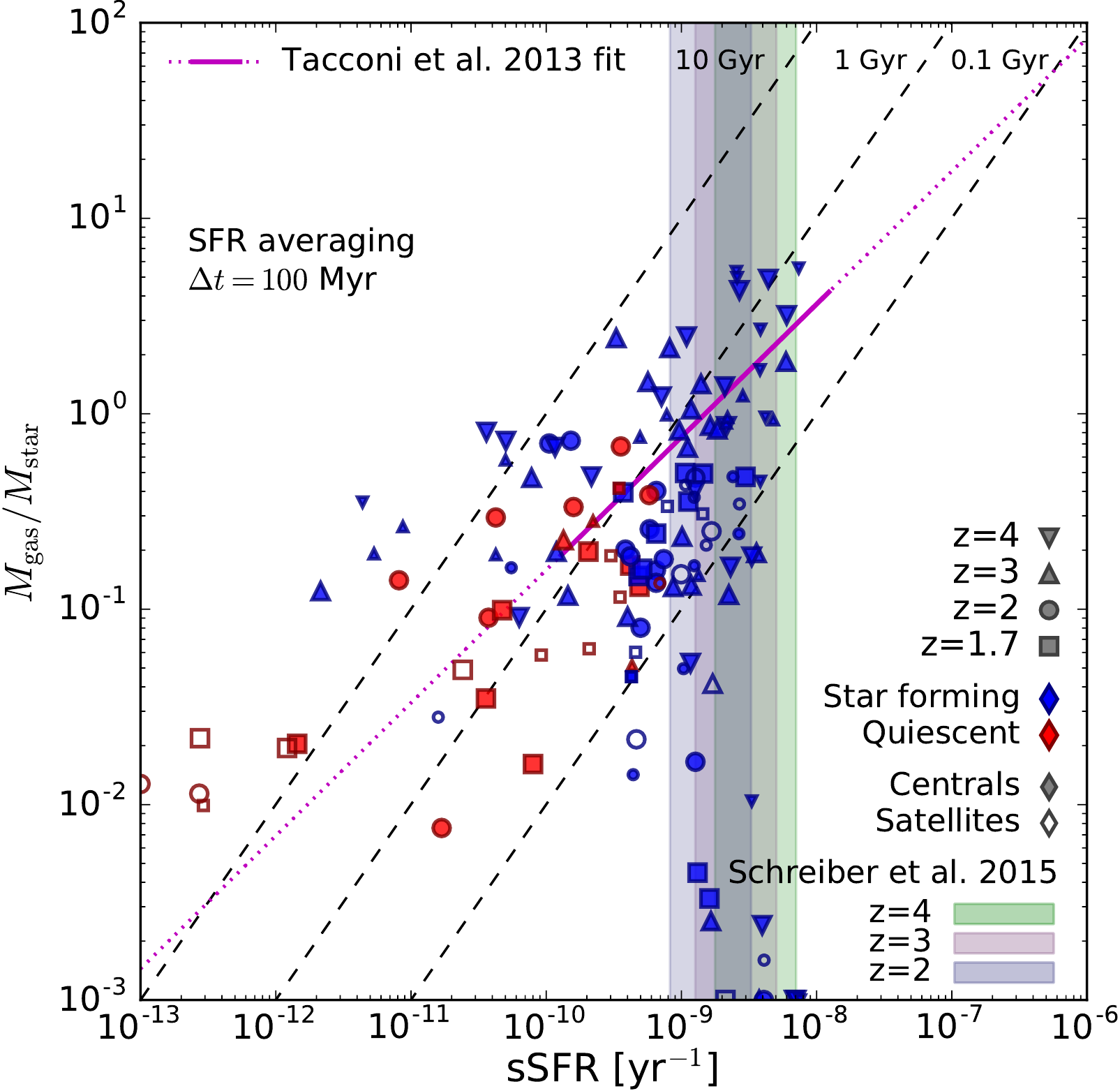}
\end{tabular}
\caption{\emph{Gas fraction ($M_{\rm gas}/M_{\rm star}$) versus sSFR}.
SFR are measured on 5 Myr (left panel) and on 100 Myr (right panel) timescales. Diagonal dashed lines indicate depletion times (SFR$/M_{\rm gas}$) of 100 Myr, 1 Gyr, and 10 Gyr (from bottom to top). The vertical shaded regions indicate the position and spread of the star forming sequence reported by \protect\cite{Schreiber2015b} for $M_{\rm star}=2\times{}10^{10}$ $M_\odot$ galaxies. A fit to the observations by \protect\cite{Tacconi2013a} is shown by the solid magenta line. Extrapolations to lower and higher sSFRs are indicated by the dotted lines. Galaxies with sSFRs above the star forming sequence have both higher gas fractions and lower depletion times than galaxies on or below the star forming sequence. Quiescent galaxies tend to have lower gas fractions and longer depletion times than star forming galaxies.}
\label{fig:fgas}
\end{figure*}

Nearby star forming and quiescent galaxies differ not only in their colours, SFRs, and dust abundances, but also in their gas fractions. Specifically, at fixed stellar mass, quiescent galaxies harbour less neutral and molecular interstellar gas (e.g., \citealt{Schiminovich2010, Saintonge2011g, Serra2012a}). A qualitatively similar result appears to hold at high redshift although perhaps with a change in the overall normalisation (e.g., \citealt{Sargent2015, Gobat2017}). Overall, star forming galaxies with SFRs above the star forming sequence tend to have higher gas fractions and lower gas depletion times than galaxies that lie below the star forming sequence \citep{Saintonge2011g, Tacconi2013a, Genzel2015, Scoville2016, Schinnerer2016}.

The gas fractions of the galaxies in our sample are shown in Fig.~\ref{fig:fgas}. The main results shown by the figure are in qualitative agreement with observations. Specifically, we find that star forming galaxies tend to have higher gas fractions and slightly shorter gas depletion times than quiescent galaxies. Furthermore, gas-to-stellar mass ratios ($\mu_{\rm gas}$) scale with sSFRs. The scaling is sub-linear, however, implying that galaxies with low sSFRs are not only gas poor but also comparably inefficient in converting the existing gas into stars.

For a quantitative analysis we perform a linear regression between $\log_{10}\mu_{\rm gas}$ and $\log_{10}\sSFR$ for all central galaxies in our $z=1.7-2$ sample with stellar masses above $10^{10}$ $M_\odot$ and with $\sSFR$ in the range $10^{-9.9} - 10^{-7.9}$ yr$^{-1}$, similar to the properties of the sample of \cite{Tacconi2013a}. We find $\log_{10}\mu_{\rm gas} = 0.34(0.23) \log_{10}(\sSFR/{\rm Gyr}) - 0.55(0.09)$. While the slope is consistent with \cite{Tacconi2013a} given statistical errors, our simulations predict at face value lower gas fractions by $\sim{}0.4$ dex. However, gas fractions depend somewhat sensitively on the radius within they are measured. For instance, increasing the radius within which $\mu_{\rm gas}$ and sSFRs are measured from 5 to 10 kpc changes the regression result to $\log_{10}\mu_{\rm gas} = 0.51(0.18) \log_{10}(\sSFR/{\rm Gyr}) - 0.34(0.08)$. The offset between simulation predictions and observations is now $\sim{}0.2$ dex. The combined systematic uncertainties of SFR, stellar mass, and especially gas mass estimates are likely significantly larger than 0.2 dex.

The relation between $\mu_{\rm gas}$ and sSFR is significantly tighter if near-instantaneous SFRs are used instead of SFRs averaged over long timescales. This is not surprising as the instantaneous SFRs in our simulations (as well as in nature) depends primarily on the presence of (dense molecular) gas. SFRs measured on 100 Myr correlate well with galaxy colours (see Fig.~\ref{fig:colorSSFR}) but lag behind short-term changes in the gas reservoirs. Consequently, long SFR averaging timescales result in transition galaxies that are still `on' the star forming sequence but have low gas fractions (galaxies leaving the star forming sequence) as well as gas-rich galaxies with low sSFRs (galaxies returning to the star forming sequence). Galaxies classified as quiescent based on their colours are much less sensitive to the choice of the SFR averaging timescale than star forming galaxies. This finding is consistent with the result presented in \S\ref{sect:GalColors} that quiescent galaxies experience prolonged periods of reduced star formation activity while the SFRs of many star forming galaxies fluctuate strongly on short timescales.

\subsection{The environments of quiescent and star forming galaxies}
\label{sect:environ}

A large body of observational evidence points to a link between the environment of galaxies and their star formation activity (e.g., \citealt{Dressler1985a, Baldry2006, Weinmann2010, Peng2010, McGee2010}). At $z\leq{}1$ the likelihood that a galaxy is `quenched', i.e., that its SFR is much lower than those of star forming galaxies of similar stellar masses, increases both with the stellar mass and the environmental overdensity \citep{Peng2010}. \citet{Kovac2014} argue that the environmental dependence has its origin in a higher satellite quenching efficiency in denser environments. Hence, `environmental quenching' is perhaps a result of environmental processes that affect primarily member galaxies of groups and clusters via, e.g., ram pressure stripping, tidal stripping, or harassment. However, environmental processes may leave an imprint out to 2-3 virial radii \cite{Balogh2000} and, hence, may  also affect a large fraction of central galaxies \citep{Cen2014a}.

\citet{Feldmann2015} and \citet{Feldmann2016} argue that a significant fraction of massive galaxies at $z\sim{}2$ are subject to an additional process related to the gravity-driven growth of dark matter haloes. In particular, they show that quiescent (star forming) galaxies at $z\sim{}2$ reside preferentially in haloes with low (high) dark matter accretion rates. This process, `cosmological starvation', does not explain why the quiescent fraction strongly increases with stellar mass at $z\lesssim{}1$ but it sheds light on the co-existence of $z\sim{}2$ star forming and quiescent central galaxies \emph{with the same stellar mass}. 

In this section we discuss the environments of quiescent and star forming galaxies in our sample. This analysis complements the work by \citet{Feldmann2016} who discuss the fractions of star forming and quiescent galaxies in \MassiveFIRE{} and their dependences on stellar mass.

We study this using two different measures of the local environment (\S\ref{sect:PostProcessing}). The local environmental density, $\rho$, includes all matter within a sphere of $5\,R_{\rm halo}$ centred on a given halo but with the halo itself excluded, while the Hill radius, $R_{\rm Hill}$, defines a minimal distance beyond which material is no longer gravitationally bound to the primary halo. The latter is also related to the maximal tidal force exerted by neighbouring haloes on the primary halo \citep{Hahn2009}.

\begin{figure}
\begin{tabular}{c}
\includegraphics[width=85mm]{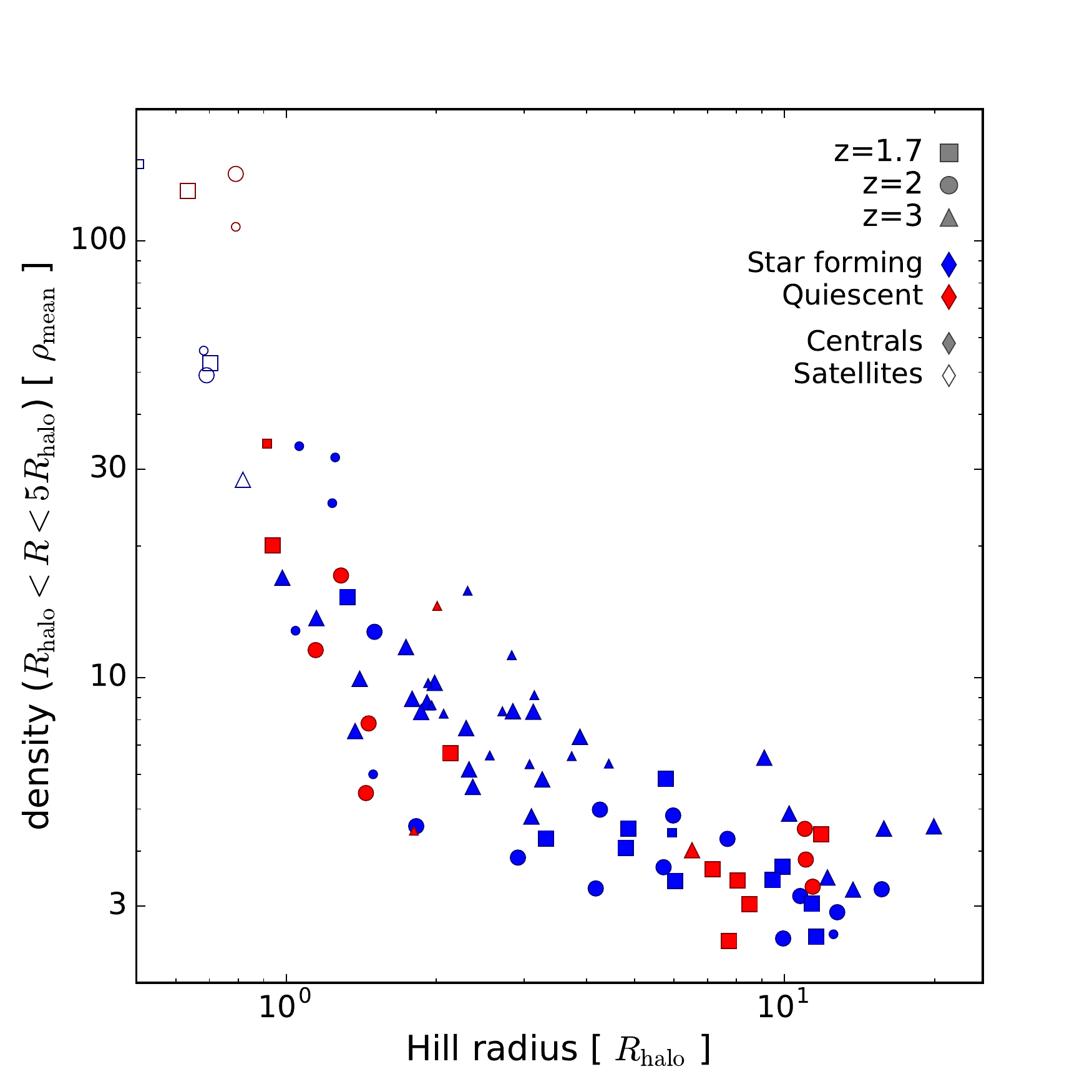}
\end{tabular}
\caption{\emph{Local environmental density as function of Hill radius.} The y-axis shows the matter density within a shell $R_{\rm halo}<R<5R_{\rm halo}$ in units of the mean density of the Universe at the given epoch. The x-axis shows the Hill radius (see text) associated with each galaxy in units of its $R_{\rm halo}$. In contrast with star forming galaxies, quiescent centrals in our sample have a bimodal distribution of Hill radii clustered around $\sim{}R_{\rm halo}$ and $\sim{}10\,R_{\rm halo}$.}
\label{fig:Hill}
\end{figure}

We plot the local environmental density of our galaxies as function of the Hill radii of their haloes in Fig.~\ref{fig:Hill}. Both measures of environments are clearly correlated. Galaxies residing in regions with a larger environmental density have shorter Hill radii. This result is expected, as the presence of neighbouring massive haloes increases both the environmental density and decreases the Hill radius \citep{Hearin2016}. For satellites, $R_{\rm Hill}$ ($\rho$) is dominated by the parent halo, hence they have the smallest (largest) values of $R_{\rm Hill}$ ($\rho$). The locus of the density -- Hill radius relation does not evolve strongly with redshift over the $z=1.7-3$ range. 

Fig.~\ref{fig:Hill} shows that star forming centrals span the whole range of Hill radii while quiescent centrals are clustered at $R_{\rm Hill}\sim{}1-2$ and $R_{\rm Hill}\sim{}7-11$. This bimodal distribution is statistically significant (according to a dip test, \citealt{Hartigan1985a}) even given our small sample size.

Quiescent centrals with large $R_{\rm Hill}$ are more massive ($M_{\rm star}\sim{}2-9\times{}10^{10}$) than those with small $R_{\rm Hill}$ ($M_{\rm star}\sim{}1-2\times{}10^{10}$) adding further evidence that they belong to two different classes of galaxies. All quiescent galaxies with large $R_{\rm Hill}$ reside at the centres of specifically targeted (`primary') \MassiveFIRE{} haloes, see \S\ref{sect:Setup}. In contrast, most of the quiescent centrals with low $R_{\rm Hill}$ reside in isolated haloes that happen to lie in one of the highly-resolved zoom-in regions.

Quiescent centrals with small $R_{\rm Hill}$ values are likely tidally affected by the presence of nearby massive haloes \citep{Wang2007} and, hence, need to compete with those haloes for the gas in their vicinity. We thus confirm that environmental effects can alter the SFRs and colours of galaxies at several kpc from massive haloes. The low $R_{\rm Hill}$ quiescent galaxies appear to form a continuous sequence matching quiescent satellites. The physics driven the SFR and colour evolution might be the same in both cases, low $R_{\rm Hill}$ quiescent galaxies just have not yet crossed the halo to become satellites.

\subsection{Stellar mass -- Halo mass relation}
\label{sect:SHMR}

Stellar masses of galaxies do not scale linearly with the masses of their parent dark matter haloes (e.g., \citealt{More2011, Yang2012, Moster2013, Behroozi2013c, Shankar2014a, VanUitert2016}). Cosmological simulations of galaxies have been struggling to reproduce this observation quantitatively (e.g., \citealt{Scannapieco2012, Martizzi2012d}), although significant advances have been made in recent years in the modelling galaxies of low to moderate masses (e.g., \citealt{Guedes2011, Aumer2013, Munshi2013, Hopkins2014}). Much of the recent progress stems from the realisation by many research groups that stellar feedback has to be modelled more accurately in order to retain its effectiveness in regulating star formation and in generating galactic outflows. Some challenges remain, however, as it is unclear, for instance, whether stellar feedback is sufficient to reduce the gas-to-star conversion efficiency in \emph{massive} galaxies to the observed level. Simulations without AGN feedback typically predict stellar masses of massive galaxies that are at least a factor 2 too high (e.g., \citealt{Feldmann2010a, Martizzi2012d}). However, in contrast with the stellar feedback model adopted here, see \S\ref{sect:PhysicsNumerics}, many of the previous models do not account for the full energy and momentum input from stellar sources. 

\begin{figure}
\begin{tabular}{c}
\includegraphics[width=85mm]{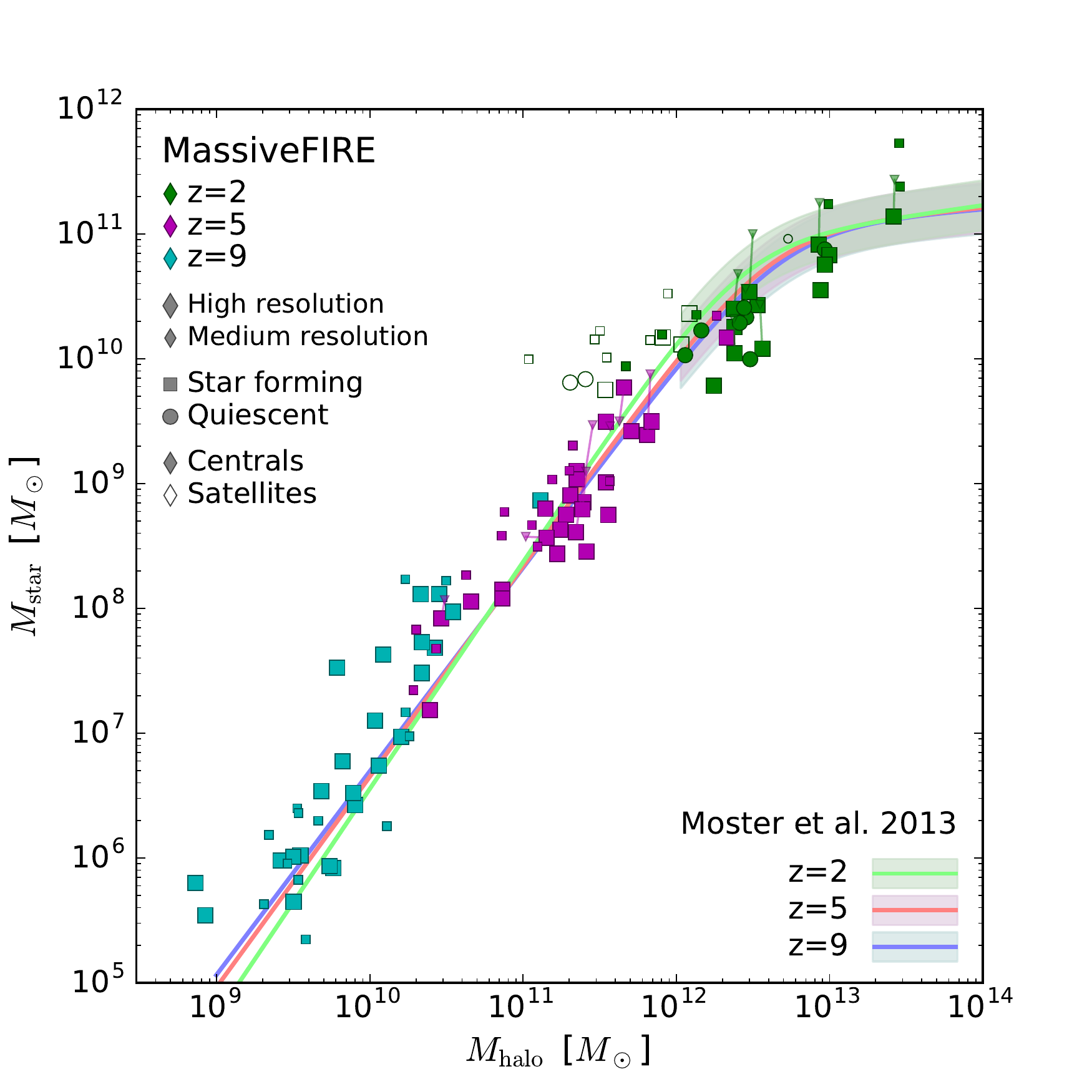}
\end{tabular}
\caption{\emph{Stellar mass -- halo mass relation (SHMR) of the galaxies in our sample.} The plot shows stellar and halo masses at $z=2$ (green symbols, top right), $z=5$ (purple, middle), and $z=9$ (cyan, bottom left). Circles and squares indicate whether galaxies are classified as quiescent and star forming, respectively, according to the UVJ criterion \protect\citep{Whitaker2011d}. Filled and empty symbols distinguish central from satellite galaxies. Large (small) symbols show simulations run at high (medium) resolution. Small triangles show MR re-simulations of HR runs (solid lines connect the corresponding runs). The three dotted lines show the abundance matching estimate \protect\citep{Moster2013} of the SHMR and its extrapolation to high redshifts and low stellar masses. Halo masses by \protect\cite{Moster2013} are converted to match the definition of \S\ref{sect:Setup} using an average halo mass -- concentration relation, see \protect\cite{Kravtsov2014}. The overlapping semi-transparent patches show the expected scatter of 0.2 dex from the mean relation in $M_{\rm halo}\gtrsim{}10^{12}$ $M_\odot$ haloes \protect\citep{Moster2013, Behroozi2013c, Reddick2013}. The simulations agree well with predictions from abundance matching for haloes in the $M_{\rm halo}\sim{}10^{10}-10^{13}$ $M_\odot$ range.}
\label{fig:Mhalo_Mstar}
\end{figure}

Fig.~\ref{fig:Mhalo_Mstar} shows that the relation between $M_{\rm star}$ and $M_{\rm halo}$ for our sample is in good agreement with the empirically derived relation based on abundance matching \citep{Moster2013}. The agreement holds over five orders of magnitude in halo mass and over a large redshift range\footnote{The \citealt{Moster2013} predictions are derived based on stellar mass functions that span the $z=0-4$ range and for halo masses $>10^{11}$ $M_\odot$, i.e., the $z=9$ curve is an extrapolation. Our predicted SHMR is also in agreement with the fit from \citealt{Behroozi2013c} that incorporates stellar mass function estimates at higher $z$. It should be noted that abundance-matching estimates at high redshift suffer from significant systematic uncertainties.}. The predicted stellar masses at $z\sim{}2$ are somewhat on the lower side. There are several possible explanations, such as stellar feedback being too efficient in our simulations or our simulations underestimating cooling from the halo, perhaps due to uncertainties in how gas mixing is handled.

Table~\ref{tab:Res} in the Appendix provides more detail on the SHMR of central galaxies in the \MassiveFIRE{} sample for different halo mass and redshift ranges. The logarithmic slope changes from about two for $10^{9}-10^{11}$ $M_\odot$ haloes at $z=9$, to less than one for $10^{12}-10^{14}$ $M_\odot$ haloes at $z=1.7-2$. We do not find strong evidence that the SHMR is significantly different for star forming and quiescent galaxies at $z=2$. We note that the limited size of our sample prevents us from detecting small differences with statistical significance.

We calculate the scatter of the relation between $\log_{10}M_{\rm star}$ and $\log_{10}M_{\rm halo}$ via a least squares linear regression, excluding satellite galaxies. The scatter is about 0.2 dex for our sample at $z\sim{}2$. This amount of scatter agrees with estimates based on abundance matching constraints, kinematic measurements, and weak lensing maps for galaxies in $\gtrsim{}10^{12}$ $M_\odot$ haloes at lower redshifts (e.g., \citealt{More2009, Leauthaud2012a, Behroozi2013c, Reddick2013}). The scatter is larger for the less massive progenitor galaxies at higher $z$ (e.g., $\sim{}0.25$ dex at $z\sim{}7$) but at $z=0$ such low mass haloes also exhibit larger scatter (the precise value is currently not well constrained, e.g.,  \citealt{Brook2015, Garrison-Kimmel2016}.)

The parent haloes of galaxies in our sample span a broad range of halo masses, up to $10^{13.5}$ $M_\odot$. If we assume that the SHMR is not strongly evolving with redshift at $z\geq{}2$, e.g., \citet{Moster2013}, but cf. \citet{Behroozi2013c}, then we can combine our data from different redshift epochs ($z\sim{}1.7-9$) and fit the SHMR over the large mass range sampled by \MassiveFIRE{}. A quadratic dependence
\begin{multline*}
\log_{10}(M_{\rm star}/M_{\rm halo}) = \\
-0.089^{+0.019}_{-0.015} [\log_{10} (M_{\rm halo} / M_{\odot}) - 12.91^{+0.41}_{-0.23}]^2 - 2.20^{+0.05}_{-0.03} 
\end{multline*}
is visually a good fit to the \MassiveFIRE{} data (reduced $\chi^2$ of 1.39 if we assume a fixed dispersion of 0.25 dex). The ratio between stellar mass of the galaxy and halo mass reaches a maximum for haloes with $M_{\rm halo}\sim{}10^{12.9}$ $M_\odot$, perhaps slightly higher than, but still in approximate agreement with, estimates based on abundance matching at $z\sim{}2$. We can also fit $M_{\rm star}/M_{\rm halo}$ with the double power law suggested by \citet{Moster2013} that has 4 free parameters (their equation 2). The fitted slope in the low mass regime, $\beta=0.63^{+0.18}_{-0.11}$, is consistent with the $\beta$ value reported by \citet{Moster2013} for redshift 2.

\section{Summary and Conclusions}
\label{sect:Summary}

This paper analyses the global properties of over 30 massive galaxies at $z\sim{}2$ that were simulated in high-resolution, cosmological zoom-in simulations (\MassiveFIRE{}). We focus in particular on the differences between star forming and quiescent galaxies residing in or near group-sized haloes ($M_{\rm halo}\sim{}10^{12.5}-10^{13.5}$ $M_\odot$) at $z\sim{}2$. The resolution, numerical methods, and the modelling of star formation and stellar feedback match previous runs by the FIRE project \citep{Hopkins2014, Faucher-Giguere2015} but here we study galaxies residing in order-of-magnitude more massive haloes. The modelling approach adopted by FIRE has been validated against observations in a number of published works (see introduction) allowing us to explore the evolution of massive galaxies in the absence of AGN feedback.

In this paper, we focus on the colours, masses, SFRs, and environments of star forming and quiescent galaxies. Additional properties, such as galaxy sizes and ISM properties, will be discussed in future work. Our simulations include galaxies with a range of morphologies, including dusty disk galaxies, star forming irregular galaxies, and early type galaxies with low dust content and reduced SFRs. About 1/3 of the galaxies in the sample have an extended ($>$kpc) stellar disk, see Figs.~\ref{fig:UVJ_images} and \ref{fig:UVJ_images_nd}.

Galaxies are divided into star forming or quiescent based on rest-frame U, V, and J colours. Furthermore, galaxies are classified as centrals or satellites depending on whether they reside at the centres of isolated haloes or sub-haloes, respectively. Quiescent galaxies make up about half the sample at $z\sim{}2$ and are comprised in about equal parts of centrals and satellites. The quiescent fraction of our sample decreases with increasing redshift reaching zero at $z=4$. However, because we follow the evolution of a specific set of galaxies, their $z\sim{}4$ progenitors are not massive galaxies (none with $M_{\rm star}\gtrsim{}10^{10}$ $M_\odot$). Two-thirds of the star forming galaxies at $z\sim{}2$ in our sample are centrals and one-third are satellites. 

Various properties of observed galaxies are reproduced reasonably well in these simulations despite the absence of AGN feedback. Simulated galaxies show a similar mix of morphologies (e.g., star forming disk galaxies, irregular galaxies, dust-poor early type galaxies; Fig.~\ref{fig:UVJ_images}), a reasonable scaling of dust extinction with stellar mass and SFR (Fig.~\ref{fig:MstarAV}), and both a stellar mass -- sSFR relation (Fig.~\ref{fig:Mstar_sSFR}) and a stellar mass -- halo mass relation (Fig.~\ref{fig:Mhalo_Mstar}) with normalisation and scatter in broad agreement with observations. Gas fractions in the simulations are in rough agreement with observations although perhaps somewhat biased low (Fig.~\ref{fig:fgas}). On the other hand, the simulations do not account for the observed population of massive galaxies with the reddest U-V colours (Fig.~\ref{fig:UVJ_z}). Furthermore, they do not show clear evidence for a colour bimodality at $z\sim{}2$.

Our main findings include the following:

\begin{enumerate}

\item Galaxies migrate between the star forming and quiescent populations as the colours of their stellar populations change and evolve. Our sample includes both galaxies that are quiescent for only a brief period of time ($<100$ Myr) as well as those that become quiescent and remain so for up to 1 Gyr, in roughly comparable numbers. The classification of galaxies into quiescent and star forming is not sensitive to the chosen line-of-sight.

\item The broad-band colours of our simulated galaxies reasonably overlap with those of observed galaxies at $z\sim{}2$. However, there appear to be significant differences in detail. In particular, our simulations do not produce galaxies with very red U--V colours (${\rm U-V}>1.6$) and they also predict a significant fraction of star forming galaxies somewhat closer to the star forming vs. quiescent separation than observed.

\item We do not find statistically significant evidence for a colour bimodality in our sample. However, our simulation sample is likely too small to detect a colour bimodality even if present (see Appendix \ref{sect:ColorBimodality}).

\item Galaxies at $z\sim{}2$ that are classified as quiescent based on their U--V and V--J colours come in two varieties, see Fig.~\ref{fig:MstarAV}. The first class consists of galaxies with low sSFR and relatively low amounts of dust extinction ($A_{\rm V}\sim{}0.3$ mag). The second class includes galaxies with somewhat reduced sSFR (compared to the star forming sequence) that harbour significant amounts of dust ($A_{\rm V}\sim{}1.5$ mag). Light from on-going star formation in these galaxies is largely blocked by the dust and the stellar colours are dominated by older stars.

\item The star forming galaxies in our sample reproduce the observed relation between stellar masses and sSFRs at $z\sim{}1.7-2$, i.e., the star forming (`main') sequence, including normalisation ($1-2\,$Gyr$^{-1}$) and scatter ($\sim{}0.3$ dex). The SFR histories of individual galaxies are bursty, with starbursts followed by a brief ($\lesssim{}100$ Myr) suppression of star formation activity, and subsequent return to the star forming sequence. In most cases, these starbursts are not triggered by galaxy mergers. 

\item The SFRs of high redshift galaxies are affected by both internal processes (starbursts and outflows) that change the SFR on short ($<100$ Myr) timescales, see Fig.~\ref{fig:sSFR_evolution_examples}, as well as external processes (e.g., cosmological starvation) that determine the average SFRs and galaxy colours over longer time scales, see Fig~\ref{fig:sSFR_evolution}.

\item The median dust extinction of our sample increases with time as galaxies become more massive and metal-rich. The $A_{\rm V}$ distributions of star forming centrals and star forming satellites are statistically indistinguishable. At $z\sim{}1.7-2$, the dust extinction is significantly higher in star forming galaxies than in quiescent galaxies.

\item The Hill radius of a halo is a good proxy for the local environmental density. Galaxies residing in denser (less dense) environments have smaller (larger) Hill radii. The Hill radius distribution of quiescent \emph{centrals} is bimodal (Fig.~\ref{fig:Hill}), suggesting perhaps different pathways for reducing the star formation activity in central galaxies depending on environment.

\item The predicted stellar mass -- halo mass relation (SHMR) agrees with the observed relation over 5 orders of magnitude in halo mass, see Fig.~\ref{fig:Mhalo_Mstar}. This finding extends the result previously found for lower-mass galaxies simulated with FIRE physics \citep{Hopkins2014}. The scatter of the SHMR is $\sim{}0.2$ dex for massive galaxies at $z\sim{}2$. 

\item The average stellar masses (and halo masses) of star forming and quiescent central galaxies in our sample at $z\sim{}2$ are comparable. However, the $z\geq{}4$ progenitors of $z\sim{}2$ star forming centrals are typically less massive than the progenitors of quiescent centrals. Hence, stellar and halo masses of star forming centrals grow faster between $z\sim{}2-4$ than those of quiescent centrals.

\end{enumerate} 

\begin{figure}
\begin{tabular}{c}
\includegraphics[width=85mm]{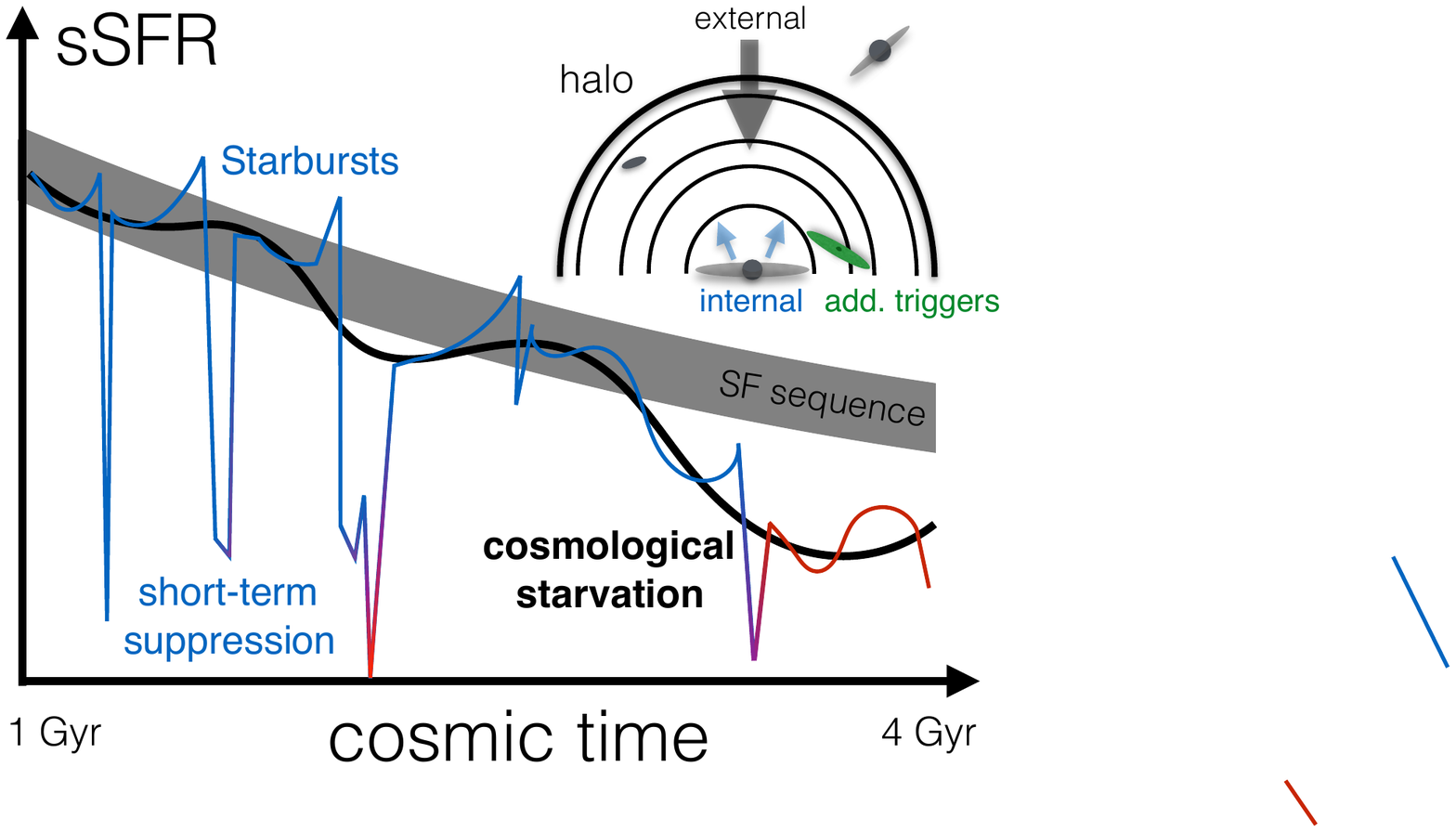}
\end{tabular}
\caption{\emph{Schematic representation of the SFR evolution of simulated central galaxies with moderate masses at the Cosmic Noon (neglecting AGN feedback).} At early times ($z\gtrsim{}3$), the sSFRs of central galaxies co-evolve with the star forming sequence. However, starbursts and outflows triggered by various internal and external processes can result in brief, but severe, interruptions. Processes external to galaxies, such as cosmological starvation, resulting from, e.g., entering a denser environment, being accreted as a satellite, or residing in a low density region, drive long-term changes to the star formation history and, hence, modulate galaxy colours.}
\label{fig:Cartoon}
\end{figure}

Our findings suggest that, in our simulations, the SFRs and colours of moderately massive galaxies at $z\sim{}2$ and their higher redshift progenitors are affected by two separate processes, see Fig.~\ref{fig:Cartoon}.
First, starbursts and outflows triggered by various internal and external mechanisms frequently lead to brief, but severe, deviations from the star forming sequence. These deviations occur less often at later times when galaxies are more massive and cosmological accretion and merger rates are declining. In many cases, expelled or consumed gas is replenished quickly at high redshift (e.g., \citealt{Tacchella2016}) and, hence, these ``quenching'' events tend to be short-lived. Secondly, absent additional mechanisms of removing gas from galaxies (see below), galaxies likely need to reduce the rate at which they \emph{accrete gas} from their surroundings if their star formation activity is to decrease for extended periods of time  ($>100$ Myr). Various pathways to reduce gas accretion onto galaxies have been studied in the literature (e.g., halo quenching, environmental processes) and they likely all play a role depending on the circumstances. Additional quenching processes are also likely necessary to explain the observed scaling of quiescent fraction with stellar mass at $z\lesssim{}1$.

We clearly identify both traditional satellites and a significant population of formally `central', quiescent galaxies within a few $R_{\rm halo}$ of more massive haloes, where environmental processes (e.g., ram pressure and tidal forces) likely reduce gas accretion and dominate the suppression of star formation. For the sub-population of our sample that are truly isolated centrals with more sustained ($\gtrsim{}100$ Myr) quiescent periods, we argue that their reduced growth at $z\sim{}2$ is a consequence of assembling a large fraction of their final halo and stellar mass at earlier times.

The stellar masses and star formation rates of our simulated galaxies at $z\sim{}2$ are in broad agreement with observations. We attribute this success largely to the detailed and accurate modelling of stellar feedback processes and to the high numerical resolution ($\sim{}10$ pc, $\sim{}10^{4.5}$ $M_\odot$) of our simulations. However, our sample of galaxies does not show a colour bimodality at $z\sim{}2$. Furthermore, the U--V colours of our simulated galaxies are not as red as those of many observed quiescent galaxies at those redshifts.

The absence of a clear colour bimodality is likely related to the large number of galaxies with intermediate colours in our simulations. Stellar feedback driven outflows typically do not suppress star formation long enough to turn colours sufficiently red (Fig.~\ref{fig:colorSSFR}), resulting in many galaxies with intermediate ("green") colours. Cosmological starvation helps to redden galaxy colours by reducing the overall level of star formation activity but may also move galaxies from the star forming sequence into the green valley.

AGN feedback (and/or other processes) may be important in resolving these differences. For example, the energy and momentum injection from AGNs could potentially fully suppress any residual star formation. AGN feedback may thus help to produce (and extend the duty cycle of) a larger number of galaxies with zero SFR at high z \citep{Kriek2006} as well as redden galaxy colours further, creating a visible colour bimodality. In addition, AGN feedback is likely an effective mechanism to reduce the cooling rate of gas, and thus the SFR, in massive galaxies at late times \citep{Croton2006}. Finally, galactic outflows powered by AGN could affect the overall content, chemistry, and structure of the CGM. We plan to study the role and impact of AGN feedback for massive galaxies in future work.

\acknowledgements

RF thanks the referee for suggestions that helped to improve the quality of the paper. RF was supported in part by NASA through Hubble Fellowship grant HF2-51304.001-A awarded by the Space Telescope Science Institute, which is operated by the Association of Universities for Research in Astronomy, Inc., for NASA, under contract NAS 5-26555, in part by the Theoretical Astrophysics Center at UC Berkeley, and by NASA ATP grant 12-ATP-120183. RF also acknowledges financial support from the Swiss National Science Foundation (grant no 157591). EQ was supported by NASA ATP grant 12-ATP-120183, a Simons Investigator award from the Simons Foundation, and the David and Lucile Packard Foundation. Support for PFH was provided by an Alfred P. Sloan Research Fellowship, NASA ATP Grant NNX14AH35G, and NSF Collaborative Research Grant \#1411920 and CAREER grant \#1455342. CAFG was supported by NSF through grants AST-1412836 and AST-1517491, and by NASA through grant NNX15AB22G. DK was supported by NSF grant AST-1412153 and by a Cottrell Scholar Award.
Simulations were run with resources provided by the NASA High-End Computing (HEC) Program through the NASA Advanced Supercomputing (NAS) Division at Ames Research Center, proposal SMD-14-5492. Additional computing support was provided by HEC allocations SMD-14-5189, SMD-15-5950, and NSF XSEDE allocations AST120025, AST150045. This work made extensive use of the NASA Astrophysics Data System and arXiv.org preprint server.\\

\appendix

\section{Colour bimodality}
\label{sect:ColorBimodality}

\subsection{Comparison with UltraVISTA}
\label{sect:CompUltraVISTA}

We make use of the UltraVISTA survey \citep{McCracken2012} to support our statement that the colours of the simulated galaxies are in reasonable (but far from perfect) agreement with observations. Specifically, we use version 4.1 of the K-band selected catalog\footnote{\url{http://www.strw.leidenuniv.nl/galaxyevolution/ULTRAVISTA/Ultravista/K-selected.html}} \citep{Muzzin2013d}. The full catalog contains 262615 sources. We discard stars and objects that are near bright sources (${\rm USE}=1$, ${\rm nan\_contam}=0$, ${\rm K\_flag}\leq{} 2$), and only use sources with a total K band magnitude brighter than 23.4 corresponding to the 90\% flux completeness limit. To approximately match the properties of the \MassiveFIRE{} sample we only select UltraVISTA galaxies with $1.6 < z < 2.1$ and with stellar masses in the range $10^{10} - 10^{11}$ $M_\odot$. This selection results in 7493 galaxies. While UltraVISTA is, e.g., 0.6 mag deeper than NMBS \citep{Whitaker2011d}, it is not quite mass-complete down to $M_{\rm star}\sim{}10^{10}$ $M_\odot$ at $z\sim{}2$. In fact, the 95\% mass completeness limit is $1.3\times{}10^{10}$ $M_\odot$ at $z=1.6$ and $3.1\times{}10^{10}$ $M_\odot$ at $z=2.1$. However, as we show below the large sample size of UltraVISTA helps in detecting a bimodal signal with high statistical significance. 

For the bimodality test we further restrict galaxies to fall within a range of rest-frame $U-V$ and $V-J$ colours, effectively excluding the dustiest galaxies with extremely red V-J colours that are absent in the \MassiveFIRE{} sample. This final selection ($0.2<V-J<1.2$, $0.5<U-V<2$) brings the number of UltraVISTA galaxies down to 4821.

\begin{figure}
\begin{tabular}{c}
\includegraphics[width=85mm]{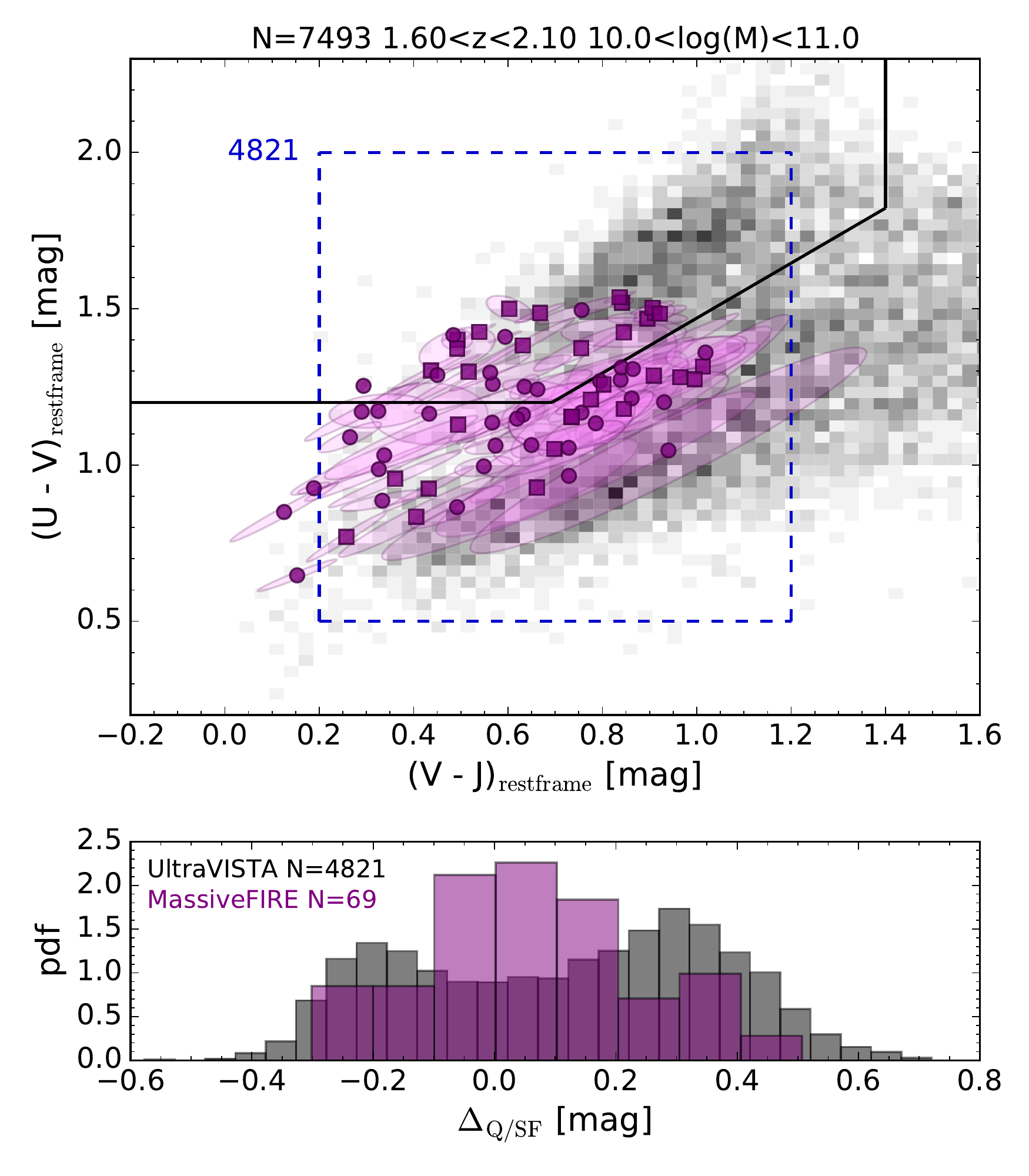}
\end{tabular}
\caption{\emph{Distribution of galaxy colours in UltraVISTA and in \MassiveFIRE{}.} (Top) Restframe U--V and V--J diagram. 
The shaded background shows the distribution of $1.6<z<2.1$ galaxies selected from the UltraVISTA catalog (see text) on a linear grey scale. Squares and circles denote the average rest-frame colours of \MassiveFIRE{} galaxies at $z=1.7$ and $z=2$ respectively. Semi-transparent ellipses show the $1-\sigma$ variations of the mean colours with changing lines-of-sights. The solid line is the dividing line between quiescent (to the top and left) and star forming (to the bottom and right) galaxies \citep{Whitaker2011d}. (Bottom) Normalized distribution of $\Delta{}_{\rm Q/SF}$, 
the distance of a source in U--V, V--J space to the quiescent vs. star forming dividing line. We use the 4821 UltraVISTA galaxies with colours that place them inside the region enclosed by the dashed line in the top panel. For \MassiveFIRE{} we 
compute the median value of $\Delta{}_{\rm Q/SF}$ (over 50 random lines of sight) for each each galaxy and we combine the $z=1.7$ and $z=2$ snapshots resulting in 69 sources. While the colours of the simulated galaxies generally overlap with the colours of observed galaxies, their statistical distributions appear to be different, see text for a more detailed discussion. In addition, UltraVISTA galaxies show a clear bimodal signal in $\Delta{}_{\rm Q/SF}$, while no such signal is apparent in our simulations. However, the lack of a strong bimodal signal is expected given the modest size of our sample, see text.}
\label{fig:UltraVistaVsMassiveFIRE}
\end{figure}

We compare the UltraVISTA and \MassiveFIRE{} samples in Fig.~\ref{fig:UltraVistaVsMassiveFIRE}. While the colours of the simulated galaxies generally overlap with the colours of observed galaxies, they are distributed differently. In particular, our simulations lack galaxies with U--V rest-frame colours above 1.6. Also, the colours of the star forming galaxies are somewhat closer ($\sim{}0.2$ mag) to the star forming vs. quiescent separation line in our simulations than in observations. 

We note, however, that this ``bye-eye'' comparison is simplistic and suffers from important systematics. First, there are systematic uncertainties of $\sim{}0.1$ mag regarding the zero-points in the various observational filters \citep{Muzzin2013d}. In addition, rest-frame fluxes are derived from integrating the best-fit spectral energy distribution. Hence, systematic effects enter via photo-metric redshift estimates and template selections. Second, our approach of accounting for dust absorption is relatively simplistic and likely introduces a non-significant systematic error. Third, we note that the U--V and V--J colours of our simulated galaxies are not precisely matched to the UltraVISTA observations. There are differences in the aperture size\footnote{Colours in UltraVISTA are measured within a 2.1 arcsecond diameter aperture (aperture radius of $\sim{}9$ kpc for galaxies in the considered redshift range), while we measure colours within a circular aperture of 5 kpc radius.}, in the adopted extinction law, and in the precise shape of the U, V, and J filter transmission curves. Fourth, our sample was selected to cover a broad range of halo growth histories (\S\ref{sect:SampleSelection}) and is thus not necessarily representative of a purely mass-selected sample. Clearly, additional work is required to fully determine the importance of the apparent differences in the colour distributions.

The presence or absence of a colour-bimodality should be somewhat less dependent on systematic colour shifts. The bottom panel in Fig.~\ref{fig:UltraVistaVsMassiveFIRE} compares the distributions of $\Delta{}_{\rm Q/SF}$, 
the distance of a source in U--V, V--J space to the quiescent vs. star forming dividing line, in our simulations with those in UltraVista. The latter show a clear bimodality, while no such bimodality is apparent in our sample of simulated galaxies. We note, however, that the UltraVISTA sample (after selecting galaxies with masses and redshifts similar to our simulated sample) is 2 orders of magnitude larger. Hence, a reasonable question is whether we should expect to see a bimodality signal given the modest size of the \MassiveFIRE{} sample.

\subsection{Testing for multi-modality in the colour distributions}
\label{sect:TestMultiModal}

We apply the dip test \citep{Hartigan1985a} to the distribution of $\Delta{}_{\rm Q/SF}$ and to the distribution of the U--V rest-frame colour in UltraVISTA. Specifically, we use the subset of 4821 UltraVISTA galaxies with redshifts, stellar masses, and colours in broad agreement with those of our simulated set of galaxies. The dip test allows us to decide whether the distributions are significantly different from unimodal distributions. The dip test compares favourably with other approaches aimed at quantifying multi-modality, see, e.g., \cite{Freeman2013}. It is available as a package\footnote{\url{https://cran.r-project.org/web/packages/diptest/index.html}} for the R Project for Statistical Computing\footnote{\url{https://www.r-project.org}}.

We first choose a sample size (ranging from 8 to 4821 objects) and then generate 1000  samples of this size drawn randomly with replacement from the selected UltraVISTA subset. We run the dip test on each of the 1000 samples and record how many samples show evidence of multi-modality, i.e., the number of samples that have a $p$ value below $\alpha=0.05$.

\begin{figure}
\begin{tabular}{c}
\includegraphics[width=85mm]{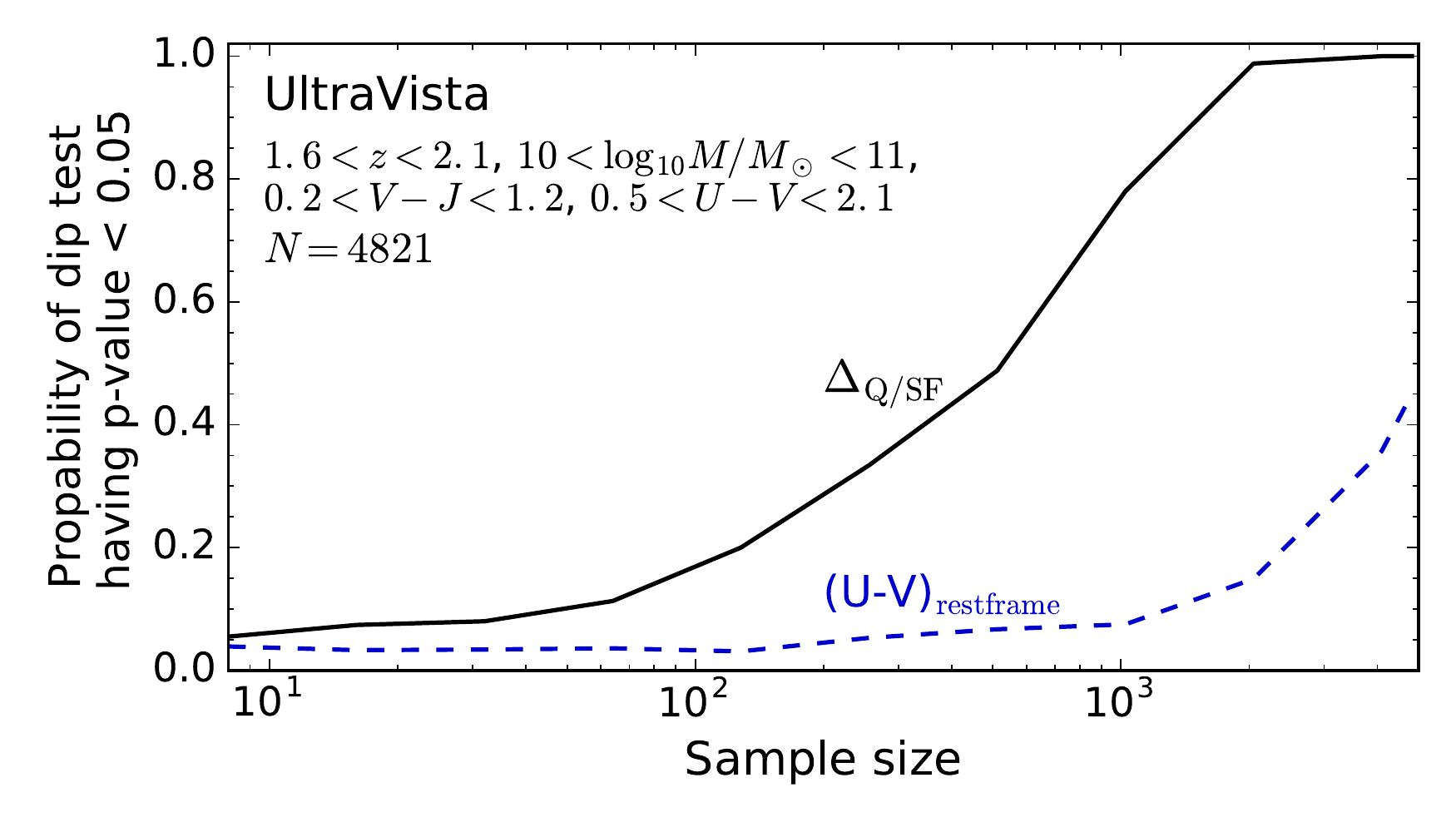}
\end{tabular}
\caption{\emph{Probability of detecting bimodality in the $\Delta{}_{\rm Q/SF}$ and U--V distributions at a significance level of $\alpha=0.05$ as function of sample size.} It requires many hundreds (for $\Delta{}_{\rm Q/SF}$) or thousands (for U--V) of galaxies drawn randomly from the UltraVISTA parent distribution to find a deviation from uni-modality with high statistical significance (for samples with masses, colours, and redshifts similar to our simulated sample).}
\label{fig:UltraVistaBimodality}
\end{figure}

Fig.~\ref{fig:UltraVistaBimodality} shows how the sample size affects the likelihood of detecting a multi-modality in the $\Delta{}_{\rm Q/SF}$ and U--V distributions. Samples containing several hundreds (several thousands) of galaxies offer a 50\% chance of confirming a multi-modal $\Delta{}_{\rm Q/SF}$  (U--V) distribution at the $\alpha=0.05$ significance level. The size of the \MassiveFIRE{} sample ($\sim{}70$ galaxies if we combine the $z=1.7$ and $z=2$ redshifts) is too small to reliably detect a bimodality, even if our simulated galaxies had colour distributions that mirrored those of UltraVISTA.

According to Fig.~\ref{fig:UltraVistaBimodality}, it is significantly easier to detect bimodality in the $\Delta{}_{\rm Q/SF}$ distribution than in the U--V rest-frame colour distribution. Clearly, a careful choice of the observable can significantly boost the likelihood of detecting a bimodality in the underlying data set.

\section{Additional Tables}
Table~\ref{tab:GalPropsz17} is similar to Table~\ref{tab:GalPropsz20} but shows the properties of our simulated galaxies at $z=1.7$. Table~\ref{tab:Res} provides results from a linear regression analysis of the stellar mass -- halo mass relation, supplementing the discussion in \S\ref{sect:SHMR}. 

\begin{table*}
\begin{center}
\begin{tabular}{lccccccccr}
\tableline\tableline
Name & Central/Satellite & $\log_{10} M_{\rm 180m}$  & $R_{\rm halo}$ & $\log_{10} M_{\rm star}$ & SFR & $\log_{10}$ sSFR & U-V & V-J & $f_{\rm Q}$ \\
          &  & ($\log_{10} M_\odot$)            & (kpc)                 &  ($\log_{10} M_\odot$)    & ($M_\odot$ yr$^{-1}$) & ($\log_{10}$ yr$^{-1}$) & (mag) & (mag) & (\%) \\  \tableline  \vspace{-0.4cm}  \\ 
\tableline
A1:0 & Central & 12.44 & 171.1 & 10.47 & 11.5 & -9.31 & 1.467 & 0.913 & 84 \\
A2:0 & Central & 12.62 & 197.1 & 10.66 & 19.2 & -9.19 & 1.314 & 1.018 & 0 \\
A3:0 & Central & 12.45 & 173.0 & 10.31 & 0.5 & -10.45 & 1.297 & 0.422 & 100 \\
A4:0 & Central & 12.52 & 182.4 & 10.46 & 6.0 & -9.43 & 1.259 & 0.785 & 38 \\
A5:0 & Central & 12.45 & 172.2 & 10.34 & 0.0 & -11.84 & 1.422 & 0.534 & 100 \\
A6:0 & Central & 12.58 & 190.9 & 10.47 & 8.6 & -9.38 & 1.478 & 0.917 & 100 \\
A7:0 & Central & 12.85 & 235.3 & 10.55 & 16.2 & -8.97 & 1.315 & 0.994 & 0 \\
A8:0 & Central & 12.82 & 229.9 & 10.48 & 13.2 & -8.84 & 0.895 & 0.650 & 0 \\
A9:0 & Central & 12.87 & 238.7 & 10.07 & 1.6 & -9.69 & 1.373 & 0.613 & 100 \\
A9:1 & Satellite & 11.99 & 75.7 & 10.40 & 42.3 & -8.67 & 0.732 & 0.223 & 0 \\
A10:0 & Central & 12.77 & 220.9 & 10.64 & 30.3 & -8.79 & 0.928 & 0.318 & 0 \\
B1:0 & Central & 12.96 & 255.7 & 11.00 & 27.0 & -9.32 & 1.285 & 0.897 & 8 \\
B2:0 & Central & 13.09 & 283.1 & 10.92 & 3.9 & -10.10 & 1.519 & 0.844 & 100 \\
B3:0 & Central & 13.16 & 297.1 & 10.91 & 51.5 & -8.88 & 0.891 & 0.387 & 0 \\
B3:1 & Satellite & 10.79 & 21.9 & 9.62 & 0.0 & -11.92 & 1.409 & 0.484 & 100 \\
B4:0 & Central & 12.92 & 248.3 & 10.68 & 22.8 & -8.95 & 1.294 & 1.022 & 0 \\
B4:1 & Central & 12.20 & 142.4 & 10.14 & 12.3 & -8.52 & 0.844 & 0.390 & 0 \\
B4:2 & Central & 12.02 & 124.5 & 10.19 & 0.5 & -10.32 & 1.304 & 0.512 & 86 \\
B4:3 & Satellite & 11.90 & 51.0 & 10.06 & 0.3 & -10.62 & 1.477 & 0.649 & 100 \\
B5:0 & Central & 13.01 & 265.0 & 10.84 & 17.5 & -9.29 & 1.171 & 0.843 & 0 \\
B5:1 & Satellite & 11.12 & 21.8 & 9.64 & 0.0 & -12.56 & 1.512 & 0.590 & 98 \\
Cm3:0 & Central & 13.78 & 478.6 & 11.68 & 112.7 & -9.37 & 1.164 & 0.698 & 12 \\
Cm3:1 & Satellite & 12.48 & 67.2 & 11.28 & 35.4 & -9.68 & 1.428 & 0.817 & 84 \\
Cm3:3 & Satellite & 11.96 & 94.9 & 10.67 & 57.7 & -8.84 & 1.213 & 0.768 & 4 \\
Cm3:4 & Satellite & 11.53 & 35.7 & 10.61 & 15.5 & -9.45 & 1.505 & 0.905 & 100 \\
Cm3:5 & Satellite & 11.84 & 62.5 & 10.21 & 11.9 & -9.10 & 1.054 & 0.654 & 0 \\
Cm3:6 & Satellite & 11.50 & 51.8 & 10.06 & 5.1 & -9.34 & 1.027 & 0.336 & 22 \\
Cm3:7 & Central & 11.64 & 92.8 & 10.07 & 2.7 & -9.46 & 1.380 & 0.764 & 100 \\
Cm3:8 & Satellite & 11.32 & 72.2 & 10.32 & 1.9 & -10.04 & 1.536 & 0.837 & 100 \\
Cm3:9 & Satellite & 11.56 & 87.3 & 10.50 & 8.0 & -9.52 & 1.484 & 0.924 & 100 \\
Cm3:10 & Satellite & 10.78 & 27.4 & 10.00 & 0.0 & -12.54 & 1.404 & 0.485 & 98 \\
\tableline
\end{tabular}
\caption{\emph{Properties of \MassiveFIRE{} galaxies at $z=1.7$.} Columns refer to the same quantities as in Table~\ref{tab:GalPropsz20}.}
\label{tab:GalPropsz17}
\end{center}

\begin{center}
\begin{tabular}{cccccccc}
\tableline\tableline
 redshift &  $\log_{10} M_{\rm 180m}$  & $\langle{}\log_{10} M_{\rm 180m}\rangle{}$ & $\langle{}\log_{10} M_{\rm star}\rangle{}$ & slope &  scatter & remark\\
              & ($\log_{10} M_\odot$)          &  ($\log_{10} M_\odot$)                            & ($\log_{10} M_\odot$)                &         &  (dex) & \\  \tableline  \vspace{-0.4cm}  \\ 
\tableline
9.2       & $9.0 - 11.0$ & 9.90 & 6.76 & 1.96$^{+0.18}_{-0.15}$ & 0.42$^{+0.06}_{-0.04}$ &  \\
7.1       & $9.5- 11.5$ & 10.50 & 7.87 & 1.58$^{+0.13}_{-0.11}$ & 0.26$^{+0.04}_{-0.01}$ &  \\
5.1       & $10.0 - 12.0$ & 11.15 & 8.66 & 1.26$^{+0.11}_{-0.10}$ & 0.26$^{+0.03}_{-0.02}$ & \\
3.0       & $10.5 - 13.0$ & 12.06 & 9.89 & 0.89$^{+0.15}_{-0.13}$ & 0.25$^{+0.04}_{-0.02}$ & \\
2.0       & $11.5 - 14.0$ & 12.59 & 10.51 & 0.91$^{+0.13}_{-0.10}$ & 0.23$^{+0.05}_{-0.02}$ & \\
2.0       & $11.5 - 14.0$ & 12.47 & 10.32 & 0.93$^{+0.13}_{-0.10}$ & 0.15$^{+0.12}_{-0.02}$ & quiescent \\
2.0       & $11.5 - 14.0$ & 12.63 & 10.57 & 0.89$^{+0.13}_{-0.10}$ & 0.24$^{+0.05}_{-0.02}$ & star forming \\
1.7       & $12.0 - 14.0$ & 12.80 & 10.66 & 0.88$^{+0.08}_{-0.08}$ & 0.19$^{+0.08}_{-0.04}$ & \\
1.7       & $12.0 - 14.0$ & 12.56 & 10.39 & 0.42$^{+0.52}_{-0.31}$ & 0.21$^{+0.15}_{-0.02}$ & quiescent \\
1.7       & $12.0 - 14.0$ & 12.95 & 10.81 & 0.97$^{+0.07}_{-0.04}$ & 0.12$^{+0.03}_{-0.01}$ & star forming \\
\tableline
2.0       & $12.3 - 12.6$ & 12.45 & 10.28 & -- & 0.17$^{+0.04}_{-0.02}$ & \\
2.0       & $12.3 - 12.6$ & 12.45 & 10.26 & -- & 0.16$^{+0.11}_{-0.02}$ & quiescent \\
2.0       & $12.3 - 12.6$ & 12.45 & 10.29 & -- & 0.18$^{+0.05}_{-0.02}$ & star forming \\
2.0       & $12.8 - 13.2$ & 12.97 & 10.86 & -- & 0.21$^{+0.09}_{-0.04}$  & \\
\tableline
\end{tabular}
\caption{\emph{Properties of the SHMR of central galaxies for various redshifts and halo mass ranges (first two columns).} Columns 3 and 4 show the average of the logarithms of the halo and stellar masses, respectively. The fifth column shows the slope of the linear regression: $\log_{10} M_{\rm star} = {\rm slope}\,(\log_{10} M_{\rm 180m} - \langle{}\log_{10} M_{\rm 180m}\rangle{}) + \langle{}\log_{10} M_{\rm star}\rangle{}$. No regression is performed for the last four rows. The penultimate column denotes the scatter of the logarithm of the stellar mass for the given halo mass range. The scatter in the top ten rows is measured \emph{relative to the linear regression line}, while the bottom four rows report the sample standard deviation of $\log_{10} M_{\rm star}$ in narrow bins of halo masses at $z=2$. The final column denotes whether results are shown for a subset (quiescent, star forming) of the total set of central galaxies. Confidence intervals (1-$\sigma$) are computed via bootstrapping.}
\label{tab:Res}
\end{center}
\end{table*}

\clearpage

\end{document}